\documentclass[%
 aip,
 jmp,%
 amsmath,amssymb,
 reprint,%
]{revtex4-2}

\usepackage{graphicx}
\usepackage{dcolumn}
\usepackage{bm}
\usepackage{newcomh}
\usepackage[caption=false]{subfig}
\usepackage{float}

\begin{document}

\preprint{arxiv}

\title[New physics in double Higgs production at NLO]{New physics in double Higgs production at NLO}

\author{Bo-Yan Huang}%
 \email{bhuang9@uic.edu}
\affiliation{ 
Physics Department, University of Illinois at Chicago.
}

\date{\today}

\begin{abstract}
After observing the Higgs boson by the ATLAS and CMS experiments at the LHC, accurate measurements of its properties, which allow us to study the electroweak symmetry breaking mechanism, become a high priority for particle physics. The most promising of extracting the Higgs self-coupling at hadron colliders is by examining the double Higgs production, especially in the $\bbgg$ channel.
In this work, we presented full loop calculation for both SM and New Physics effects of the Higgs pair production to next-to-leading-order (NLO), including loop-induced processes $gg\to HH$, $gg\to HHg$, and $qg \to qHH$. We also included the calculation of the corrections from diagrams with only one QCD coupling in $qg \to qHH$, which was neglected in the previous studies. 
With the latest observed limit on the HH production cross-section, we studied the constraints on the effective Higgs couplings for the LHC at center-of-mass energies of 14 TeV and a provisional 100 TeV proton collider within the Future-Circular-Collider (FCC) project.
To obtain results better than using total cross-section alone, we focused on the $\bbgg$ channel and divided the differential cross-section into low and high bins based on the total invariant mass and $p_{T}$ spectra. The new physics effects are further constrained by including extra kinematic information. However, some degeneracy persists, as shown in previous studies, especially in determining the Higgs trilinear coupling. Our analysis shows that the degeneracy is reduced by including the full NLO corrections.
\end{abstract}

\maketitle


\section{\label{sec:intro}Introduction}
In 2012, a new scalar resonance \cite{Aad:2012tfa,Chatrchyan:2012ufa} with a mass of $125.09\pm0.24$ GeV
\cite{Khachatryan:2016vau} was discovered at the Large Hadron Collider (LHC). After analyzing all the Run I data, the Standard Model (SM) Higgs boson provides best explanation for the measured properties of the new particle~\cite{Higgs:1964ia,Higgs:1964pj,Higgs:1966ev,Englert:1964et,Guralnik:1964eu,Kibble:1967sv}. Since then, high priority analyses at the Large Hadron Collider (LHC) always include the detailed study of the properties of this particle.

Theoretical uncertainties limit the reachable accuracies at the LHC. However, a wider range of Higgs couplings investigated at the LHC, and the increase of the variety of processes that involves the Higgs boson can partially compensate for this restriction. Currently, the most constrained condition is the gauge-Higgs coupling $C_v \equiv g = 0.94 +0.11$, which is very close to the SM expectation. Furthermore, due to the fact that the observed Higgs candidate particle is produced at roughly the SM rate, the extensions of the Higgs sector beyond the Standard Model are extremely constrained.
A simple model with a fourth generation of heavy quarks, for example, is excluded by the limits on Higgs production for any Higgs mass below around 600~GeV~\cite{ATLAS-CONF-2011-135,SM4-LHC} since such model predicts large deviations in the Higgs production rates from SM value~\cite{Anastasiou:2010bt,Anastasiou:2011qw,Djouadi:2012ae,Denner:2011vt,Eberhardt:2012gv}.
Unlike $C_v$, the Yukawa couplings of top-Higgs and bottom-Higgs are not constrained precisely by the data up to date. Moreover, they are within $30-40\%$ of the SM expectations \cite{hdecay,hdecay_2}.

Testing the Higgs boson’s self-interactions is particularly interesting. It is the only unmeasured experimentally property of the Higgs boson and provides the only window to probe the Higgs scalar potential, which is the origin of spontaneous symmetry breaking of the gauge symmetry and the origin of the particle masses in the Standard Model.


One of the most promising probes for LHC is the Higgs pair production. These processes provide direct measurements of the trilinear Higgs self-couplings at leading-order. Moreover, these processes complement indirect effects caused by the self-interactions of Higgs bosons in single-Higgs processes and radiative corrections to electroweak observables~\cite{Degrassi:2016wml,Degrassi:2017ucl} contaminated by possible interference effects with different models of New Physics.

Unfortunately, the Standard Model expectation of this production rate is only 0.034 pb at the Large Hadron Collider with CM energy equal to 14 TeV \cite{Baglio:2020ini}. One of the reasons for such a low rate is that the SM contributions from the box diagram and the triangle diagram (shown in Fig.~\ref{fig:gghh_feyndiagrams}(a) and (b), respectively) interfere destructively near kinematic threshold \cite{Li:2013rra}. However, the Standard Model cross-section rises dramatically to 1.54 pb at a future 100 TeV proton-proton collider since the luminosity increases in the parton distribution function of gluon at lower $x$, the Bjorken scale, this provides a chance to measure the Higgs self-couplings precisely \cite{Baglio:2012np,Yao:2013ika}.
 
Four major classes of processes are responsible for the production of Higgs pair at hadron colliders. First, we have $gg \to HH$, the gluon fusion process, with a loop of heavy quark, which has a strong coupling to the Higgs boson~\cite{pp-ggHH-LO00,Glover:1987nx,pp-ggHH-LO,Plehn:1996wb}. The second class is $qq' \to qq'V^* V^* \to qq'HH$ ($V=Z, W$), the vector bosons fusion (VBF) processes, which generate two jets and two Higgs bosons in the final state~\cite{pp-ggHH-LO00,pp-VVHH,pp-VVHH_2,pp-VVHH_3,pp-VVH-Abas}. The third class is $q\bar{q}' \to V^* \to VHH$ ($V=Z,W$), the double Higgs--strahlung process, where a vector boson, $W$ or $Z$, radiates the Higgs bosons~\cite{Barger:1988}. The last one has associated producing a pair of top quarks with two Higgs bosons, $pp \to t\bar t HH$~\cite{pp-HHtt}.

Compared to single Higgs production, these processes have at least two orders of magnitude smaller production cross-sections as the phase space is small since the final state consists of two heavy particles. They have electroweak couplings of higher-order. Besides, other topologies which are irrelevant to the trilinear Higgs coupling, where the gauge boson or fermion lines radiate both Higgs bosons, which produce the same final state as the diagrams with $H^* \to HH$ splitting. Thus, these topologies pollute the correlation between the $g_{H^3}$ coupling and the double Higgs production rate. It is extremely difficult to measuring the trilinear Higgs coupling, and very high energies along with very high collider luminosities are therefore required.

\subsection{Next-to-Leading Order and Beyond Standard Model}
It is almost impossible to measure the quartic Higgs coupling, $g_{H^4}$, in the near future as an extra $v$ further suppresses it in the denominator compared to the trilinear Higgs self-coupling, and the smallness of the triple-Higgs production rate prohibit it from being probed directly~\cite{Plehn:2005nk,Binoth:2006ym,Fuks:2015hna,deFlorian:2016sit,deFlorian:2019app}\footnote{The quartic Higgs coupling is indirectly constrained by Higgs pair production ~\cite{Liu:2018peg,Bizon:2018syu,Borowka:2018pxx}.}. We can directly measure the trilinear Higgs coupling through Higgs-pair production, where Higgs pairs are dominantly produced in the gluon-fusion process mediated mainly by top-quark loops while the contribution of $b$-quark loops is negligible. Two types of diagrams, triangle, and box, contribute to the gluon-fusion process $gg \to HH$, where the triangle diagrams involve the trilinear Higgs coupling, and the interference between the one-loop box and triangle diagrams are destructive~\cite{Glover:1987nx,Plehn:1996wb}. The dominant contributions to the cross-section come from the box diagrams. The approximate relation, $\Delta\sigma/\sigma \sim -\Delta g_{H^3}/g_{H^3}$, gives a rough estimate of the correlation of the size of the trilinear Higgs self-coupling in the vicinity of the SM value of $g_{H^3}$ and the cross-section. Therefore, small uncertainties of the relevant cross-section, which can be achieved by calculating higher-order corrections, are required to determine the trilinear Higgs coupling. The next-to-leading order (NLO) QCD corrections~\cite{Borowka:2016ehy,Borowka:2016ypz,Baglio:2018lrj} and next-to-next-to-leading order
(NNLO) corrections, which adopt heavy top quark approximation~ \cite{deFlorian:2013uza,deFlorian:2013jea,Grigo:2014jma}, are fully known. The NLO corrections are significant, and therefore must be included. In comparison with NLO corrections, the NNLO contributions are much smaller but still considerable. The QCD next-to-next-to-next-to-leading order (N$^3$LO) corrections to the effective couplings of Higgs and Higgs-pair to gluons are recently computed in heavy top quark approximation limit~\cite{Spira:2016zna} and lead to a minor modification to the cross-section \cite{Banerjee:2018lfq,Chen:2019lzz,Chen:2019fhs}. The LO contributions and the higher-order corrections contribute equally to the total production rate. Lately, the NLO results have been matched to parton showers \cite{Heinrich:2017kxx,Jones:2017giv}, and the NLO mass effects with the additional top-mass effects in the double-real corrections have been merged with the full NNLO QCD results in the heavy-top limit~\cite{Grazzini:2018bsd}. The full NLO QCD corrections to the Higgs-pair production rate with the anomalous trilinear Higgs self-coupling effects have been calculated in  Ref.~\onlinecite{Baglio:2020ini}. In this work, we calculated full NLO results, including weak interaction contributions from $qg\to HHq$.

Although verifying that a scalar vev spontaneously breaks the electroweak symmetry is crucial, discovering new physics beyond the SM is always the final goal.
Multiple new physics that could potentially affect this specific channel must be considered while analyzing the double Higgs production. One possible new physics from a new diagram involving the anomalous quartic coupling, $HHt\bar{t}t$, as shown in Fig.~\ref{fig:gghh_feyndiagrams}(c) could give significant effect \cite{Dib:2005re,Grober:2010yv, 1506.03302}. The presence of this quartic coupling makes the total production rate insensitive to the Higgs self-coupling and makes measuring this coupling incredibly difficult~\cite{Contino:2012xk}.

\subsection{Recent Searches in a Rare Particle Decay}
The most significant double Higgs decay channel in the important low-mass region is the bottom quark pair plus photons pair channel, $HH \to \gamma\gamma b\bar{b}$. Recently, new analysis techniques for searching this rare process have been developed by physicists in ATLAS collaboration. To optimize the sensitivity to the self-coupling of Higgs bosons, they first split the $pp$ collision events into low and high invariant mass groups. After that, they used a multivariate discriminant (Boosted Decision Tree) to separate the events that can be categorized as the $HH \to \gamma\gamma b\bar{b}$ process from those that can not. Finally, the Higgs-pair production rate is determined first, and then they observed how the production rate varies as a function of the Higgs self-coupling to its SM value ratio $\lambda / \lambda_{SM}$. By using the above procedures, the ATLAS team constrained the Higgs self-coupling and allowed it to vary between $-1.5$ and $6.7$ times the SM value. Physicists, therefore, can set a currently best limit on the Higgs pair production rate of 4.1 times the SM value. 

However, the work is far from being done. A huge amount of data is required to precisely measure the Higgs self-coupling and see if it were close to its SM value. The High-Luminosity upgrade of the LHC, scheduled to be operational in the late 2020s, is planned to operate at higher collision energy and deliver a dataset 20 times larger than used in this analysis. The Higgs pair production will be observed in this huge dataset if the Higgs pair production indeed behaves as predicted by the Standard Model, and a more quantitative statement will be made on the strength of the Higgs self-coupling.

\subsection{Overview of This Thesis}
This paper aims to learn how multiple new physics effects interplay in different kinematic distributions and the total cross-section at next-to-leading order. To research the topic more thoroughly, we also study the distributions of differential cross-sections, especially the invariant mass of the Higgs pair, $m_{HH}$, and the transverse momentum $p_T$.
We study the LHC at center-of-mass energies of 14 TeV and a planning 100 TeV $pp$ collider in the project of Future-Circular-Collider (FCC)~\cite{Abada:2019lih,Benedikt:2018csr}.

This thesis is organized as follows. 
In Section~\ref{sec:loxs}, we present the conventions and notations.  We present the details of our calculation at LO and NLO in Section~\ref{sec:nlo}. Then we study the influence of the new physics effects on the kinematic distributions in Section~\ref{chap:numerical}, and a numerical study on constraints using the kinematic information in a $100$ TeV proton-proton collider. Finally, the conclusions are given in Section~\ref{chap:conclusion}.

\section{Leading-order Cross-section}
\label{sec:loxs}
SM contributions to calculations of Higgs pair production have been made a while back in Ref.~\onlinecite{Glover:1987nx,Plehn:1996wb}. Also, the extra contribution from the anomalous $HHtt$ coupling has been studied in Ref.~\onlinecite{Dib:2005re,Grober:2010yv}. 
\begin{figure}[hbt!]
\centering
\includegraphics[width=0.95\textwidth, angle=0]{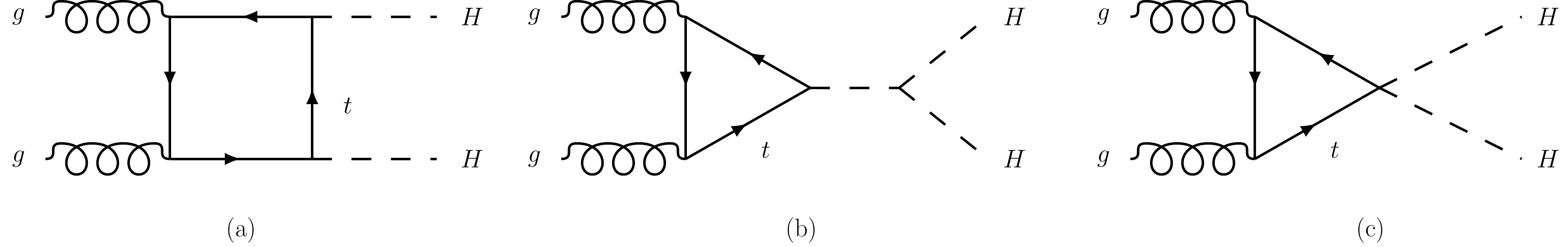} 
\caption{\label{fig:gghh_feyndiagrams}\emph{Feynman diagrams for loop-induced Higgs pair production through gluon fusion. Diagrams \emph{(a), \emph{b} are SM diagrams, where \emph{(c)} is BSM diagram with anomalous $HHtt$ coupling.}}
}
\end{figure}
At the leading-order, the production of the Higgs pair through gluon fusion is shown in Fig.~\ref{fig:gghh_feyndiagrams}, including each permutation of the external lines. The box diagram has no Higgs self-coupling, and the triangle diagram involves the Higgs trilinear coupling. We can write the matrix element of $g(p_1) g(p_2) \to H(p_3) H(p_4)$ at LO as
\begin{eqnarray}
\mathcal{M}(g^a g^b \to HH) & = & -i\,\frac{\alpha_s(\mu_R) G_F m_{HH}^2}{2\sqrt{2}\pi}
\mathcal{A}^{\mu\nu} \epsilon_{1\mu} \epsilon_{2\nu} \delta_{ab}
\nonumber \\[0.3cm]
\mbox{with} \qquad \mathcal{ A}^{\mu\nu} & = & F_1 T_1^{\mu\nu} + F_2 T_2^{\mu\nu}
\, , \nonumber \\[0.3cm]
F_1 & = & \left(g_{H^3} \frac{1}{\hat{s}-m_H^2}\ g_{Htt} + g_{HHtt} \right)\frac{v^2}{m_t} F_\triangle + g_{Htt}^2\frac{v^2}{m_t^2} F_\Box \, , \\
F_2 & = & g_{Htt}^2\frac{v^2}{m_t^2} G_\Box \, , \nonumber 
\label{eq:lomat}
\end{eqnarray}

where $m_{HH}$ is the invariant mass of the Higgs pair, $a,b$ are the color indices of the initial gluons, $\alpha_s(\mu_R)$ is the strong coupling evaluated at the renormalization scale $\mu_R$,
and $G_F$ is the Fermi constant. 

The couplings $g_{H^3}$, $g_{Htt}$ and $g_{HHtt}$ denote the trilinear Higgs self-coupling, the top-Higgs coupling, and the anomalous nonlinear $HHtt$ coupling, respectively. 
The Lagrangian that involves these couplings reads
\beq
\label{eq:lagcoupling}
 \frac1{3!} g_{H^3}\, H^3 + g_{Htt}\, H\bar{t}t + \frac1{2!} g_{HHtt}\, H^2 \bar{t}t \ .
\eeq
Therefore in the SM we have

\beq
g_{H^3}^{(SM)} = \frac{3m_H^2}{v} \ , \qquad g_{Htt}^{ (SM)} = \frac{m_t}{v}\ , \qquad g_{HHtt}^{(SM)}= 0\ ,
\eeq

where $v=246$ GeV is the vacuum expectation value of the Higgs field.

The contributions of the two tensor structures, $T_{1}^{\mu\nu}$, correspond to the
total angular-momentum states with $S_z=0$ while $T_{2}^{\mu\nu}$ corresponds to $S_z=2$,
\begin{eqnarray}
T_1^{\mu\nu} & = & g^{\mu\nu}-\frac{p_1^\nu p_2^\mu}{(p_1\cdot p_2)}\, , \nonumber \\[0.3cm]
T_2^{\mu\nu} & = & g^{\mu\nu}+\frac{M_H^2 p_1^\nu p_2^\mu}{p_T^2 (p_1 \cdot p_2)}
-2\frac{(p_2 \cdot p_3) p_1^\nu p_3^\mu}{p_T^2 (p_1\cdot p_2)}
-2\frac{(p_1 \cdot p_3) p_3^\nu p_2^\mu}{p_T^2 (p_1\cdot p_2)}
+2\frac{p_3^\nu p_3^\mu}{p_T^2} \nonumber \\[0.3cm]
\mbox{with} \quad p_T^2 & = & 2 \frac{(p_1 \cdot p_3)(p_2 \cdot p_3)}{(p_1 \cdot p_2)} -
M_H^2 \, ,
\end{eqnarray}
where $p_T$ is the transverse momentum of each Higgs boson in the final-state. 

Here we follow the notations used in Ref.~\onlinecite{Plehn:1996wb}, the form factors for $g g \rightarrow H H$ are

\begin{eqnarray*}
\label{eq:gghh_loopfn}
F_\triangle & = & \frac{2}{S} [2+(4-S) m_t^2 C_{12}] \\ [0.3cm]
F_\Box & = & \frac{1}{S^2} \big\{4\,S+8 S\, m_t^2 \, C_{12} -2S \, mt^4\, (S-8+2 R_{H/t}^2)(D_{123}+D_{213}+D_{132})\nonumber \\[0.2cm]
&&+(2 R_{H/t}^2-8)m_t^2 [\bar{T}(C_{13}+C_{24})+\bar{U}(C_{23}+C_{14}) \\[0.2cm]
&&-(
T U -R_{H/t}^4)m_t^2 D_{132}] \big\} \nonumber\\[0.3cm]
G_\Box & = & \frac{1}{S(T U - R_{H/t}^2)}\big\{(T^2-8T+R_{H/t}^2)m_t^2 (C_{12}+\bar{T}(C_{13}+C_{24})-S T m_t^2 D_{213} )\nonumber \\[0.2cm]
&&+m_t^2 (U^2 -8U +R_{H/t}^2) (S C_{12}+\bar{U}(C_{23}+C_{14})-S U m_t^2 D_{123})\\
&&-m_t^2(T^2+U^2-2R_{H/t}^2)(T+U-8)C_{cd}\nonumber\\
&&-2m_t^2(TU-R_{H/t}^2)(D_{123}+D_{213}+D_{132})
 \big\}, \nonumber
\end{eqnarray*}

where 

\begin{displaymath}
\hat s = (p_1+p_2)^2, \hspace{1.0cm}
\hat t = (p_3-p_1)^2, \hspace{1.0cm}
\hat u = (p_3-p_2)^2
\end{displaymath}
\begin{displaymath}
S = {\hat s}/m_t^2, \hspace{1.0cm}
T = {\hat t}/m_t^2, \hspace{1.0cm}
U = {\hat u}/m_t^2
\end{displaymath}
\begin{displaymath}
R_{H/t} = m_H^2/m_t^2, \hspace{1.0cm}
\bar{T} = T -R_{H/t}, \hspace{1.0cm}
\bar{U} = U -R_{H/t}, \hspace{1.0cm},
\end{displaymath}
and the scalar integrals:
\begin{eqnarray*}
C_{ij} & = & \int \frac{d^4q}{i\pi^2}~\frac{1}
{(q^2-m_Q^2)\left[ (q+p_i)^2-m_Q^2\right]
\left[(q+p_i+p_j)^2-m_Q^2\right]} \\ \\
D_{ijk} & = & \int \frac{d^4q}{i\pi^2} \frac{1}
{(q^2-m_Q^2)\left[(q+p_i)^2-m_Q^2\right]
\left[(q+p_i+p_j)^2-m_Q^2\right]\left[ (q+p_i+p_j+p_k)^2-m_Q^2\right]}
\end{eqnarray*}

Notice that the loop function of the single Higgs production from the gluon fusion appears again in both Fig.~\ref{fig:gghh_feyndiagrams}(b) and Fig.~\ref{fig:gghh_feyndiagrams}(c). Only the loops involving SM top quark are considered in this work due to the smallness of Higgs couplings to other quarks.

For the three diagrams in Fig.~\ref{fig:gghh_feyndiagrams}, we can therefore express the partonic differential cross-section as
\bea
\label{eq:amp}
&&\frac{d\hat{\sigma}(gg\to HH)}{d\hat{t}} = \frac{G_F^2 \alpha_s^2}{512(2\pi)^3}\nonumber \\
&& \qquad \times \left[ \left|\left(g_{H^3} \frac{1}{\hat{s}-m_H^2}\ g_{Htt} + g_{HHtt} \right) \frac{v^2}{m_t} F_\triangle + g_{Htt}^2\frac{v^2}{m_t^2} F_\Box \right|^2 + \left| g_{Htt}^2\frac{v^2}{m_t^2} G_\Box \right|^2 \right] \ .
\eea
In the SM Eq.~(\ref{eq:amp}) reduces to 
\bea
\label{eq:ampSM}
 \frac{G_F^2 \alpha_s^2}{512(2\pi)^3}\left[ \left|\frac{3m_H^2}{\hat{s}-m_H^2}  F_\triangle + F_\Box \right|^2 + \left| G_\Box \right|^2\right] \ .
\eea
We can parameterize Eq.~(\ref{eq:amp}) with three dimensionless coefficients
\bea
\label{eq:ampara}
\frac{d\hat{\sigma}(gg\to HH)}{d\hat{t}} &=& \frac{G_F^2 \alpha_s^2}{512(2\pi)^3}\left[ \left|\left(c_{3H} \frac{3m_H^2}{\hat{s}-m_H^2} + c_{HHtt} \right) F_\triangle + c_{Htt} F_\Box \right|^2 + \left| c_{Htt} G_\Box \right|^2\right] \ .
\eea
In SM, these coefficients reads
\beq
c_{3H}^{(SM)}= 1 \ , \qquad c_{Htt}^{(SM)} = 1 \ , \qquad c_{HHtt}^{(SM)} = 0 \ .
\eeq
The definition of these coefficients are\footnote{Comparing the notations in Ref.~\onlinecite{Chen:2014xra} to our result, we have $ c_{3H} = c_{\triangle}$, $c_{Htt} = c_{\Box}$, and $c_{HHtt} = c_{nl}$.}
\beq
 c_{3H} = g_{H^3}\,g_{Htt}\, \frac{v^2}{3m_H^2 m_t}\ , \qquad c_{HHtt} = g_{HHtt}\,\frac{v^2}{m_t} \ , \qquad c_{Htt} = \left(g_{Htt}\,\frac{v}{m_t}\right)^2 \ .
 \eeq
Only gauge-invariant operators of dimension-6 or higher lead to new physics effects of low-energy Higgs observables in the effective theory framework. We expect the importance of operators with mass dimensions greater than four to become less for lower energy scale. For dimension-6 operators, we have
\beq
\label{eq:power}
\delta c_{3H, Htt, HHtt} \sim \mathcal{O}\left(\frac{v^2}{\Lambda_{np}^2}\right) \, ,
\eeq
where $v=246$ GeV, and $\Lambda_{np}$ denotes the generic scale of new physics. A bottom-up approach is adopted in this work while $c_{3H}$, $c_{Htt}$, and $c_{HHtt}$ are allowed to vary, without the constraints of the power counting in Eq.~(\ref{eq:power}).

 Eq.~(\ref{eq:ampara}) is a quite general expression and includes new physics effects from various models. 
Provided that there are fermions with new color while coupled to the Higgs strongly. In that case, we can include the contributions from these new colors to $gg\to HH$ by applying the mass eigenvalues in the loop functions and calculating the Higgs couplings from the eigenbasis of masses.
It is known that the $m_t\to \infty$ limit gives good approximations in $F_\triangle$ but works terribly in $F_\Box$ and $G_\Box$ \cite{Gillioz:2012se,Dawson:2012mk}.
It is known that the $m_t\to \infty$ limit gives good approximations in $F_\triangle$ but works terribly in $F_\Box$ and $G_\Box$ \cite{Gillioz:2012se,Dawson:2012mk}. Roughly speaking, this is because the partonic CM energy is the Higgs pair invariant mass, $\hat{s}$, and is always above $4m_h^2$, the kinematic threshold, while the relation $\hat{s} \ll4m_t^2$ is required in the low-energy Higgs theorems~\cite{Ellis:1975ap}. 
Therefore, the complete mass dependence must be kept in the loop functions for scenarios with new colored particles, which have been studied thoroughly in Ref.~\onlinecite{Dib:2005re,Dawson:2012mk}.

\section{Corrections up to next-to-leading-order } \label{sec:nlo}
Generically, the cross-section of double Higgs production up to next-to-leading-order can be 
expressed as \cite{Baglio:2020ini} 
\begin{eqnarray}
\sigma_{NLO}(pp \rightarrow H H + X) & = &
\sigma_{LO} + \Delta
\sigma_{virt} + \Delta\sigma_{gg} + \Delta\sigma_{gq} +
\Delta\sigma_{q\bar{q}} \, . \nonumber 
\end{eqnarray}
Here we define
\begin{eqnarray}
\sigma_{LO} & = & \int_{\tau_0}^1 d\tau~\frac{d\mathcal{
L}^{gg}}{d\tau}~\hat\sigma_{LO}(Q^2 = \tau s) \, , \nonumber \\ 
\Delta \sigma_{virt} & = & \frac{\alpha_s(\mu_R)}
{\pi}\int_{\tau_0}^1 d\tau~\frac{d\mathcal{ L}^{gg}}{d\tau}~\hat
\sigma_{virt}(Q^2=\tau s) \, ,\\ 
\Delta \sigma_{ij} & = & \frac{\alpha_{s}(\mu_R)} {\pi} \int_{\tau_0}^1
d\tau~ \frac{d\mathcal{ L}^{ij}}{d\tau} \int_{\tau_0/\tau}^1 \frac{dz}{z}~
\hat\sigma_{ij}(Q^2 = z \tau s) \qquad (ij=gg,gq,q\bar q)
 \, ,\nonumber
\label{eq:nlocxn}
\end{eqnarray}
where $\hat\sigma_{LO/virt/ij}(Q^2)$ denote the partonic cross-sections. The parton-parton luminosities are denoted by $d\mathcal{L}^{ij}/d\tau~(i,j=g,q,\bar q)$.
which is defined as
\begin{eqnarray}
\frac{d\mathcal{ L}^{gg}}{d\tau} & = &\int_\tau^1 \frac{dx}{x}
\Big[ g(x,\mu_F) g\left(\frac{\tau}{x},\mu_F\right) \Big] \, , \nonumber \\
\frac{d\mathcal{ L}^{gq}}{d\tau} & = & \sum_{q,\bar q} \int_\tau^1 \frac{dx}{x}
\Big[ g(x,\mu_F) q\left(\frac{\tau}{x},\mu_F\right)
+ q(x,\mu_F) g\left(\frac{\tau}{x},\mu_F\right) \Big] \, , \nonumber \\
\frac{d\mathcal{ L}^{q\bar q}}{d\tau} & = & \sum_q \int_\tau^1 \frac{dx}{x}
\Big[ q(x,\mu_F) \bar q\left(\frac{\tau}{x},\mu_F\right)
+ \bar q(x,\mu_F) q\left(\frac{\tau}{x},\mu_F\right) \Big],
\end{eqnarray}
where $q(x,\mu_F)$ and $g(x,\mu_F)$ are the quark and gluon densities 
at the factorization scale $\mu_F$.




\subsection{$g$ $g$ $\rightarrow$ $H$ $H$ $g$}
\label{sec:ggtohhg}
\begin{figure}
\centering
\subfloat[]{\includegraphics[width=.3\linewidth]{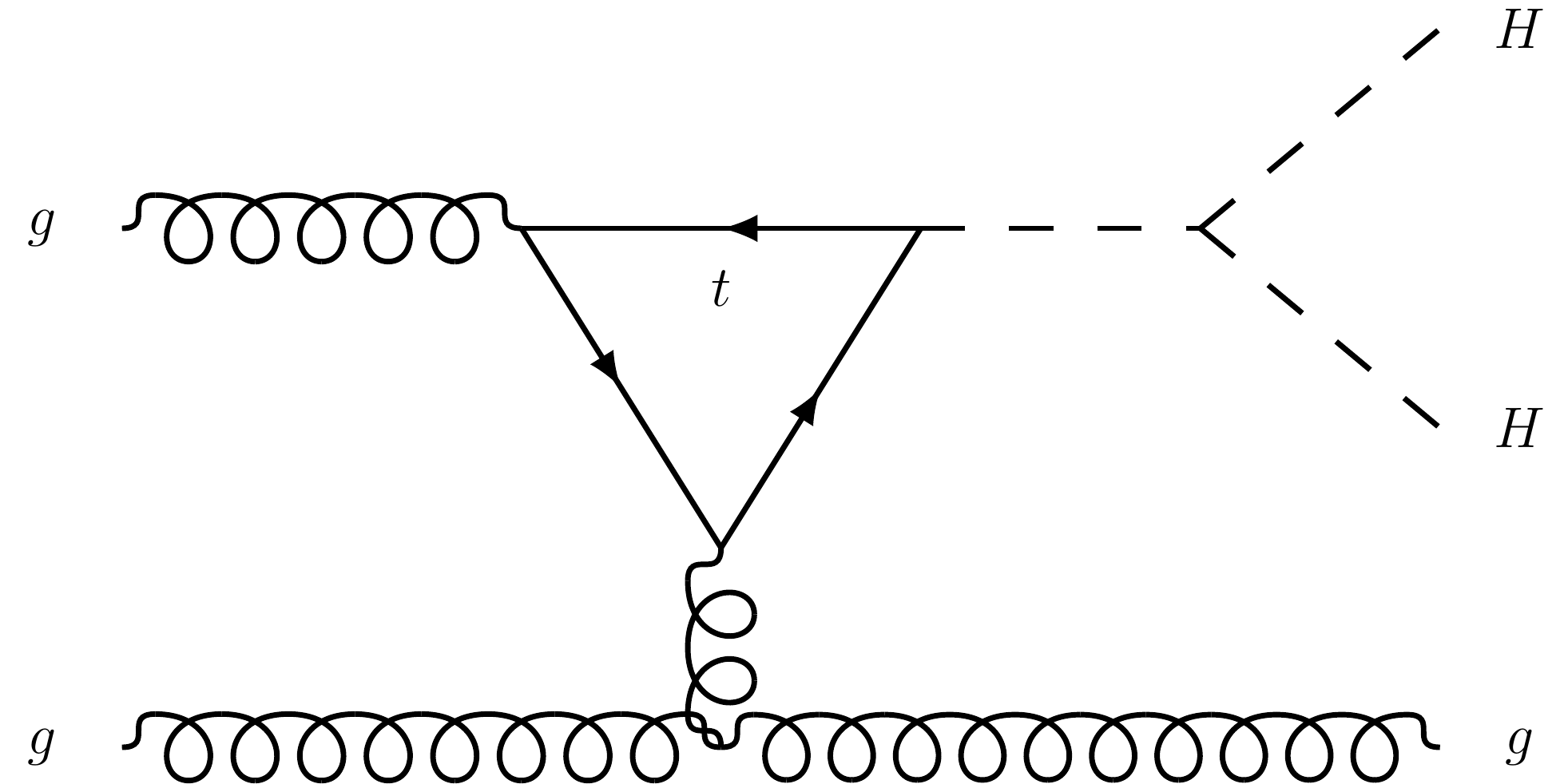}}\qquad
\subfloat[]{\includegraphics[width=.3\linewidth]{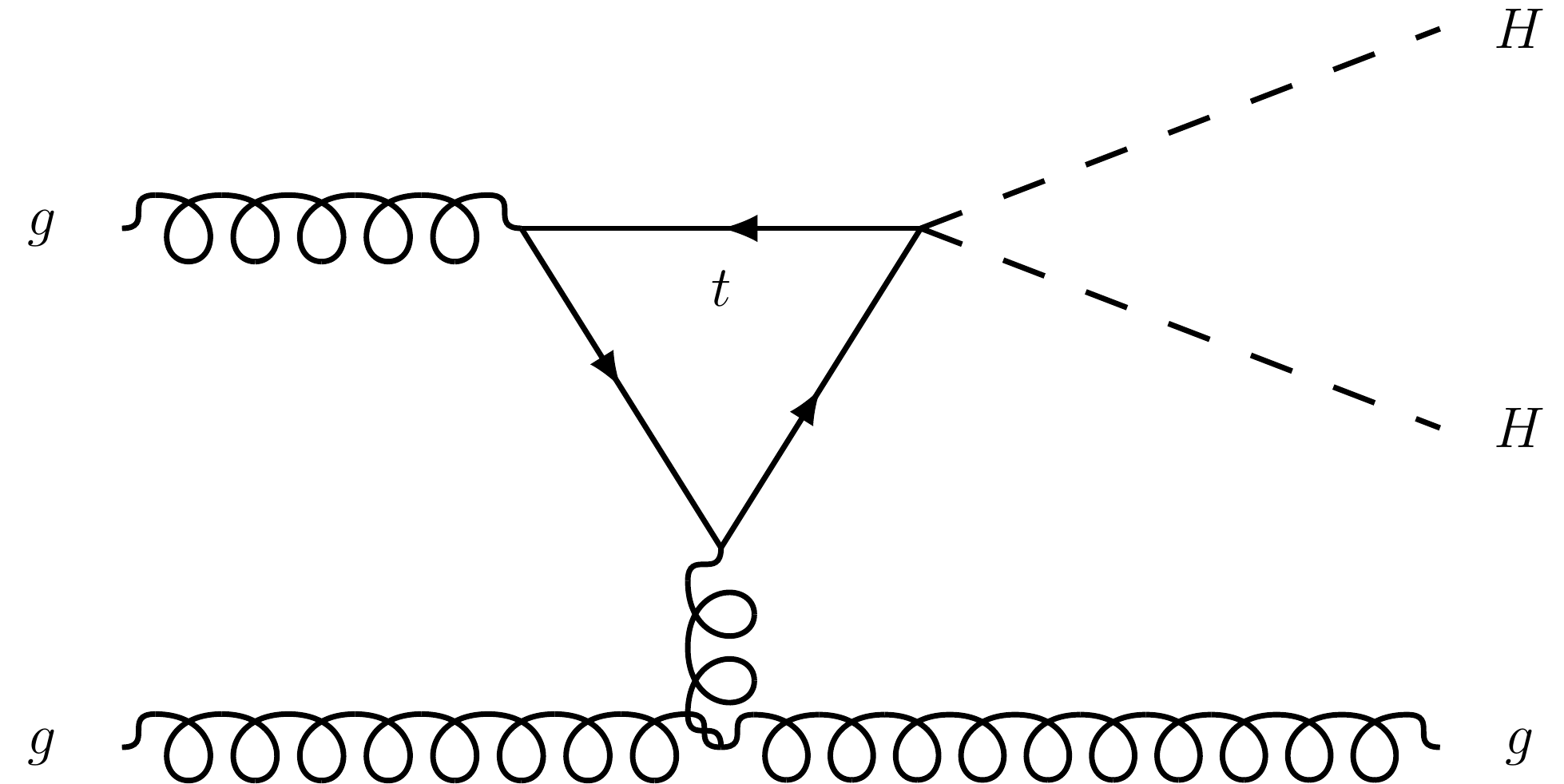}}
\caption{\label{fig:dia_t_gghhg}\emph{Generic triangle diagrams for the partonic $g g\to H H g$ channel at NLO in QCD. Diagram \emph{(a)} is SM diagrams, where \emph{(b)} is BSM diagram with anomalous $HHtt$ coupling. Each gluon can be one of the two incoming gluons or the outgoing gluon.}
}
\end{figure}

\begin{figure}[H]
\centering
\subfloat[]{\includegraphics[width=.3\linewidth]{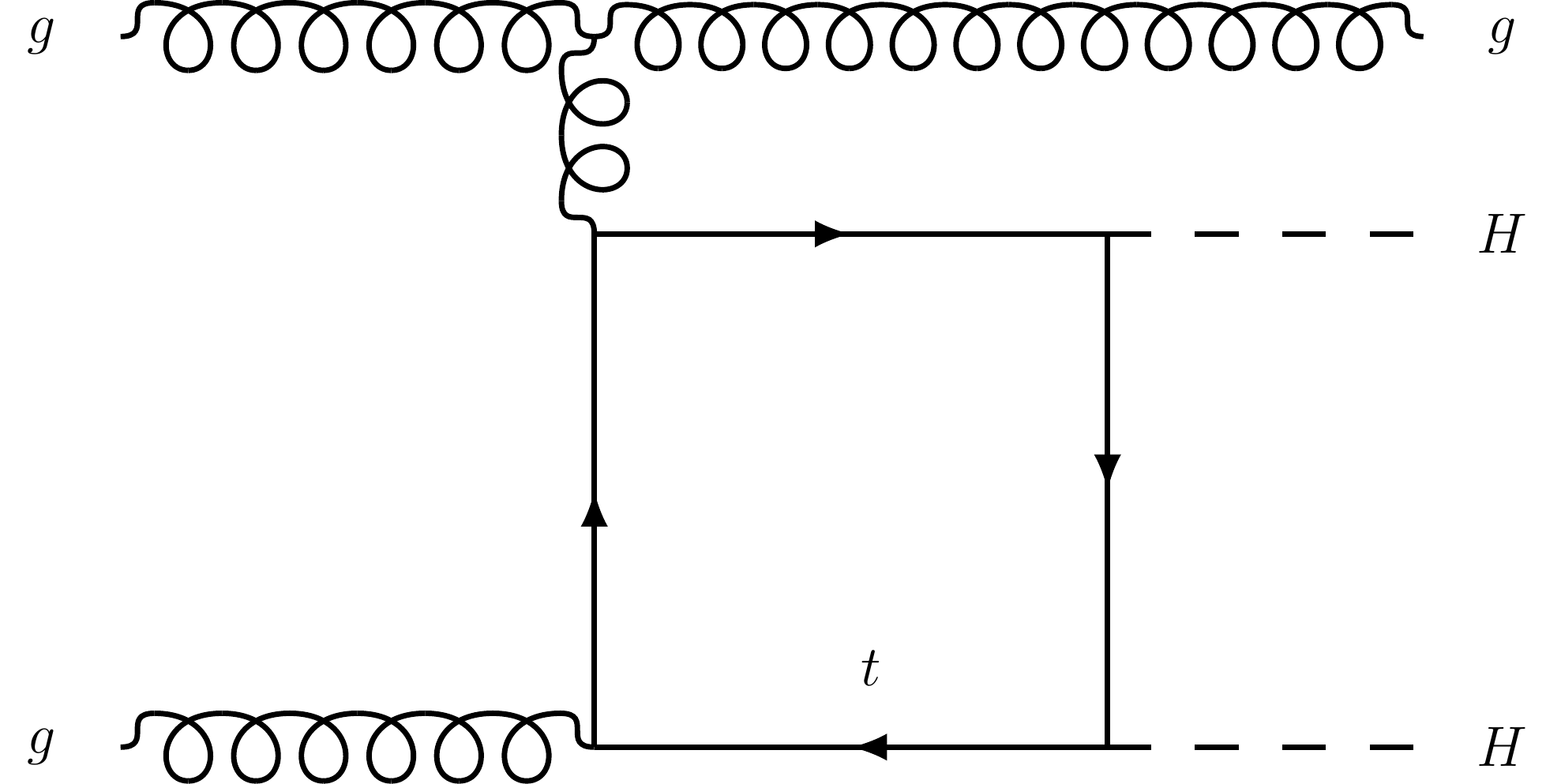}}
\subfloat[]{\includegraphics[width=.3\linewidth]{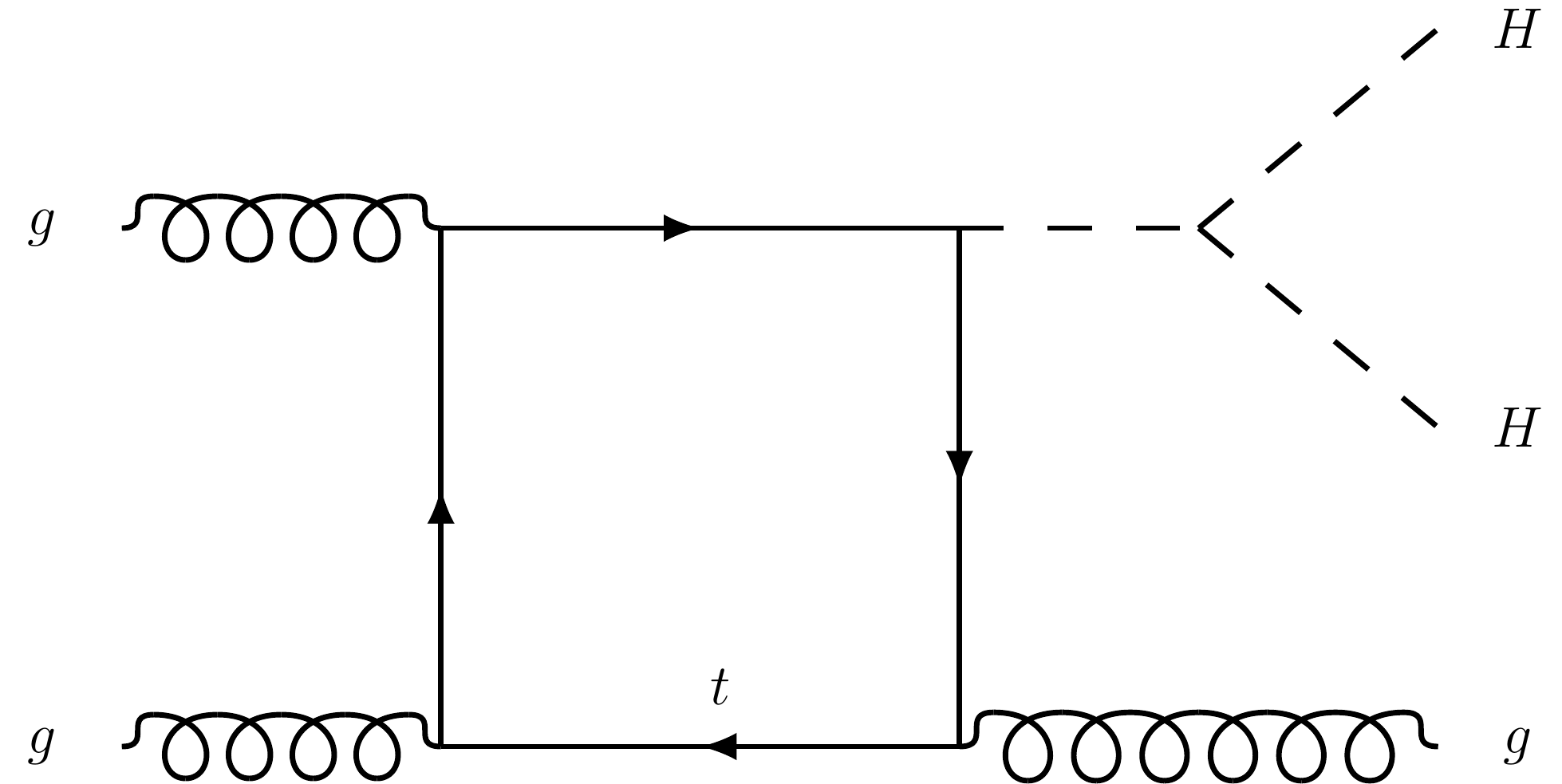}}
\subfloat[]{\includegraphics[width=.3\linewidth]{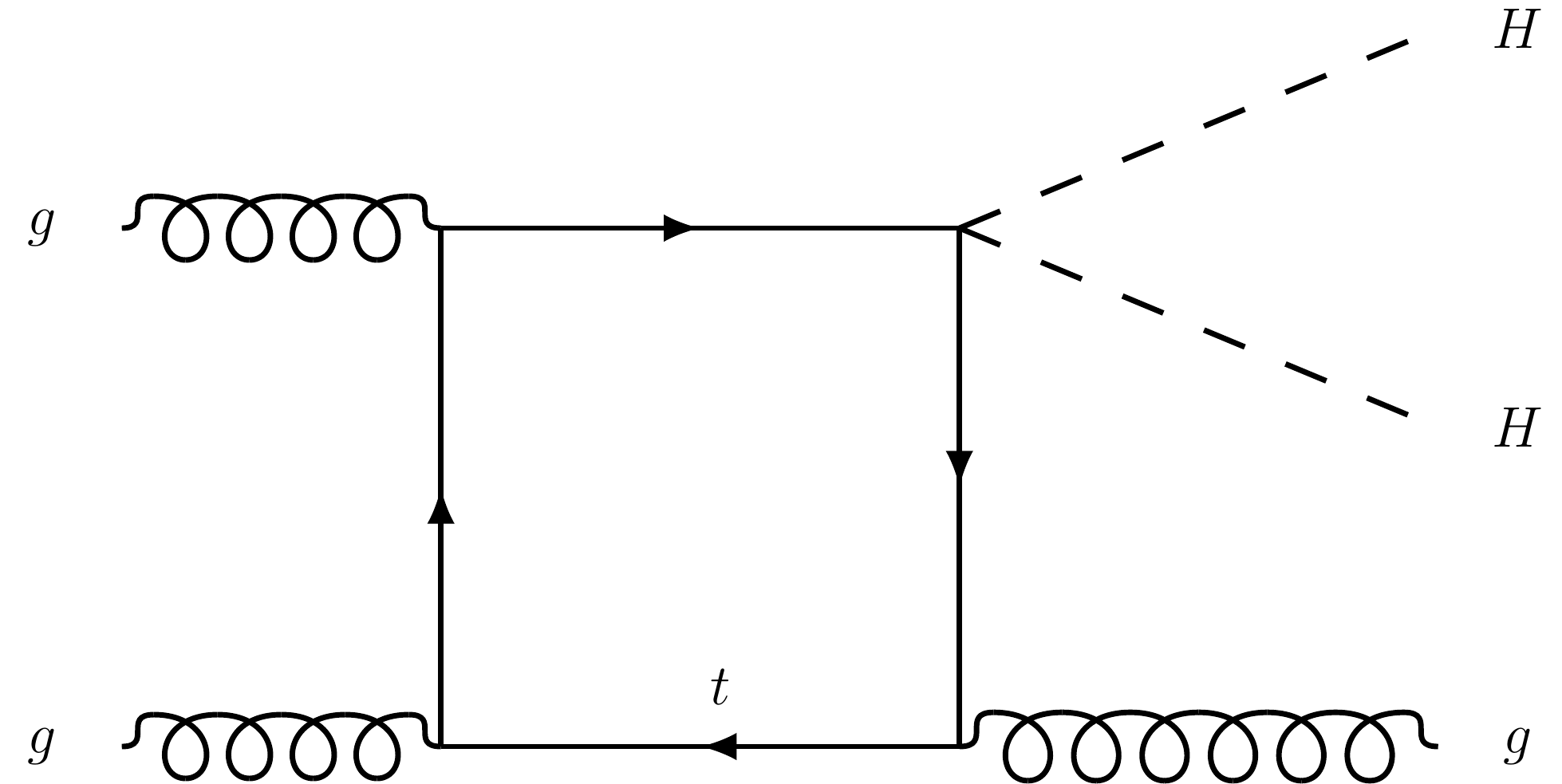}}
\caption{\label{fig:dia_b_gghhg}\emph{Generic box diagrams for the partonic $g g\to H H g$ channel at NLO in QCD. Diagrams \emph{(a)} and \emph{(b)} are SM diagrams, where \emph{(c)} is BSM diagram with anomalous $HHtt$ coupling. Each gluon can be one of the two incoming gluons or the outgoing gluon.}
}
\end{figure}

\begin{figure}[H]
\centering
\subfloat[]{\includegraphics[width=.3\linewidth]{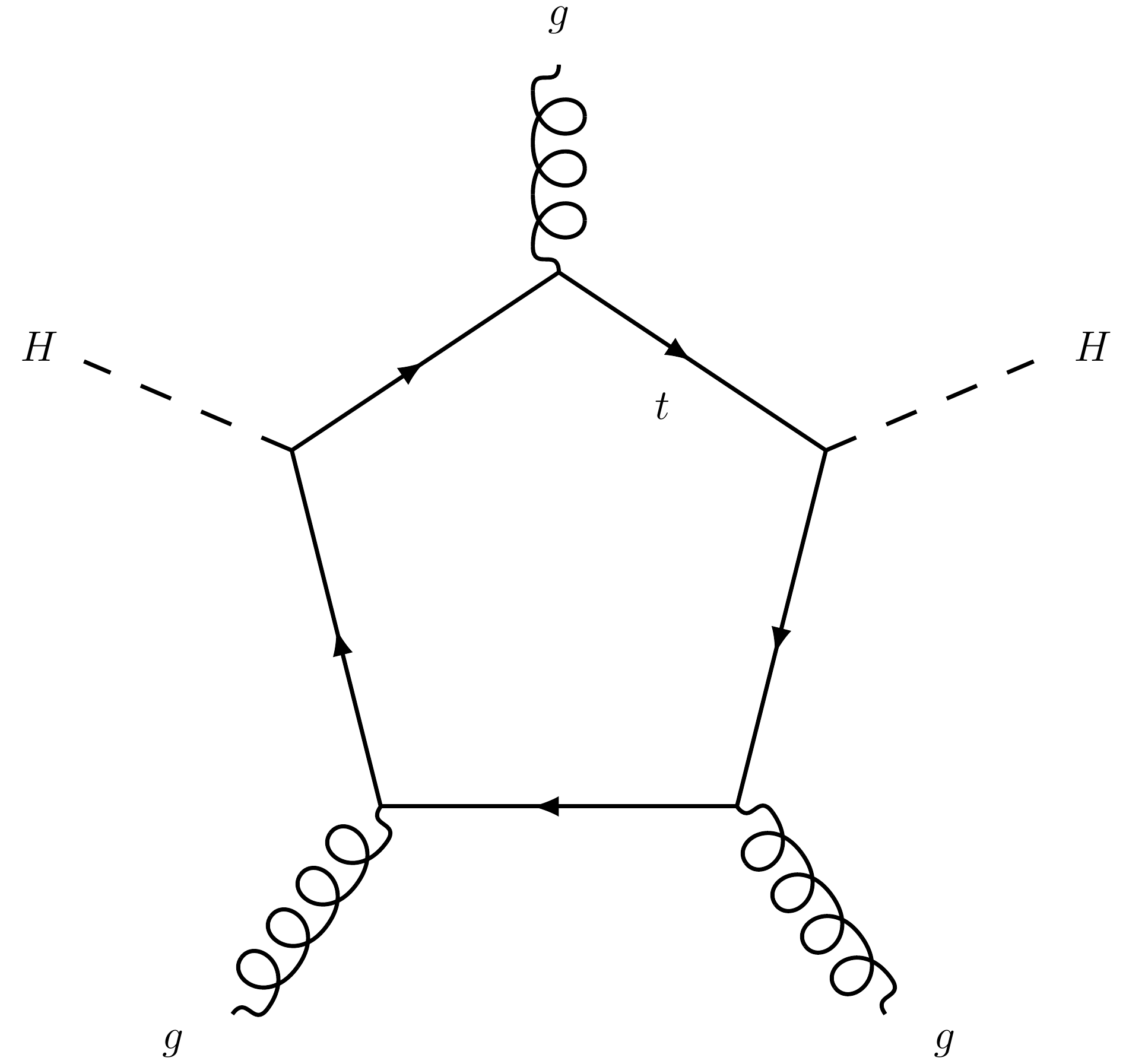}}\qquad
\subfloat[]{\includegraphics[width=.3\linewidth]{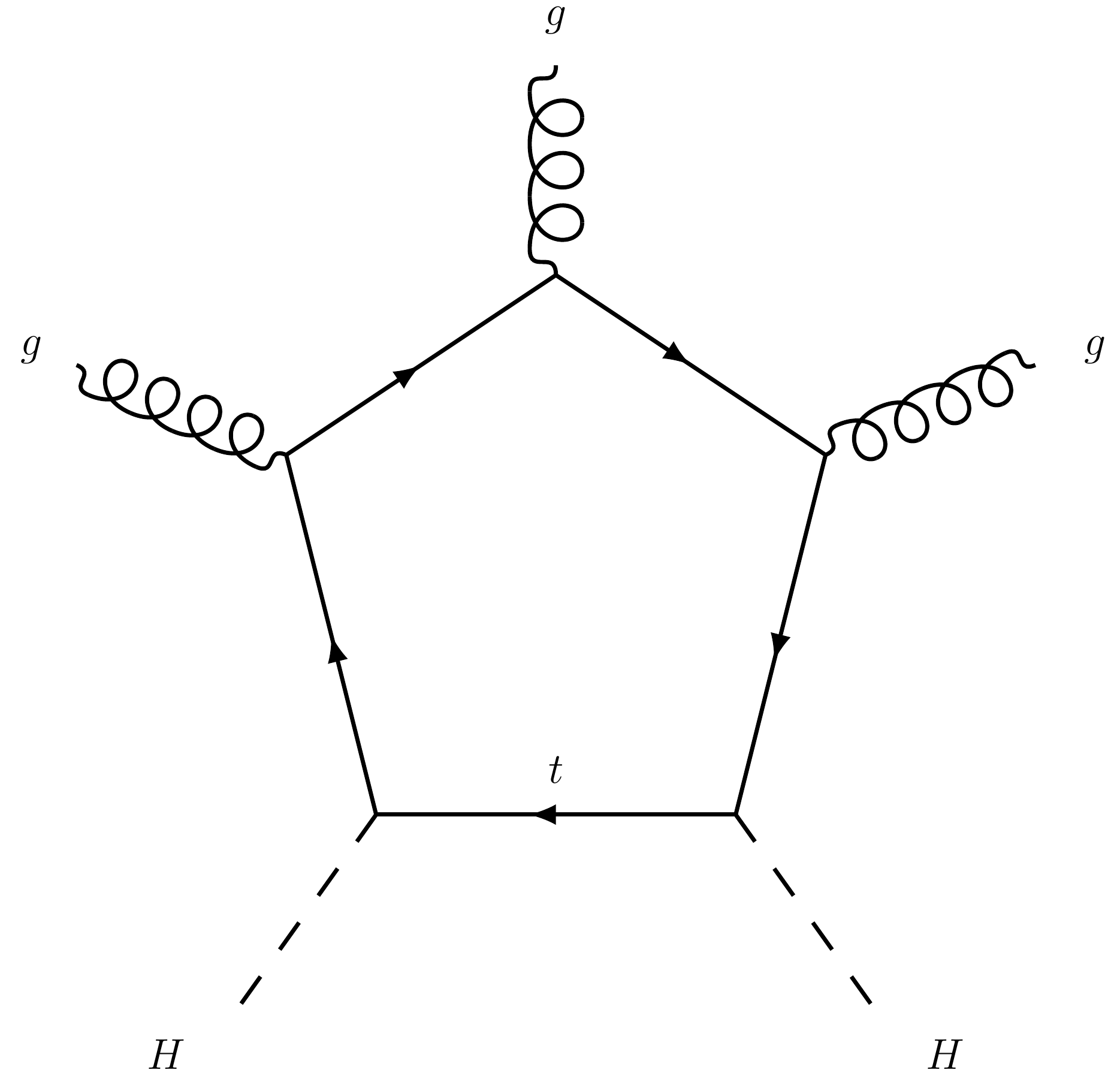}}
\caption{\label{fig:dia_p_gghhg}\emph{Generic SM one-loop pentagon diagrams for the partonic $g g\to H H g$ channel at NLO in QCD. Each gluon can be one of the two incoming gluons or the outgoing gluon.}
}
\end{figure}

 Fig.~\ref{fig:dia_t_gghhg} to Fig.~\ref{fig:dia_p_gghhg} show the generic diagrams for the $g g\to H H g$ channel.
For diagrams shown in Fig.~\ref{fig:dia_t_gghhg}, Fig.~\ref{fig:dia_b_gghhg} (a), and Fig.~\ref{fig:dia_b_gghhg} (b), the matrix elements can be easily obtained by replacing one on-shell gluon, $\epsilon_{\nu}$, with a gluon propagator and attaching the other end to a tri-gluon vertex in Eq.~(\ref{eq:lomat}). The generic amplitude for these contributions can be written as
\begin{eqnarray}
\mathcal{M}(g_{1}^{c_1} g_{2}^{c_2} \to HH g_{3}^{c_3}) & = & -\,\frac{G_F\alpha_s(\mu_R) Q^2}{2\sqrt{2}\pi} \sum_{i,j,k}
f^{c_i,c_j,c_k} \mathcal{A}^{\rho \alpha} \epsilon_{i\mu}\epsilon_{j\nu}\epsilon_{k\alpha} \frac{4\pi \alpha_s}{(p_i+p_j)^2} \nonumber \\
&&\times [g_{\mu \nu}(p_i-p_j)^{\rho}+g_{\nu \rho}(2p_j+p_i)^{\mu}-g_{\rho \mu}(2p_i+p_j)^{\mu}],
\end{eqnarray}
where gluons are labeled by $i, j, k \in {1,2,3}$ , $c_i$, $p_i$ denotes the color index and the momentum of gluon labeled by $i$ respectively and $\mathcal{A}^{\rho \alpha} $ is defined in Eq.~(\ref{eq:lomat}).

\subsection{$q$ $g$ $\rightarrow$ $H$ $H$ $q$ and $q$ $\bar{q}$ $\rightarrow$ $H$ $H$ $g$}
\label{sec:qgtoqhh}
\begin{figure}[H]
\centering
\includegraphics[width=0.95\linewidth, angle=0]{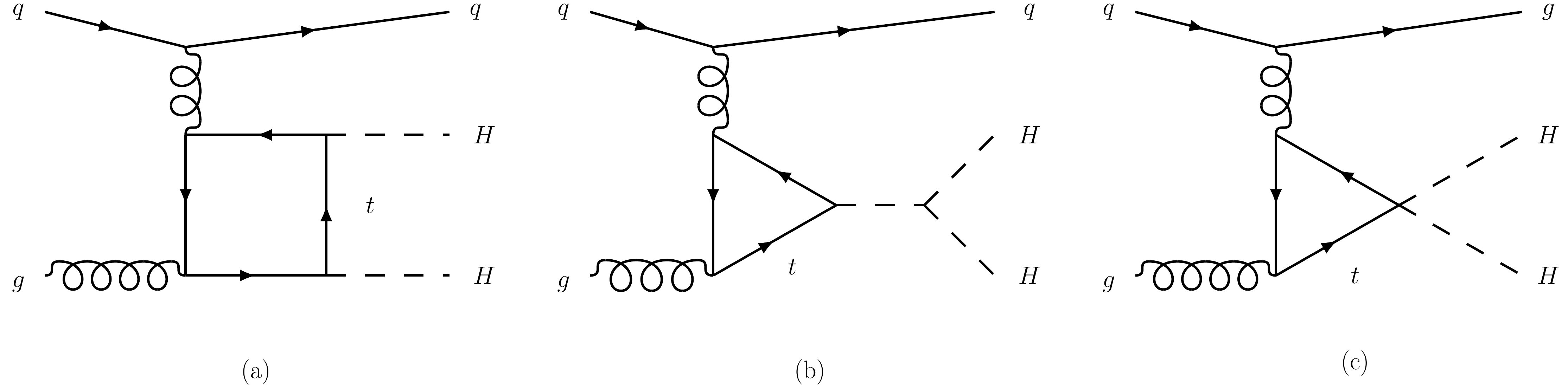} 
\caption{\label{fig:dia_qghhq_qcd}\emph{Dominant box and triangle diagrams at the one-loop level for the $qg\to HHq$ channel.}
}
\end{figure}

Although the contributions from $qg \rightarrow HHq$ and $q\bar{q} \rightarrow HHg$ are very different, as we will see in the next section, they share the same diagrams. Therefore, we only need to compute the matrix elements for $qg \rightarrow HHq$ which can be easily converted to corresponding matrix elements for $qg \rightarrow HHq$.
For contributions shown in Fig.~\ref{fig:dia_qghhq_qcd}, similar to the process $g g\to H H g$, the matrix elements can be easily obtained by replacing the only incoming gluon, $\epsilon_{\nu}$, with a gluon propagator and attaching the other end to two fermions in Eq.~(\ref{eq:lomat}). The amplitude for contributions shown in Fig.~\ref{fig:dia_qghhq_qcd} can be written as

\begin{eqnarray}
\mathcal{M}(q(p_1) g(p_2) \to q(p_3) HH) & = & -\,\frac{G_F\alpha_s(\mu_R) Q^2}{2\sqrt{2}\pi}
\mathcal{A}^{\mu\nu} \epsilon_{\mu} \bar{v}(p_3) \gamma_{\nu} u(p_1) \frac{\sqrt{4\pi \alpha_s}}{(p_3-p_1)^3} ,
\end{eqnarray}

\begin{figure}[H] 
\centering{}
\includegraphics[width=0.7\textwidth]{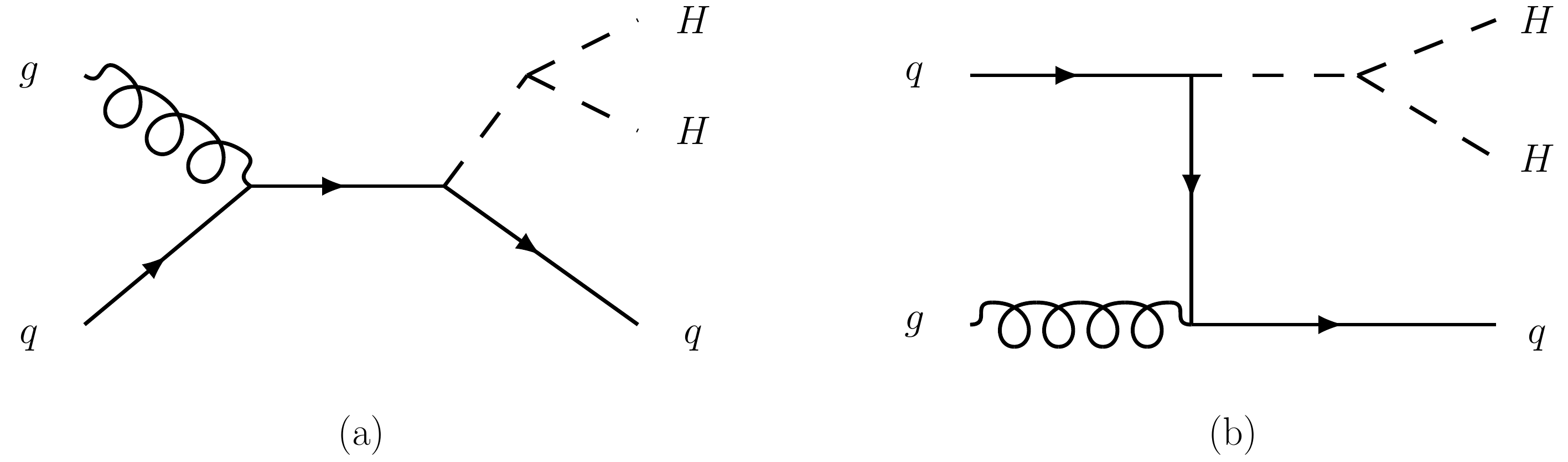}
\protect
\caption{\emph{Tree diagrams for $b(c)g \rightarrow b(c)HH$.}
\label{fig:dia_qghhq_tree}}
\end{figure}

Other than the common one-loop diagrams shown in Fig.~\ref{fig:dia_qghhq_qcd}. We also calculate the contributions from diagrams with only one strong coupling, which has not been studied before.
We denote contributions from loop diagrams with more than one strong couplings as $QCD_2$ contribution and contributions from diagrams with only one strong coupling as $QCD_1$ contribution. As we will see in the next chapter, $QCD_1$ contribution is smaller compare to $QCD_2$ contribution due to the suppression from the weak coupling.
For the production processes, $qg \rightarrow HHq$, tree-level diagrams are generally too small due to the smallness of the $qqH$ coupling. Therefore, tree diagrams are usually dropped, and only the loop diagrams are considered. However, it turns out that we still have to consider tree diagram for $b(c)g \rightarrow b(c)HH$ shown in Fig.~\ref{fig:dia_qghhq_tree} since the contributions from loop diagrams are small and the $bbH$ or $ccH$ couplings are just large enough to make these tree diagrams contribute at similar order to loop-induced $QCD_1$ contributions.

\begin{figure} 
\centering{}
\begin{tabular}{p{0.3\textwidth} p{0.3\textwidth} p{0.3\textwidth}} 
 \vspace{0pt} \includegraphics[width=0.3\textwidth]{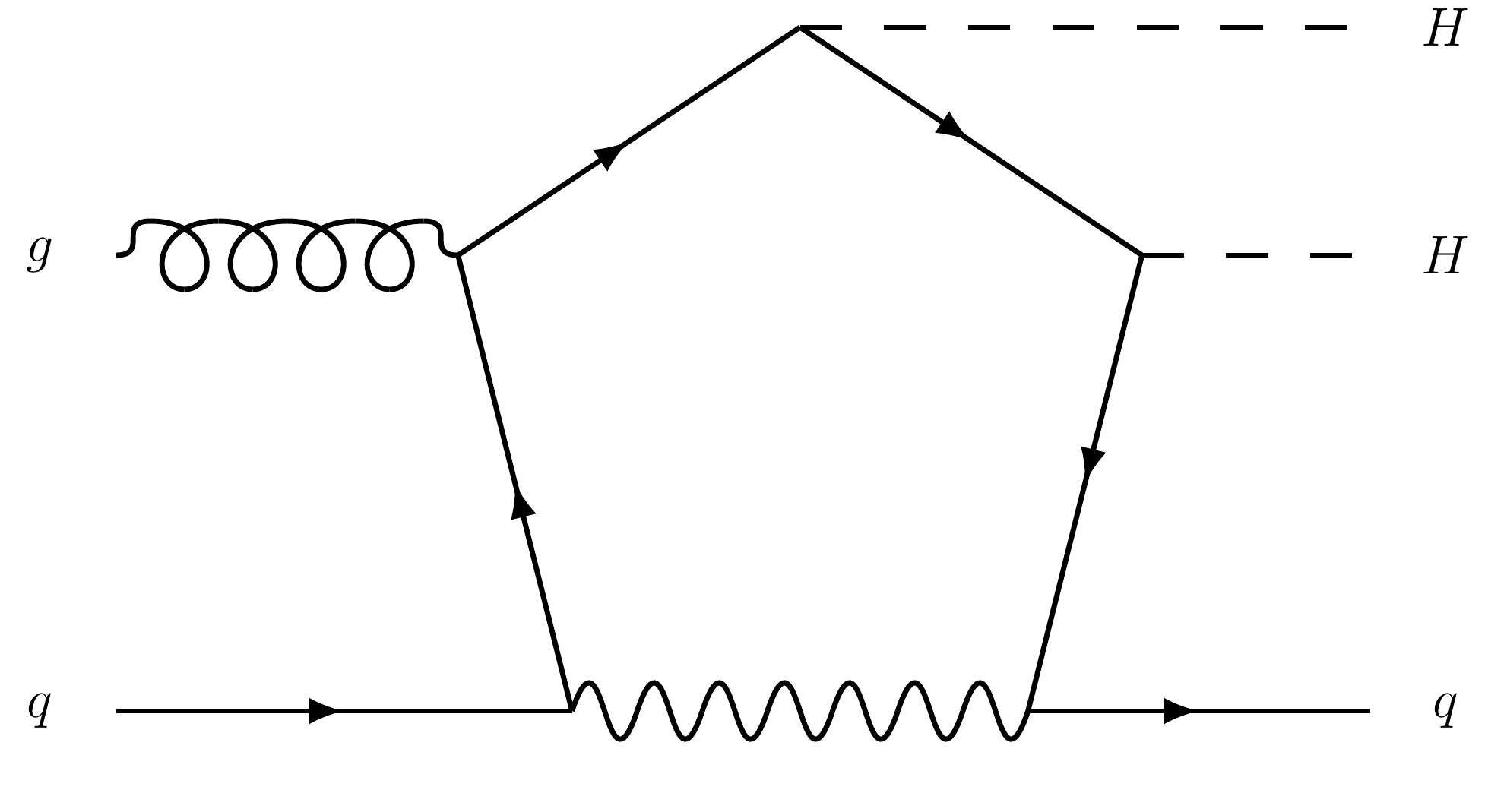} &
 \vspace{0pt} \includegraphics[width=0.3\textwidth]{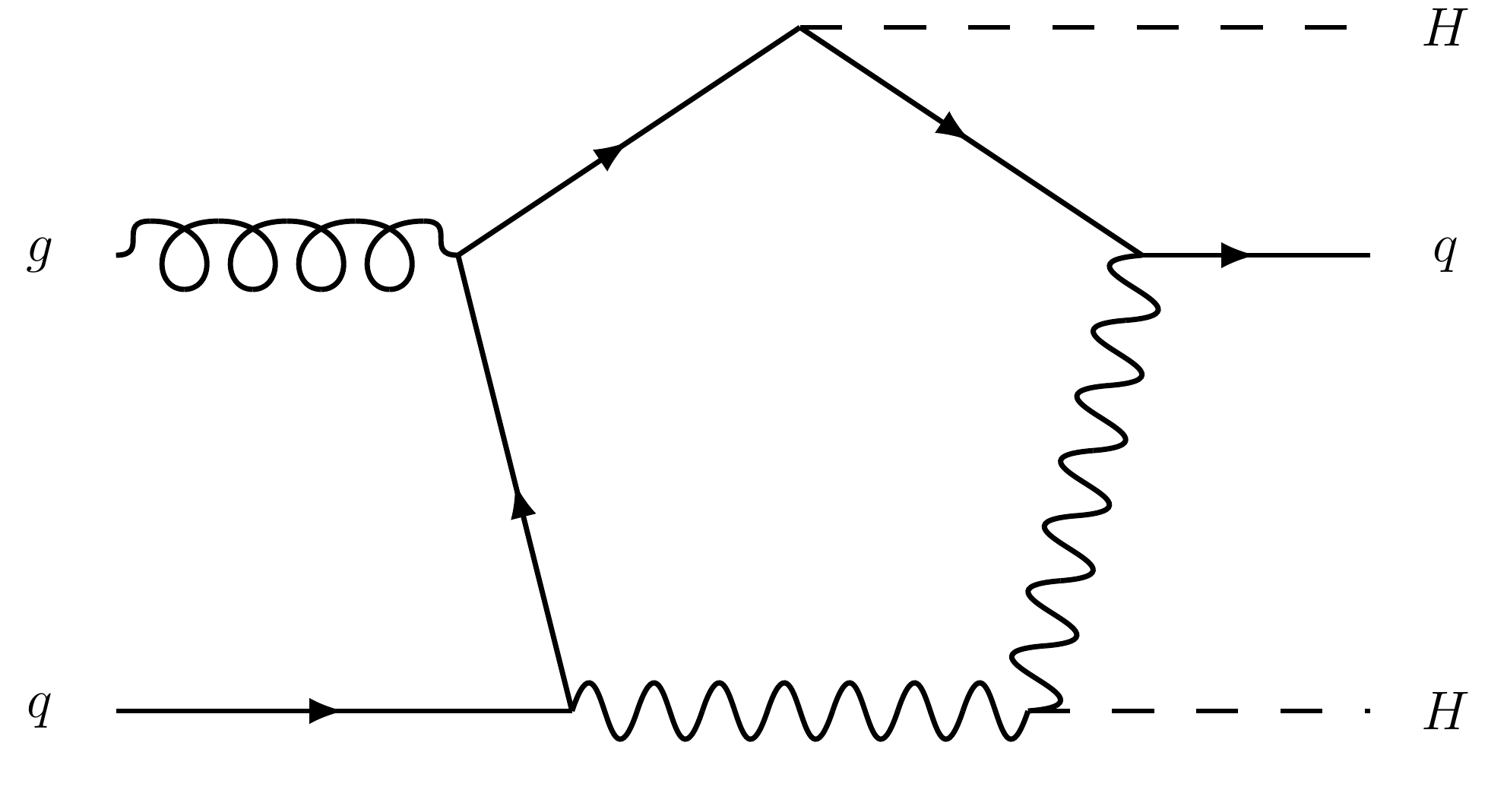} &
 \vspace{-4pt} \includegraphics[width=0.3\textwidth]{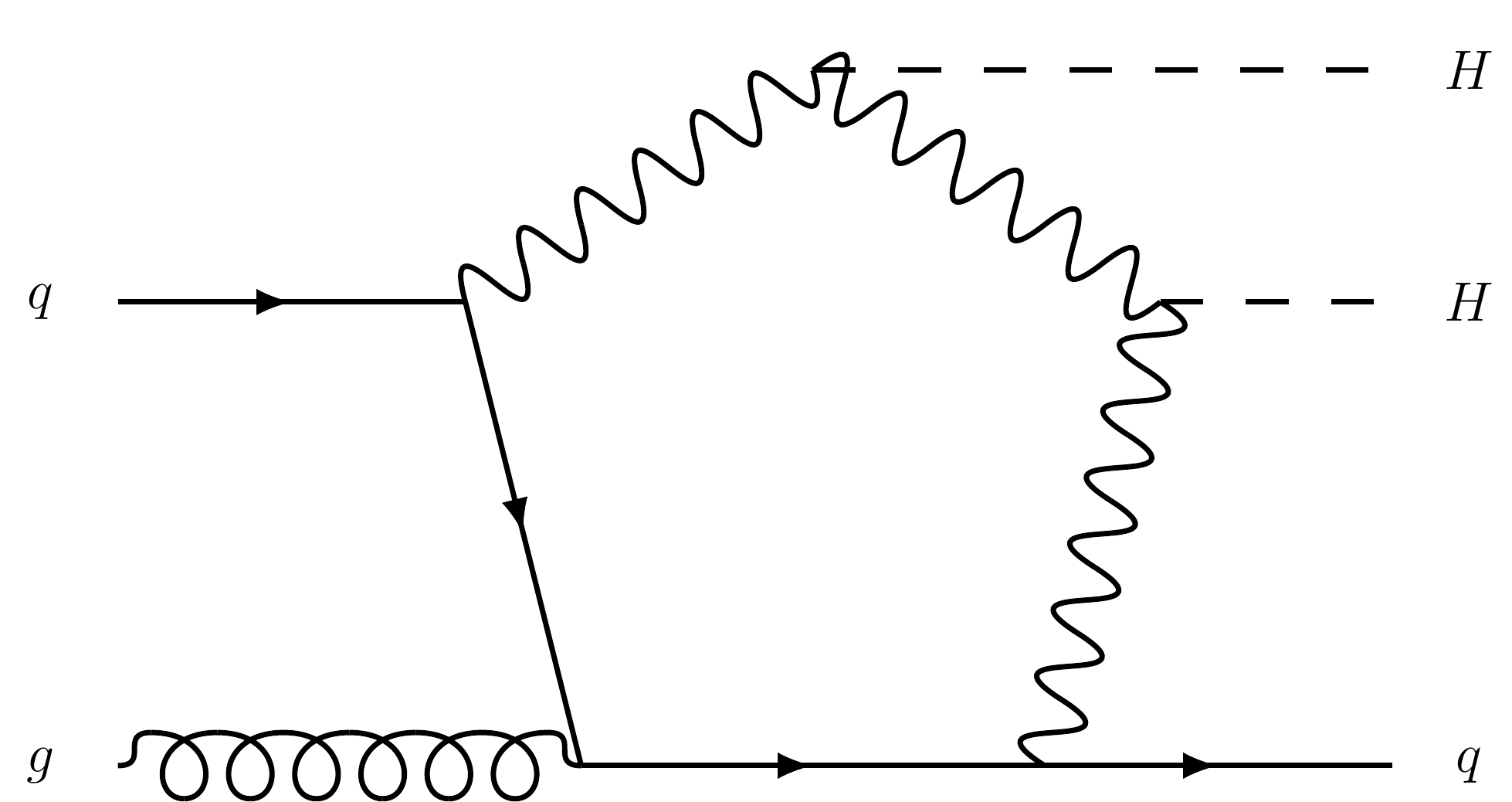}
\end{tabular}
\protect
\caption{\emph{Generic pentagon diagrams with different numbers of quarks and gauge bosons in the loop.}
\label{fig:dia_penta}}
\end{figure}

\begin{figure}[hbt!] 
\centering{}
\begin{tabular}{p{0.3\textwidth} p{0.3\textwidth} p{0.3\textwidth}} 
 \vspace{0pt} \includegraphics[width=0.3\textwidth]{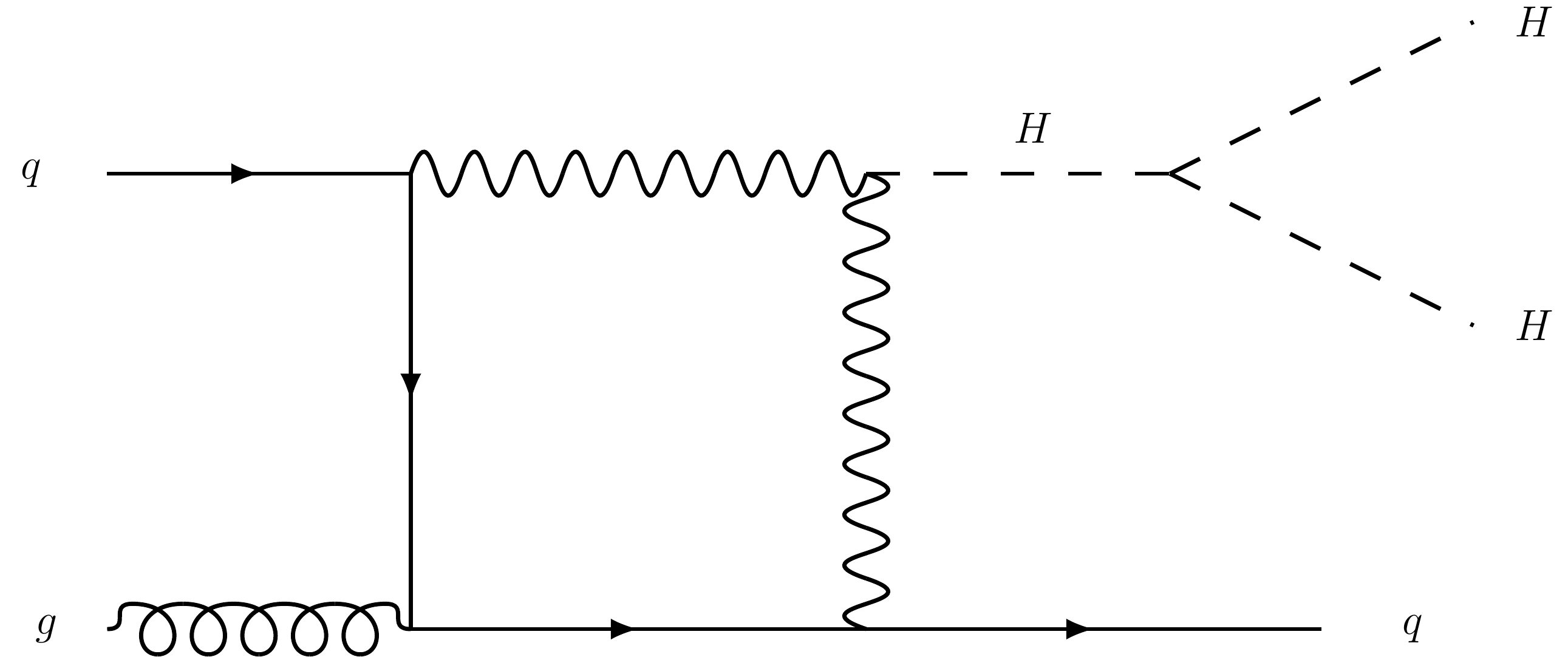} &
 \vspace{0pt} \includegraphics[width=0.3\textwidth]{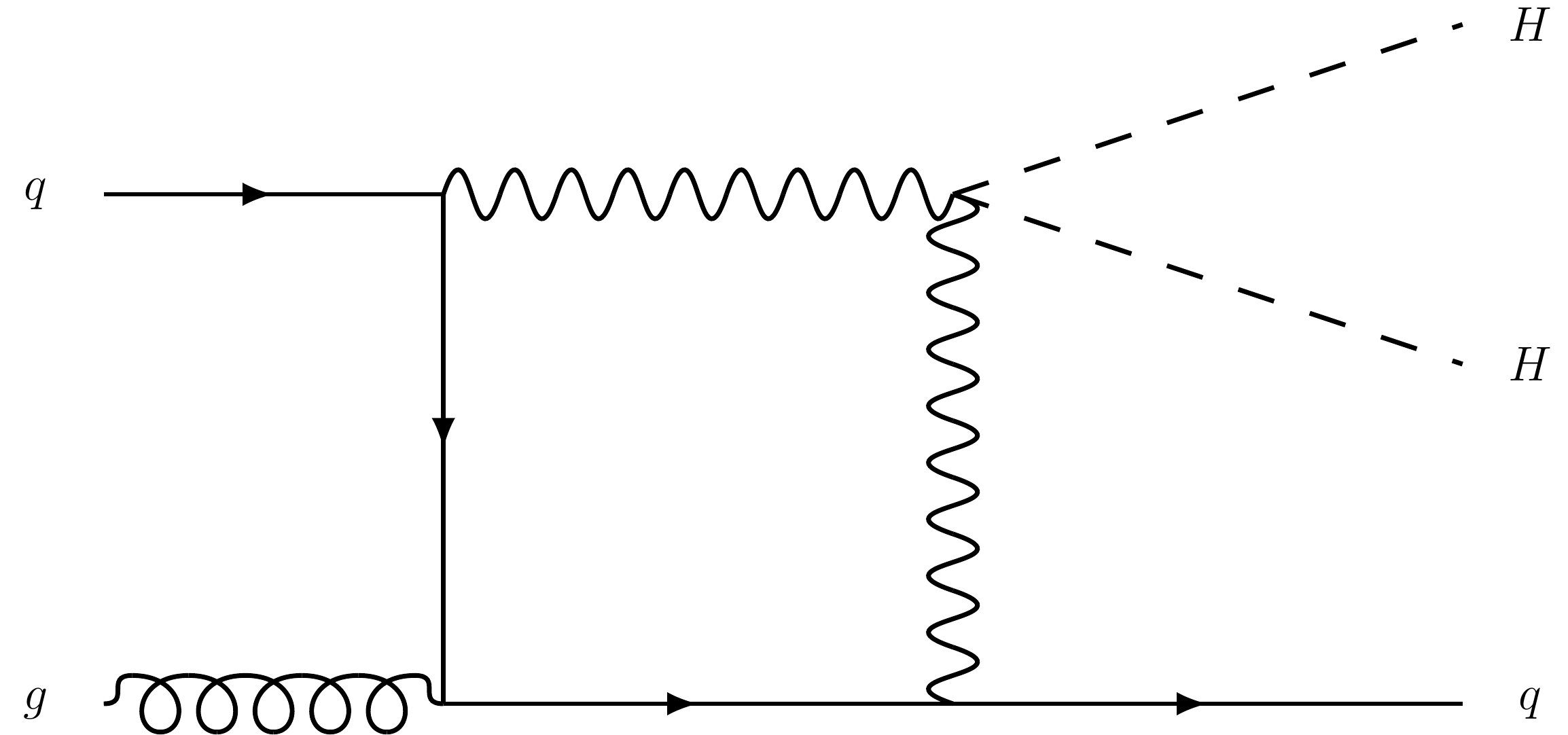} &
 \vspace{10pt} \includegraphics[width=0.3\textwidth]{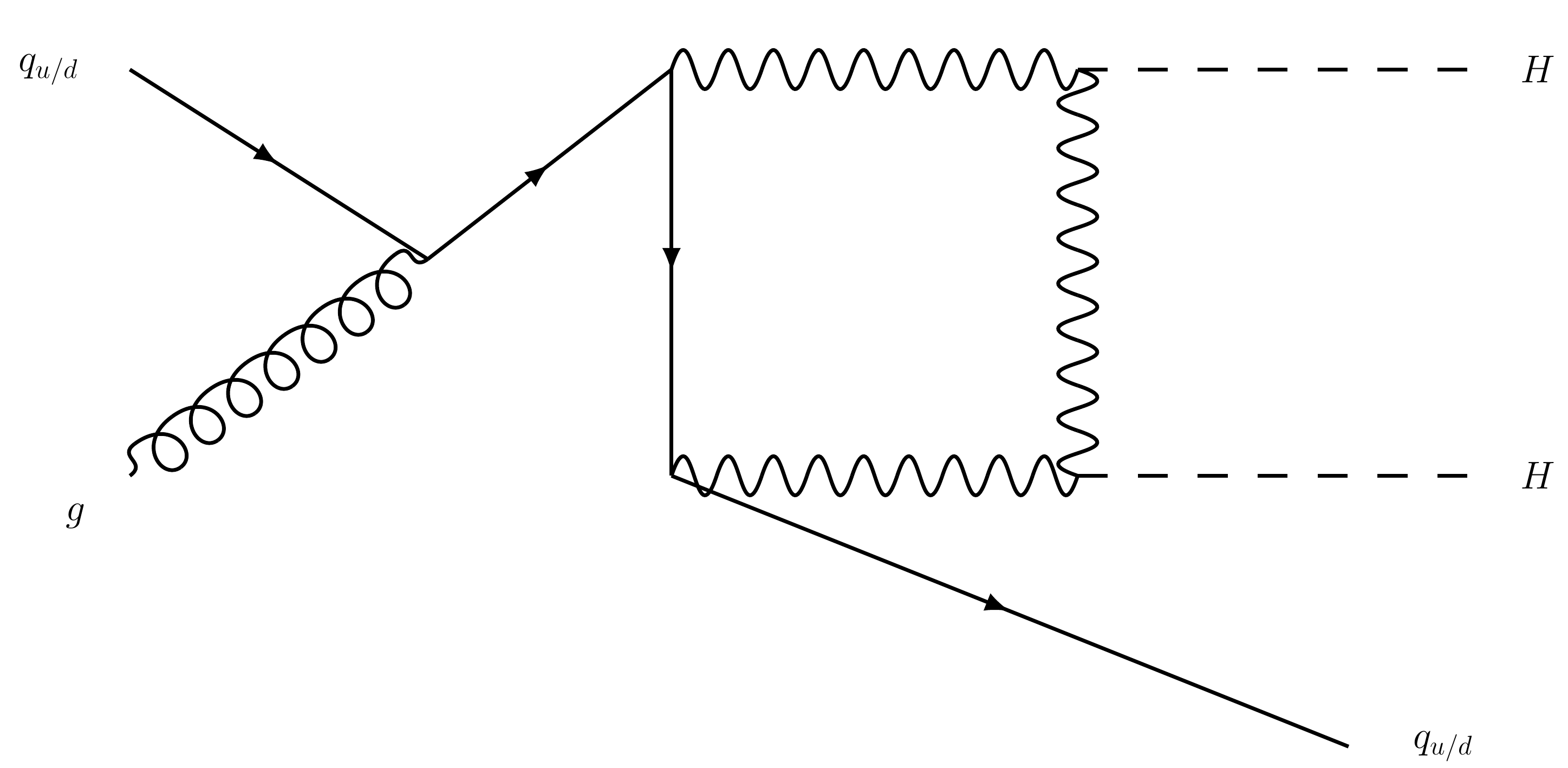}
\end{tabular}
\protect
\caption{\emph{Generic box diagrams with different numbers of quarks and gauge bosons in the loop.}
\label{fig:dia_box}}
\end{figure}

\begin{figure}[hbt!] 
\centering{}
\includegraphics[width=0.45\textwidth]{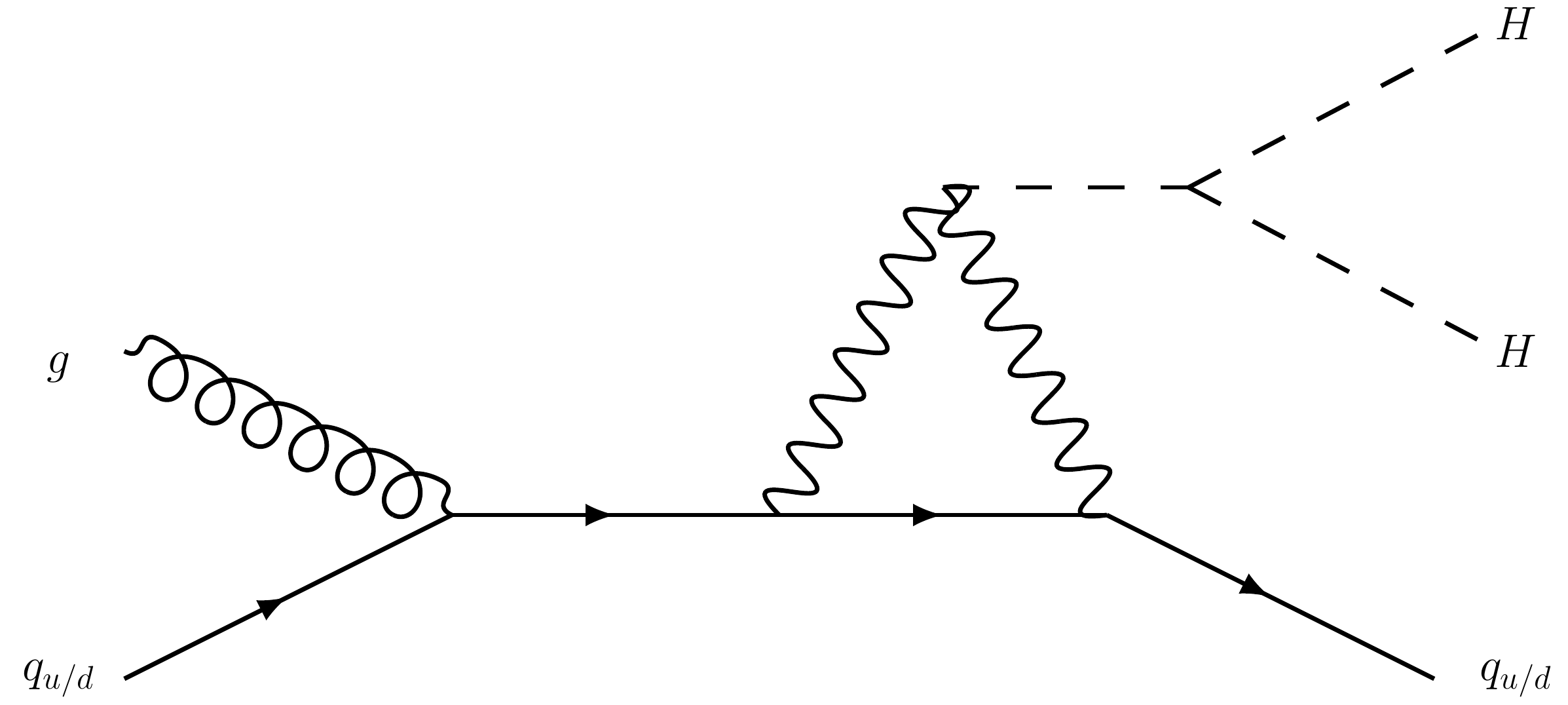}
\includegraphics[width=0.45\textwidth]{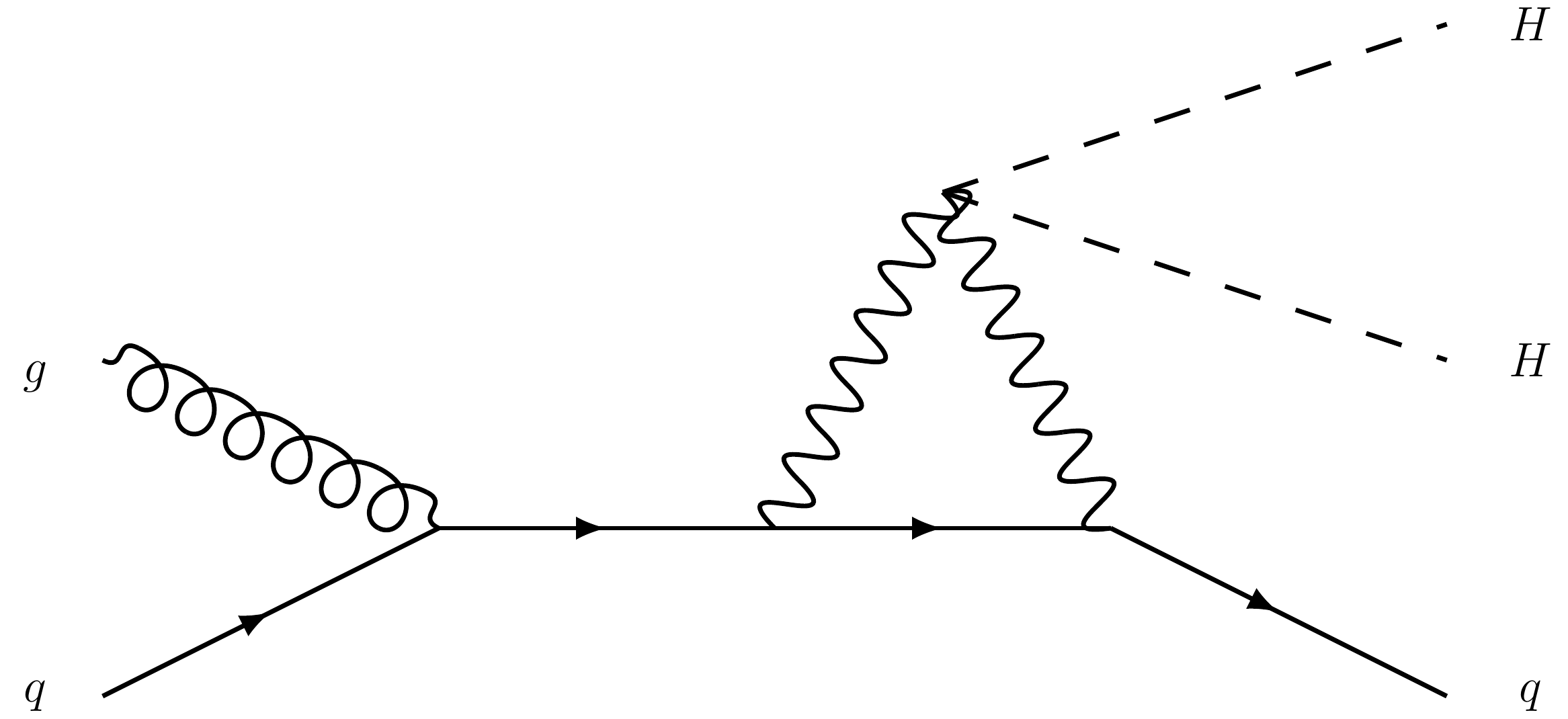}
\includegraphics[width=0.45\textwidth]{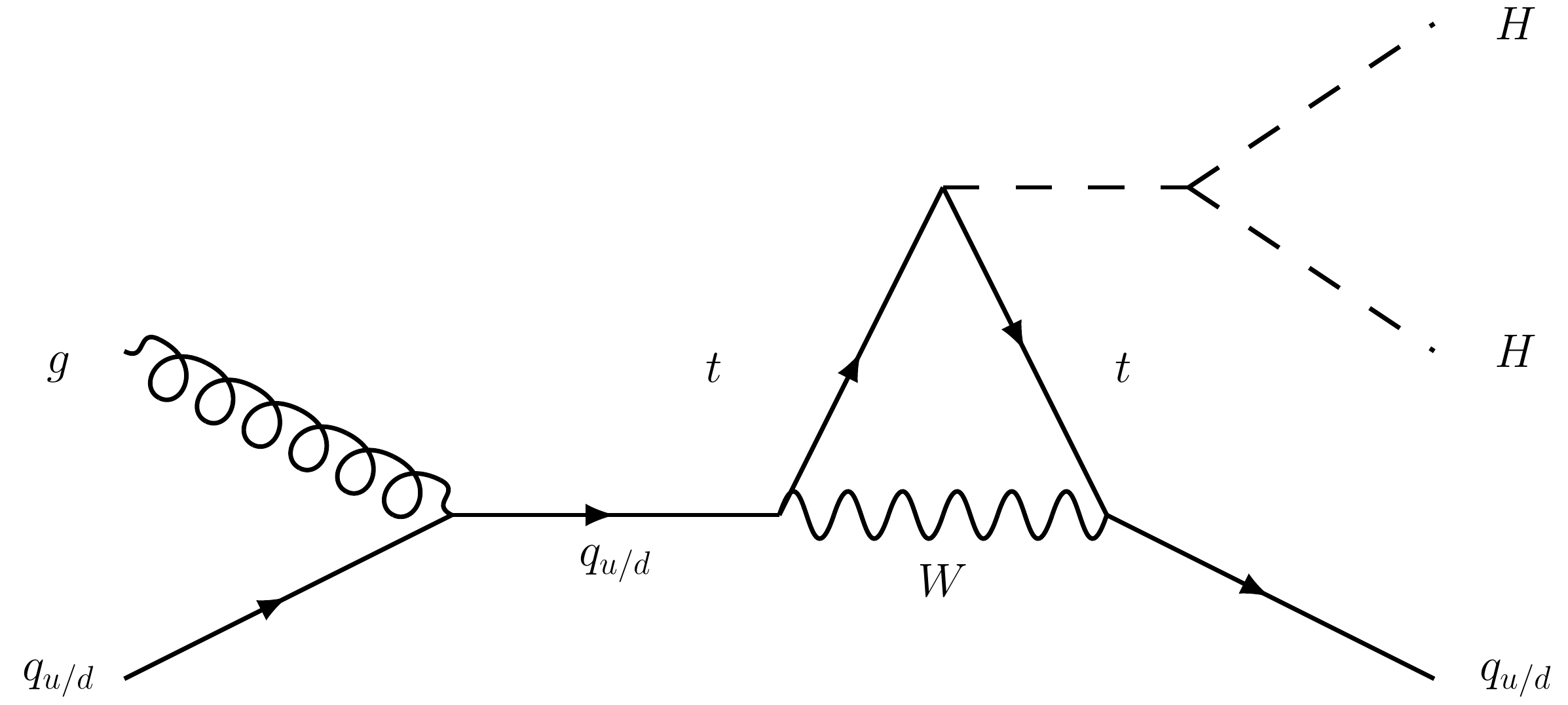}
\includegraphics[width=0.45\textwidth]{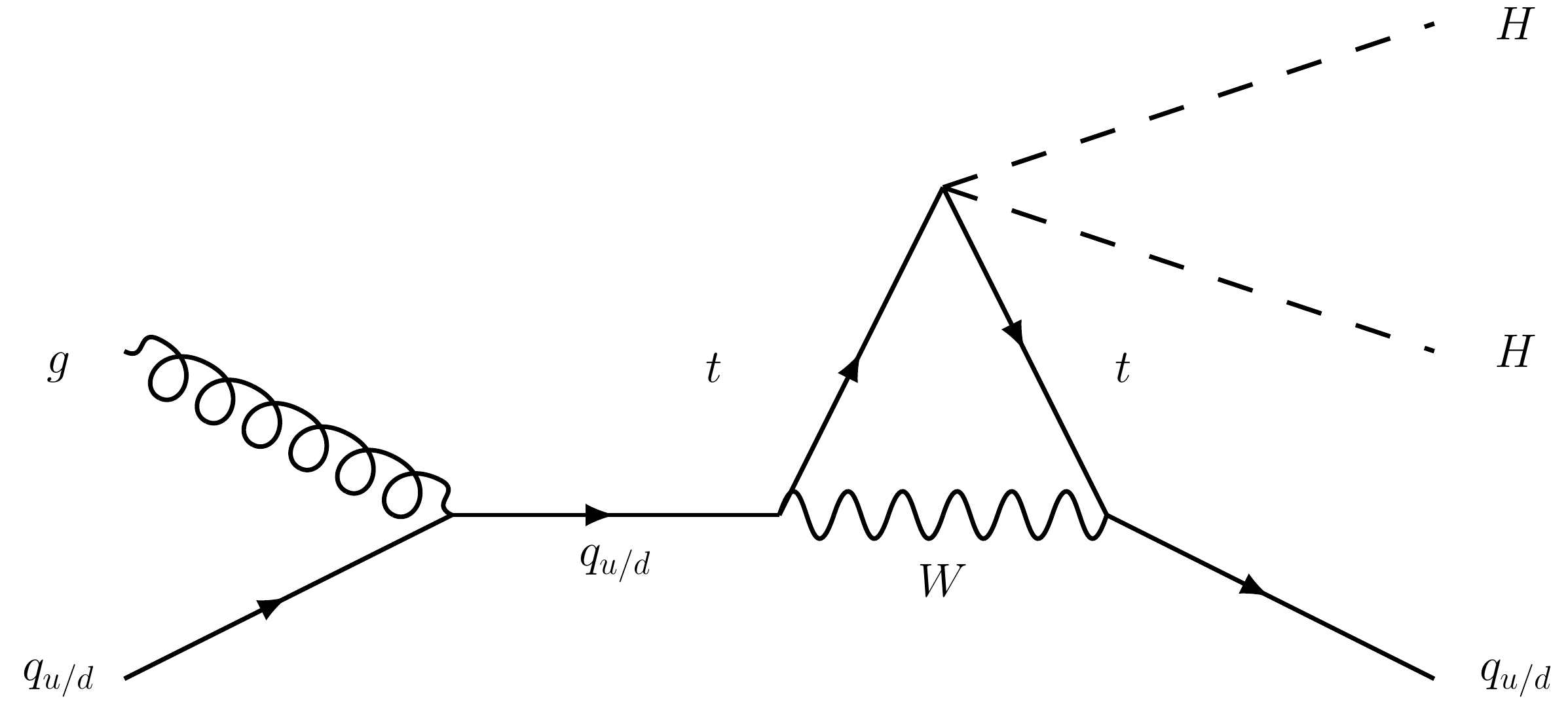}
\protect
\caption{\emph{Generic triangle diagrams with different numbers of quarks and gauge bosons in the loop.}
\label{fig:dia_tri}}
\end{figure}

Again all diagrams to one loop level for $qg \rightarrow HHq$ can be categorized into the pentagon, box, and triangle diagrams. The pentagon, box, triangle diagrams are shown in Fig.~\ref{fig:dia_penta} to Fig.~\ref{fig:dia_tri} , where solid lines and the wavy represent fermions and vector bosons ($W$, $Z$) or corresponding Goldstone boson($G^{+/-}$, $G_0$), respectively.

For contributions from $gg \rightarrow HHg$ and $qg \rightarrow HHq$, we generate the full analytical one-loop matrix elements by using \texttt{FeynArts} \cite{Hahn:2000kx} and \texttt{FormCalc} \cite{Hahn:1998yk}.
The tensor reduction performed by \texttt{FormCalc} is using the techniques developed in Ref.~\onlinecite{vanOldenborgh:1990yc,Denner:2002ii,Denner:2005nn,Denner:2010tr}, while the numerical results of the scalar integrals~\cite{tH79} are evaluated with \texttt{LoopTools} \cite{Hahn:1998yk}. 
As cross-check, the analytic matrix elements for pentagon diagrams with two quark propagators in the loop, box diagrams with one quark propagator, and triangle diagrams are also calculated by hand.
The results of matrix elements compute by hand, and matrix elements generated with \texttt{FeynArts} and \texttt{FormCalc} are in agreement with the numerical results generated by \texttt{MG5\_aMC@NLO} \cite{Mg5}, which is the primary tool for the computations of cross-sections and the generation of hard events in this article.

\section{\label{chap:numerical}Kinematic Distributions}

We use the same workflow introduced in Section~\ref{sec:gghh_kin}, and adopt only \texttt{PDF4LHC15}~\cite{PDF4LHC} to generate partonic events. For hadron level analysis, we feed the unweighted partonic events file into Pythia 8~\cite{Pythia8} to generate a large number of simulated collision events. To promptly analyze such a huge amount of simulated collision events, we adopt Delphes~\cite{Delphes}, which provides a fast multipurpose detector response simulation to reconstruct events into jets such as b jet and lepton jets, etc.

We analyzed the final results in detail for the total cross-section and the differential cross-section in the invariant mass of Higgs-pair.

For the double Higgs production with at most one extra jet, $pp\to HHj$, the SM expectation for this production cross-section is only 0.031 pb in the Large Hadron Collider with CM energy at 14 TeV. At the potential 100 TeV hadron collider, the expected SM cross-section increases significantly to 1.543 pb since the luminosity of gluon PDF at lower Bjorken scale, increases.

By convoluting the gluon and quark PDF's with the partonic cross-section in a hadron collider, we can obtain the differential cross-section in the lab frame 
\beq
\frac{d^2\sigma(pp\to HH)}{dm_{HH}\, dp_T} =\int_\tau^1 \frac{dx}{x}g(x,\mu_F) g\left(\frac{\tau}{x},\mu_F\right) \frac{2m_{HH}}{s} \frac{d\hat{\sigma}(gg\to HH)}{dp_T} \ ,
\eeq
where $p_T$ is the transverse momentum of the Higgs boson, $s$ is the CM energy of head-on hadrons, and $\tau=\hat{s}/s$, $m_{HH}=\sqrt{\hat{s}}$,
\beq
p_T^2 = \frac{\hat{u}\hat{t} - m_H^4}{\hat{s}} \ .
\ee
In this section, we firstly adopt \texttt{MG5\_aMC@NLO}~\cite{Mg5} with a custom UFO model~\cite{UFO} including the anomalous Higgs-top coupling, $HHtt$, and corresponding $R_2$ and $UV$ counterterms to generate matrix elements. We then adopt \texttt{PDF4LHC15}~\cite{PDF4LHC} to generate partonic events. In this work, we draw all plots by using the framework above with the following parameters

\beq
\label{eq:parameter}
m_t=173 {\textrm{GeV}} \ , \quad m_H=125 {\textrm{GeV}}.
\eeq
The input value $\alpha_{s}(M_Z)$ is determined by the PDF set used, where $\alpha_{s}(M_Z)=0.118$ for \texttt{PDF4LHC15}.
We set the factorization and renormalization scales to $\mu_F =\mu_R=m_{HH}$.



\subsection{Leading-order Contribution} \label{sec:gghh_kin}
\begin{figure}[H]
\centering
\subfloat[]{\includegraphics[width=.45\linewidth]{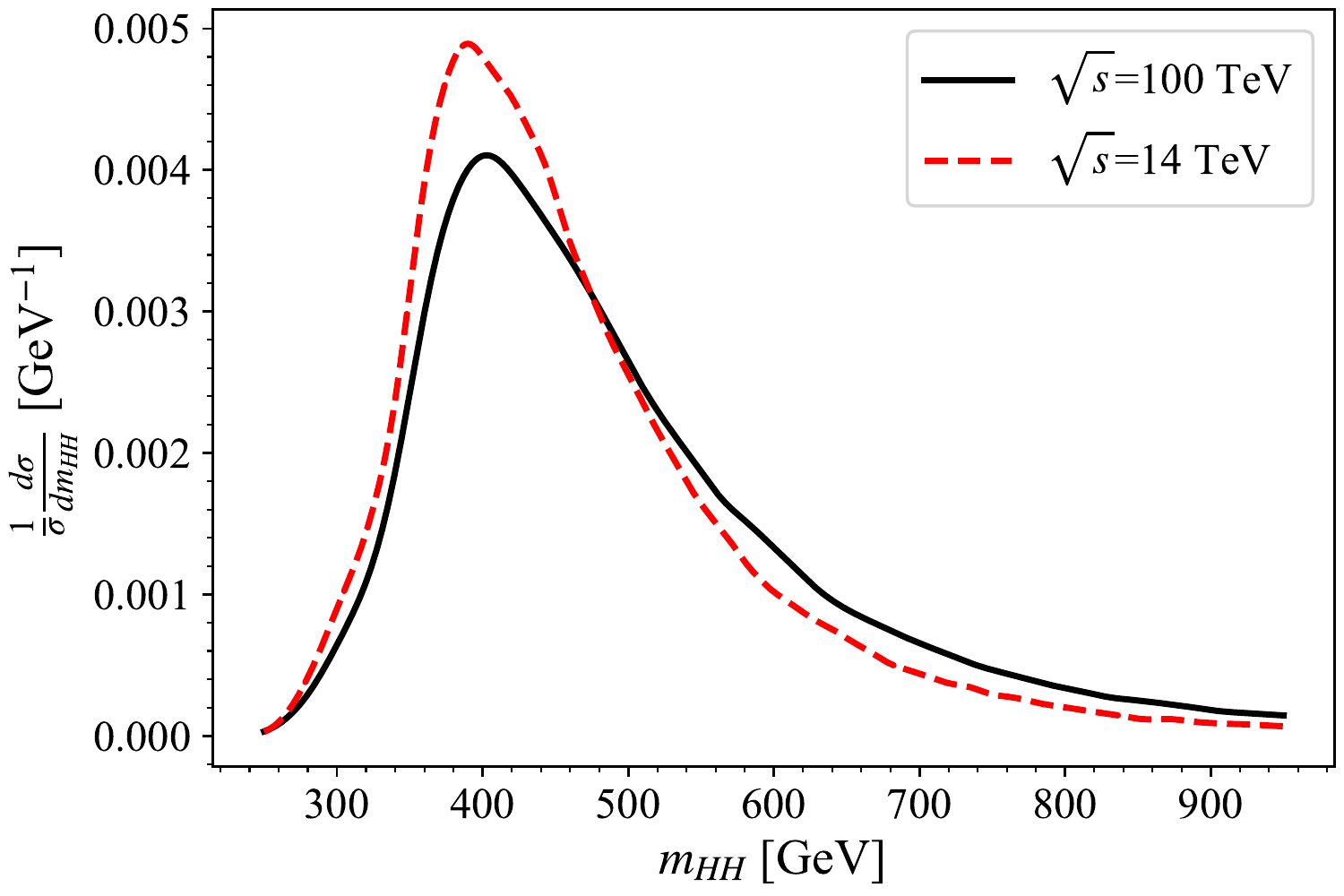}}\qquad
\subfloat[]{\includegraphics[width=.45\linewidth]{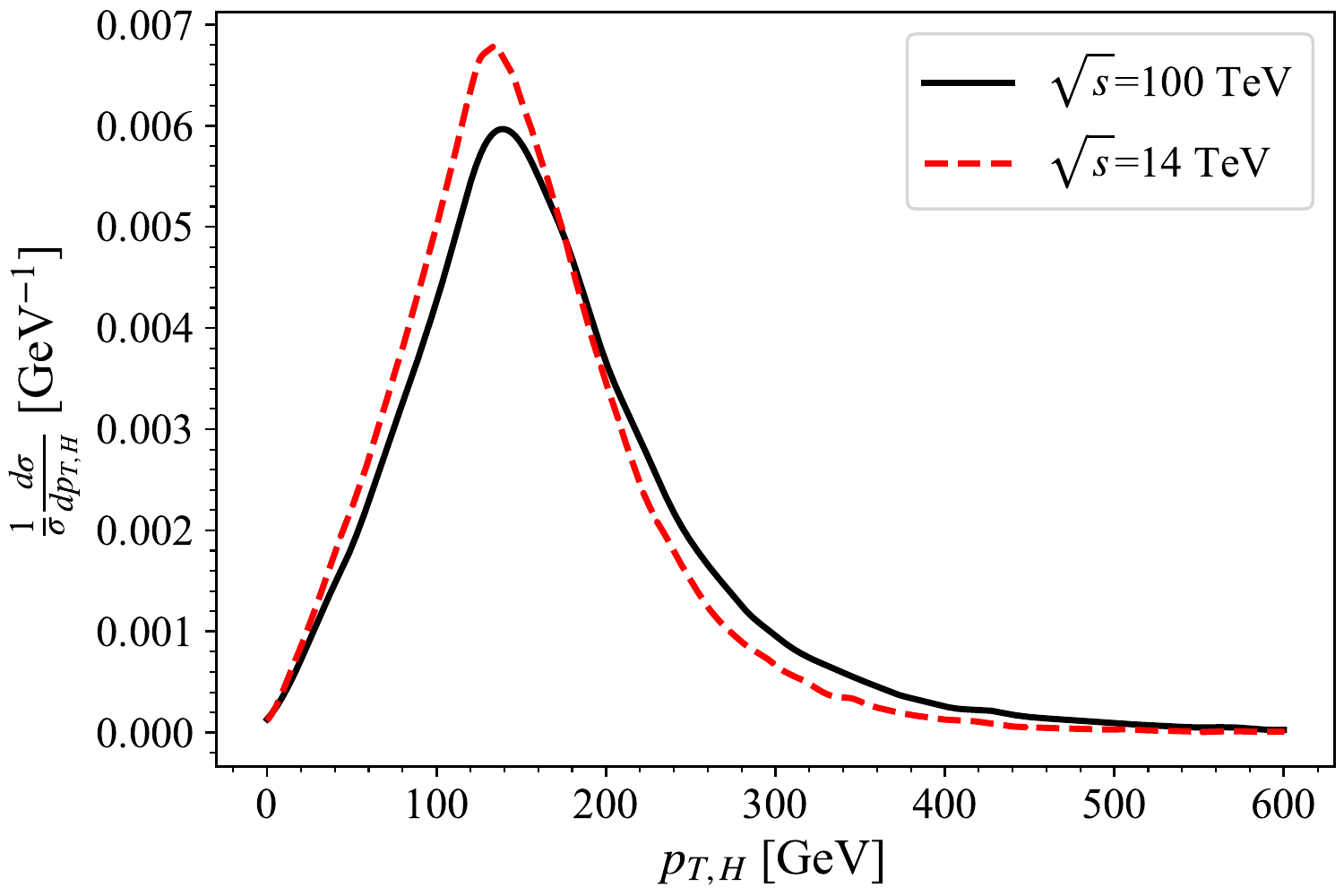}}
\caption{\label{fig:gghh_14_100TeV} \emph{The SM expectation of LO $m_{HH}$ and $p_{T,H}$ distributions at $\sqrt{s}=14$ and 100 TeV.}}
\end{figure}

in Fig.~\ref{fig:gghh_14_100TeV} we show the leading-order distributions of $m_{HH}$ and $p_T$ for the Standard Model $gg\to HH$ process in a proton-proton collider at CM energies of $14$ and $100$ TeV. We can see that the general shapes are insensitive to the CM energy of the $pp$ collider for these kinematic distributions. The kinematic distributions peak at $m_{HH} \sim 420$ GeV for the invariant mass of the Higgs pair, and $p_T \sim 150$ GeV for the transverse momentum of the single Higgs boson.

\begin{figure}[t]
\centering
\includegraphics[scale=0.45, angle=0]{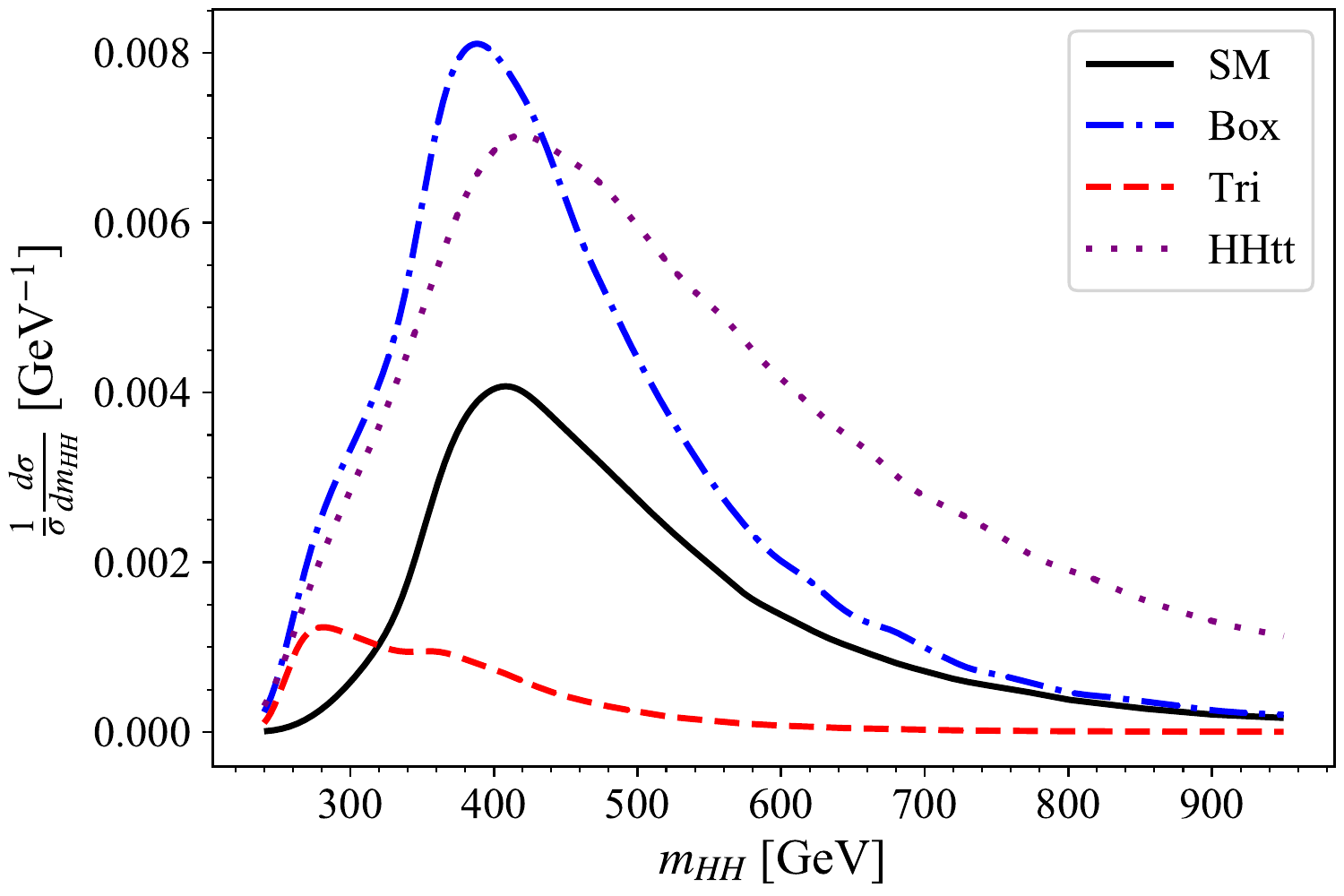}
\includegraphics[scale=0.45, angle=0]{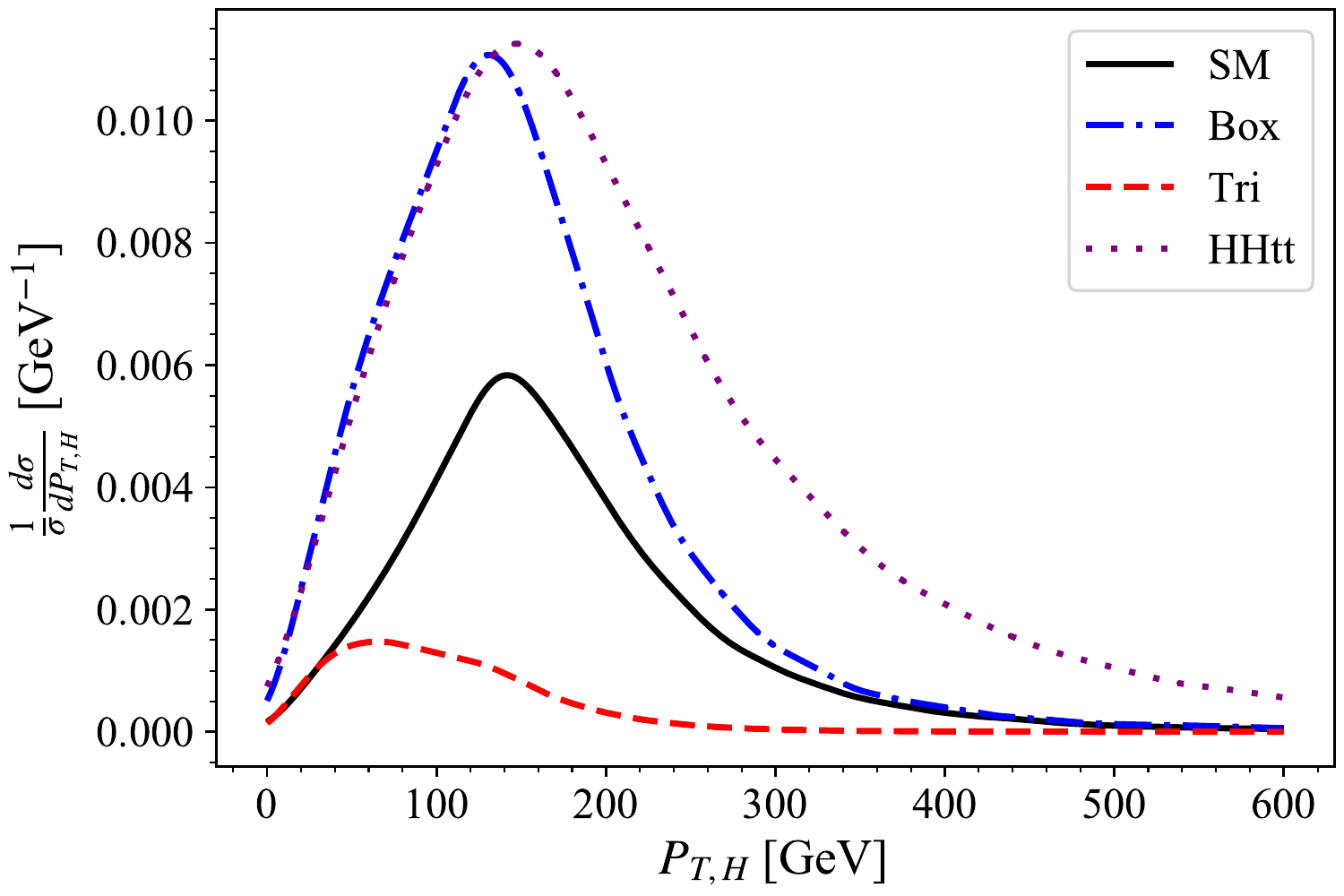}
\caption{\label{fig:gghh_100TeV_contributions_verify}\emph{Individual contribution from Triangle, Box, HHtt to the LO $m_{HH}$ and $p_{T,H}$ distributions in a hadron collider at $\sqrt{s}=100$ TeV. The results are obtained by adopting \texttt{PDF4LHC15}.}}
\end{figure}

The invariant mass of most events is remote above $2m_H$, the kinematic threshold of two outgoing Higgs bosons. Therefore, the low-energy Higgs theorem is invalid for $gg\to HH$, as mentioned at the end of Section~\ref{sec:loxs}. The contribution from $c_{3H}$ rises significantly as $m_{HH}\sim 2m_H$ since the coefficient of the loop function $F_\triangle$ is
\beq
c_{3H} \frac{3m_H^2}{\hat{s}-m_H^2} + c_{HHtt}\, .
\eeq
As a result, $c_{HHtt}$ could become dominant over $c_{3H}$ at large $m_{HH}$. Unfortunately, the total cross-section contribution from $c_{3H}$ will be suppressed since most of the events have $m_{HH} \gg 2m_H$, which was concluded in Ref.~\onlinecite{Contino:2012xk}. Therefore, it will be very difficult to measure a truly model-independent Higgs trilinear coupling solely from the total cross-section of double Higgs production. Fig.~\ref{fig:gghh_100TeV_contributions_verify} and Table \ref{tab:gghh_xs} show the individual contribution from Triangle, Box and HHtt, defined in Eq.~(\ref{eq:def_triboxbsm}), and compare them with the SM expectation. 
\bea
\textrm{Triangle} : c_{3H} =1, c_{Htt}=0, c_{HHtt}=0 \nonumber \\
\textrm{Box} : c_{3H} =0, c_{Htt}=1, c_{HHtt}=0 \\
\textrm{HHtt} : c_{3H} =0, c_{Htt}=0, c_{HHtt}=1 \nonumber
\label{eq:def_triboxbsm}
\eea
The total cross-section contribution from the SM diagrams is relatively small when $c_{3H} \sim c_{Htt}$ due to the nature of destructive interference between the triangle and box diagrams, which can be inferred from Fig.~\ref{fig:gghh_100TeV_contributions_verify}. Consequently, the measurement of $c_{3H}$ would be significantly impacted by turning on a small $c_{HHtt}$. Although there is a $30\%$ difference in the total cross-section between two PDF sets, the general shapes of the kinematic distributions are not sensitive to the PDF set we use, and we will only show the results with \texttt{PDF4LHC15} from now on.
 
\begin{table}[H]
\centering
\begin{tabular}{|l|l|l|l|l|}
\hline
$\sigma$ {[}pb{]} & Total & Triangle & Box  & HHtt \\ \hline
14 TeV      & 0.0167 & 0.004692 & 0.0349 & 0.042 \\ \hline
100 TeV      & 0.6923 & 0.146  & 1.32  & 1.98 \\ \hline
\end{tabular}
\caption{\emph{Individual contribution of cross-section from Triangle, Box and HHtt in 14 and 100 TeV hadron colliders.}}
\label{tab:gghh_xs}
\end{table}

For $p_T$ spectrum of individual contribution, $G_\Box $ has a strong $p_T$ dependence, and $p_T$ dependence in $F_\Box $ is not as strong as $G_\Box $ while $F_\triangle$ has no $p_T$ dependence. This is because the projection of the angular momentum of head-on gluons with the same helicity on the beam axis is zero, $J_z=0$, which corresponds to $F_\triangle$ and $F_\Box$. On the other hand, $G_\Box $ has $J_z=2$ on the beam axis resulted from opposite helicity gluons \cite{Glover:1987nx,Plehn:1996wb}.
Nevertheless, only the $S$-wave orbital angular momentum is contained by $F_\triangle$ since only the scalar Higgs couplings are included in Fig.~\ref{fig:gghh_feyndiagrams}(a) and Fig.~\ref{fig:gghh_feyndiagrams}(c). In other words, $F_\triangle$ is $p_T$ independent. Therefore, 
the phase space is the only source of all the $p_T$ dependence in the $c_{3H}$ and $c_{HHtt}$. However, the higher-order terms of $\hat{s}/m_t^2$ expansion for the $J_z=0$ component of the $D$-wave angular dependence leave some $p_T$ dependence for $F_\Box $~\cite{Dawson:2012mk}. $G_\Box $ has a strong $p_T$ dependence due to the $D$-wave nature.
Furthermore, the difference of the angular momentum projection between $F_\triangle$/$F_\Box$ and $G_\Box$ also explains the lack of interference between the two contributions in Eq.~(\ref{eq:amp}).
in Fig.~\ref{fig:gghh_100TeV_contributions_verify} we also show the $p_T$ spectrum of individual contribution from Triangle, Box and HHtt. The contribution from Triangle are suppressed in general, as in the $m_{HH}$ distribution since the Higgs propagator in Fig.~\ref{fig:gghh_feyndiagrams}(b) is off-shell.

\subsection{$g$ $g$ $\rightarrow$ $H$ $H$ $g$ Contribution}
Numerical calculations of $gg\to HHg$ are performed in 14 and 100 TeV hadron colliders. The standard model (SM) expectation of the cross-section at a 14 TeV $pp$ collider for this channel is 0.014 pb. At 100 TeV, the SM rate rises significantly to 0.85 pb. The contribution of the double Higgs production from $gg \rightarrow HHg$ is comparable to LO $gg \rightarrow HH$, 0.0167 pb and 0.692 pb at 14 and 100 TeV, respectively, and consist half of the total production rate.

\begin{figure}[H]
\centering
\subfloat[]{\includegraphics[width=.45\linewidth]{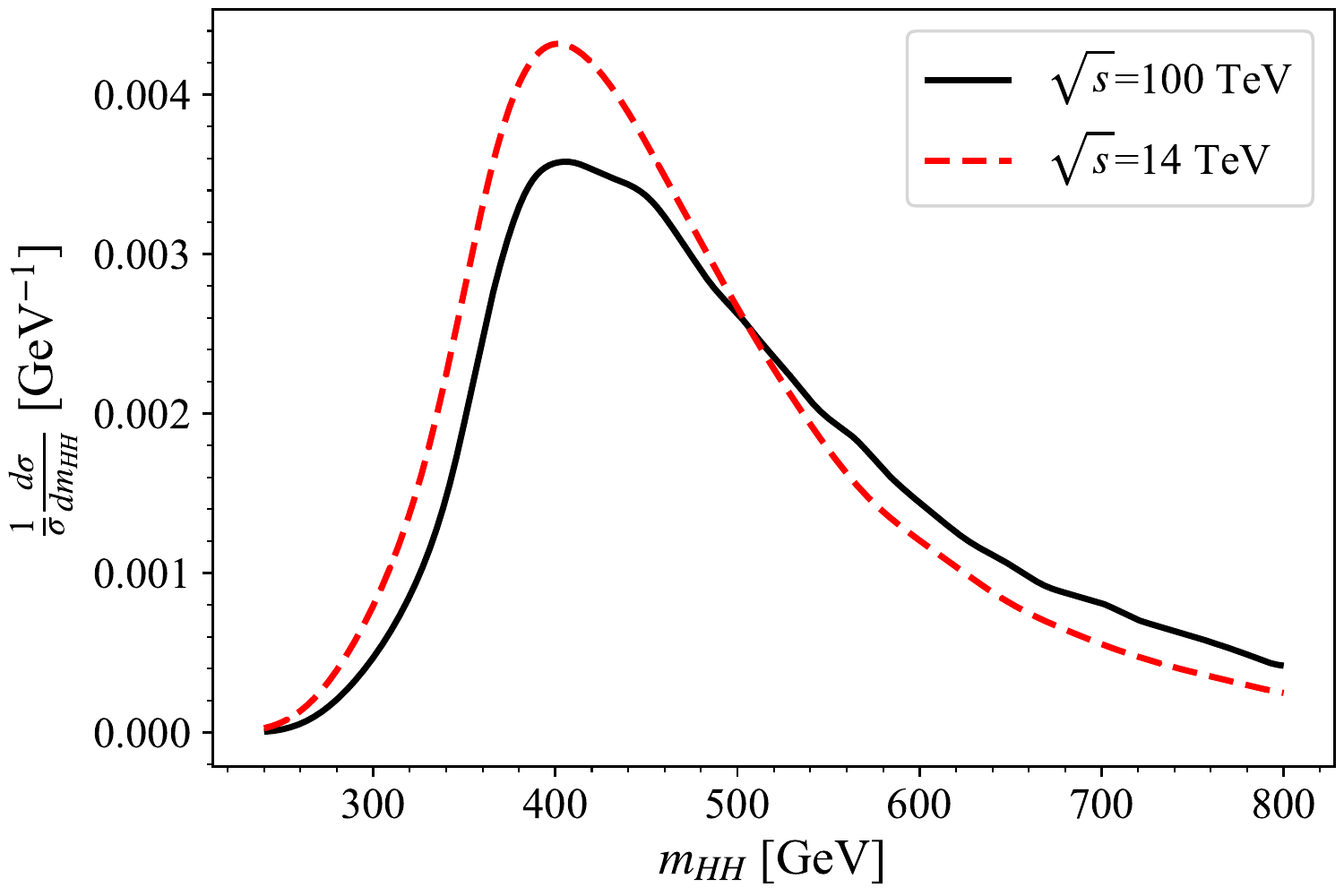}}\qquad
\subfloat[]{\includegraphics[width=.45\linewidth]{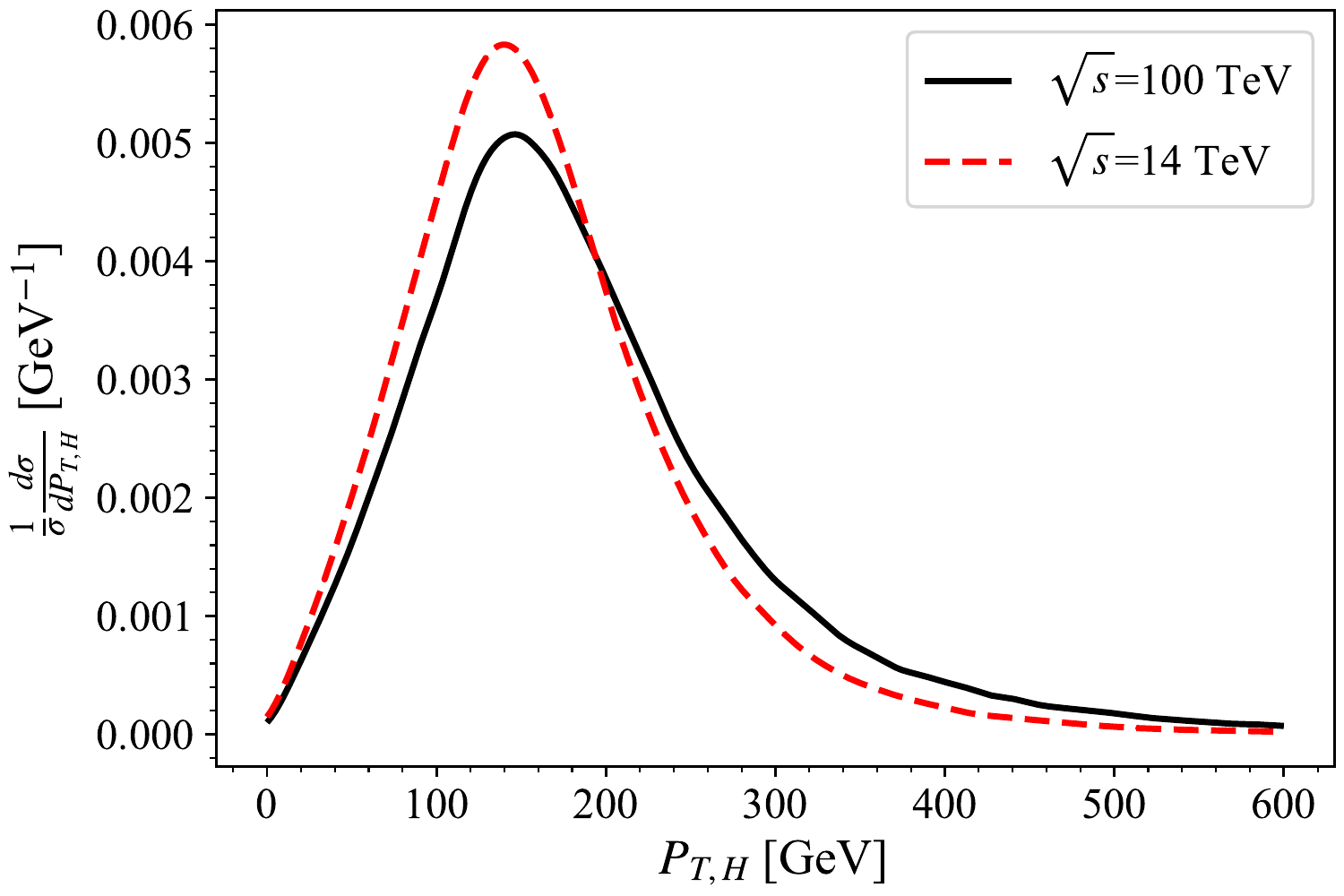}}
\caption{\label{fig:gghhg_14_100TeV} \emph{Kinematic distributions for $gg \rightarrow HHg$ in the SM at $\sqrt{s}=14$ and 100 TeV.}}
\end{figure}

In Fig.~\ref{fig:gghhg_14_100TeV}, we present the SM $m_{HH}$ and $p_T$ distributions for $gg\to HHg$ in 14 and 100 TeV hadron colliders. Since the dominant contributions come from one loop triangle and box diagrams shown in Fig.~\ref{fig:dia_t_gghhg} and Fig.~\ref{fig:dia_b_gghhg} repectively, which has the same loop function as $gg\to HH$ channel. We therefore expect the general shapes of kinematic distributions shown in Fig.~\ref{fig:qghhq_14_100TeV} are insensitive to the center-of-mass energy of the hadron collider and share some similarity to Fig.~\ref{fig:gghh_14_100TeV}. Indeed, comparing to LO counterpart, the general shapes are very similar except they are wider.

\begin{figure}[H]
\centering
\subfloat[]{\includegraphics[width=.45\linewidth]{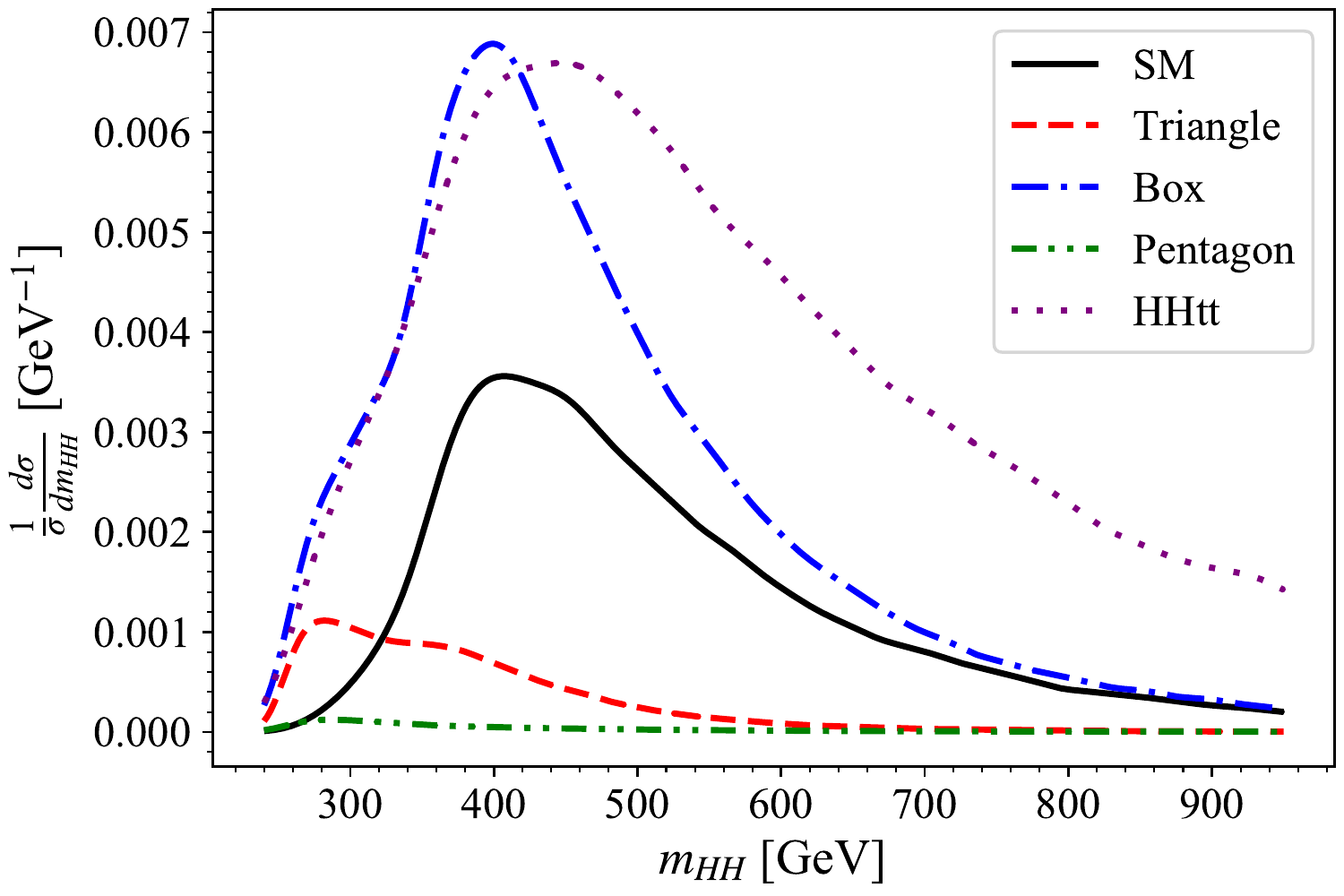}}\qquad
\subfloat[]{\includegraphics[width=.45\linewidth]{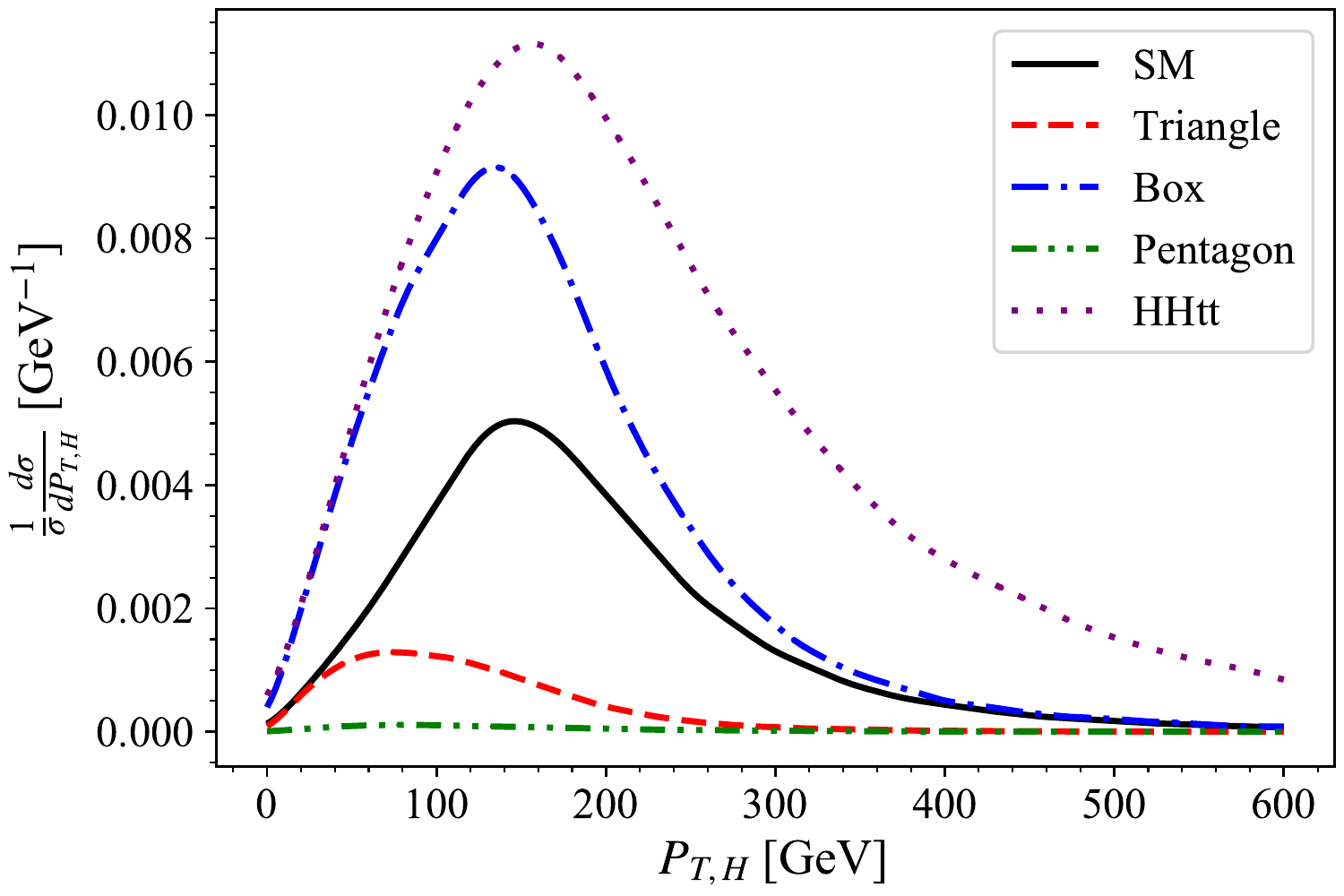}}
\caption{\label{fig:gghhg_100TeV} \emph{Individual contribution from Triangle, Box, Pentagon and HHtt to the kinematic distributions for $gg \rightarrow HHg$ in the SM at $\sqrt{s}=100$ TeV.}}
\end{figure}

The invariant mass distribution peaks at slightly lower $m_{HH} \sim 380$ GeV, while the maximum of $p_T$ distribution is still at $p_T \sim 150$ GeV. Unlike the $gg\to HH$ counterpart, we can easily see the peaks in Fig.~\ref{fig:gghh_14_100TeV} move to the right slightly and get wider as the CM energy of the head-on hadrons increases. There are even more events that have $m_{HH} \gg 2m_H$. Hence the contribution to the total cross-section from $c_{3H}$ will be further suppressed as we already discussed in Section~\ref{sec:gghh_kin}.

in Fig.~\ref{fig:gghhg_100TeV} we show the individual contribution of $gg \to HHg$ channel from Triangle, Box, Pentagon, and HHtt diagrams, shown in Fig.~\ref{fig:dia_t_gghhg} to Fig.~\ref{fig:dia_p_gghhg} , in the $m_{HH}$ and $P_T$ distribution while comparing them with the SM expectation. Triangle, box, pentagon contributions come from SM diagrams, while HHtt is the contribution from diagrams with anomalous $HHtt$ coupling.
As we can see, $gg \rightarrow HHg$ have similar kinematic distributions and similar reactions to the parameters, $c_{3H}$, $c_{Htt}$, and $c_{HHtt}$, due to the same loop function in Eq.~(\ref{eq:gghh_loopfn}). 
The $c_{3H}$ contribution to the total cross-section is still suppressed when $c_{3H} \sim c_{HHtt}$, and $c_{HHtt}$ would have a significant impact on the cross-section of $gg \rightarrow HHg$ process as we saw in the LO case.


\subsection{$q$ $g$ $\rightarrow$ $H$ $H$ $q$ Contribution}
For the $qg\to HHq$ process, we can categorize the contributions by the number of couplings to gluons. The contributions from diagrams with more than one strong couplings and contributions from diagrams with only one strong coupling are denoted as $QCD_2$ and $QCD_1$, respectively. 

\begin{figure}[!hbt]
\centering
\subfloat[]{\includegraphics[width=.45\linewidth]{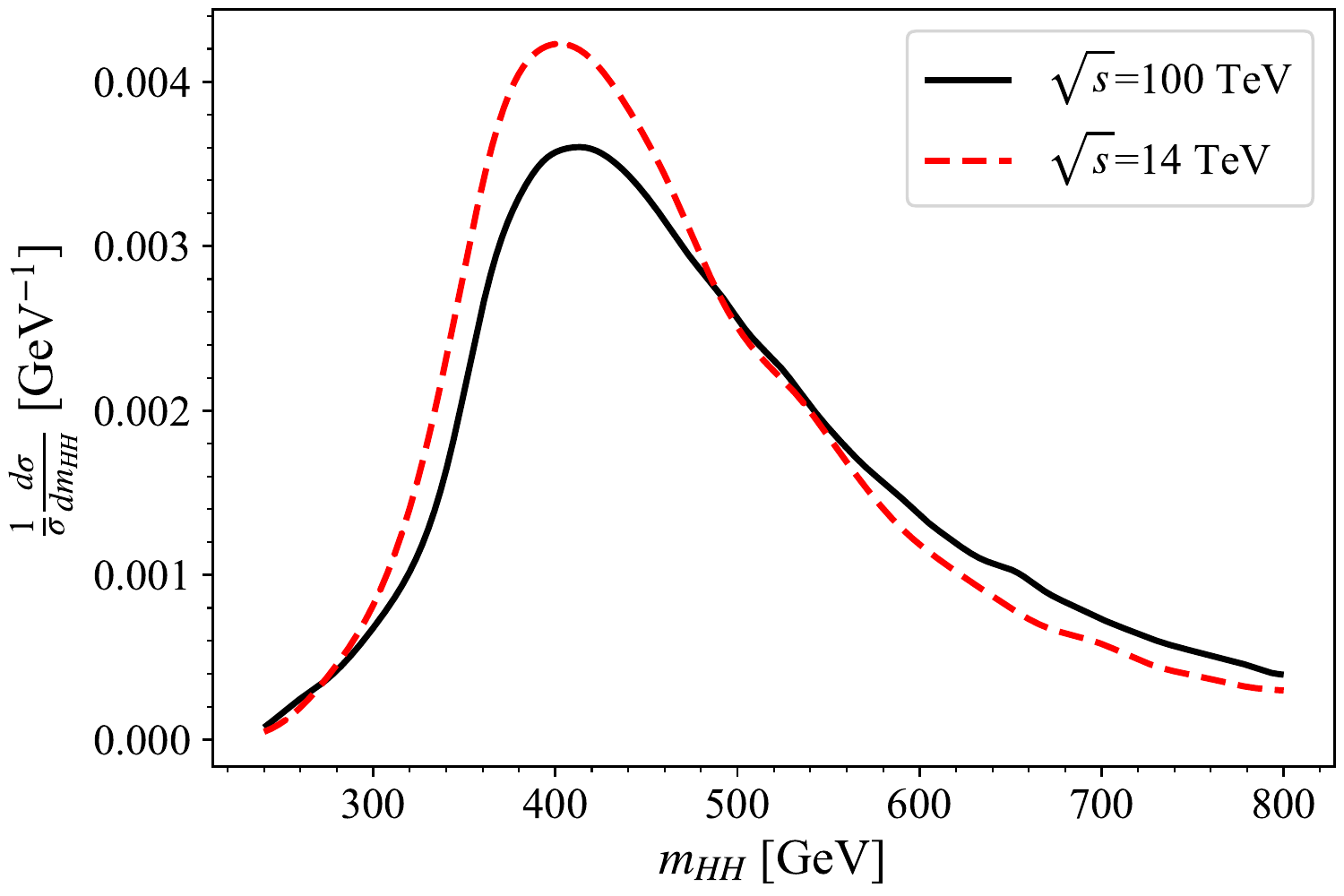}}\qquad
\subfloat[]{\includegraphics[width=.45\linewidth]{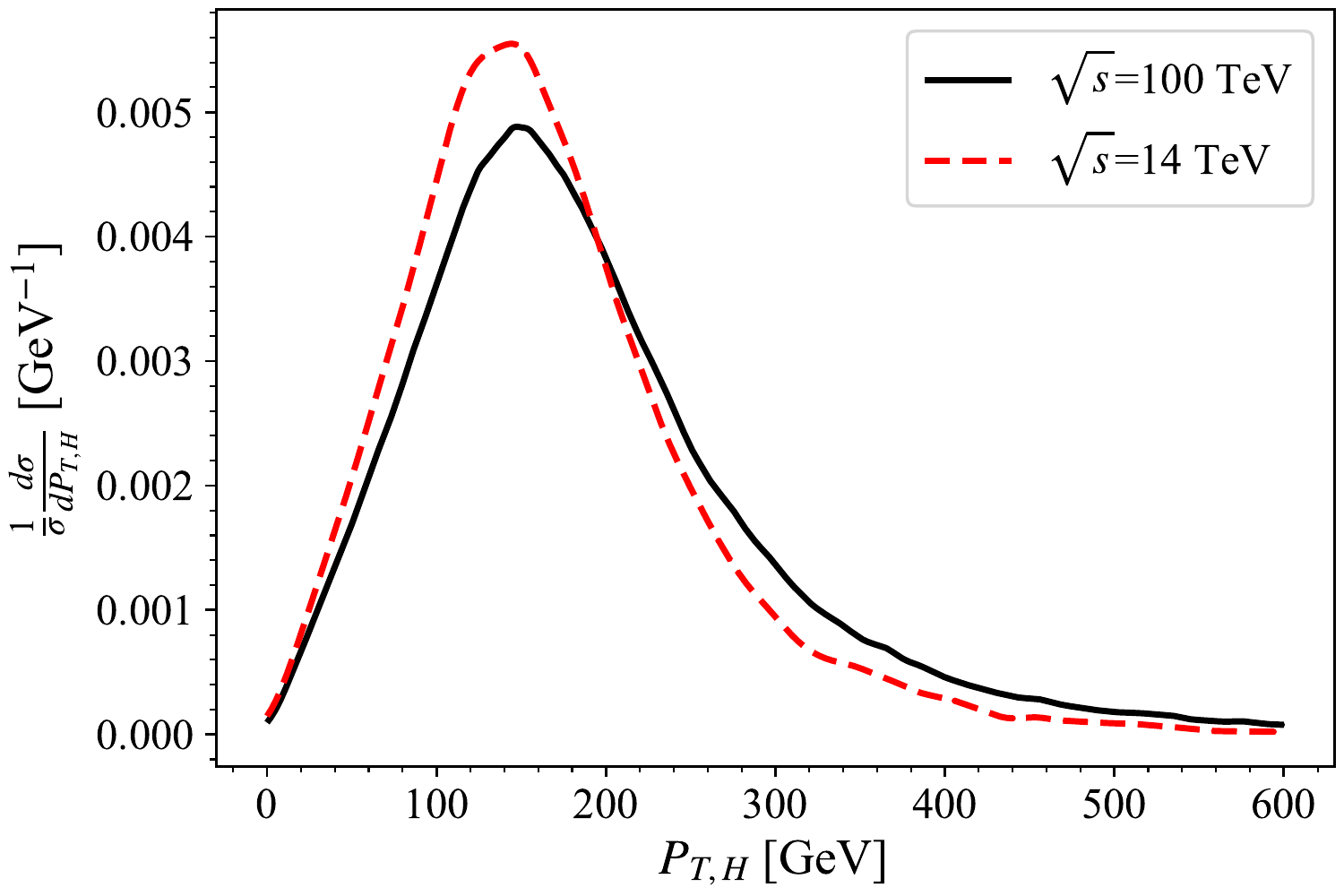}}
\caption{\label{fig:qghhq_14_100TeV} \emph{Kinematic distributions for $qg \rightarrow HHq$ in the SM at $\sqrt{s}=14$ and 100 TeV.}}
\end{figure}

\begin{figure}[!hbt]
\centering
\subfloat[]{\includegraphics[width=.45\linewidth]{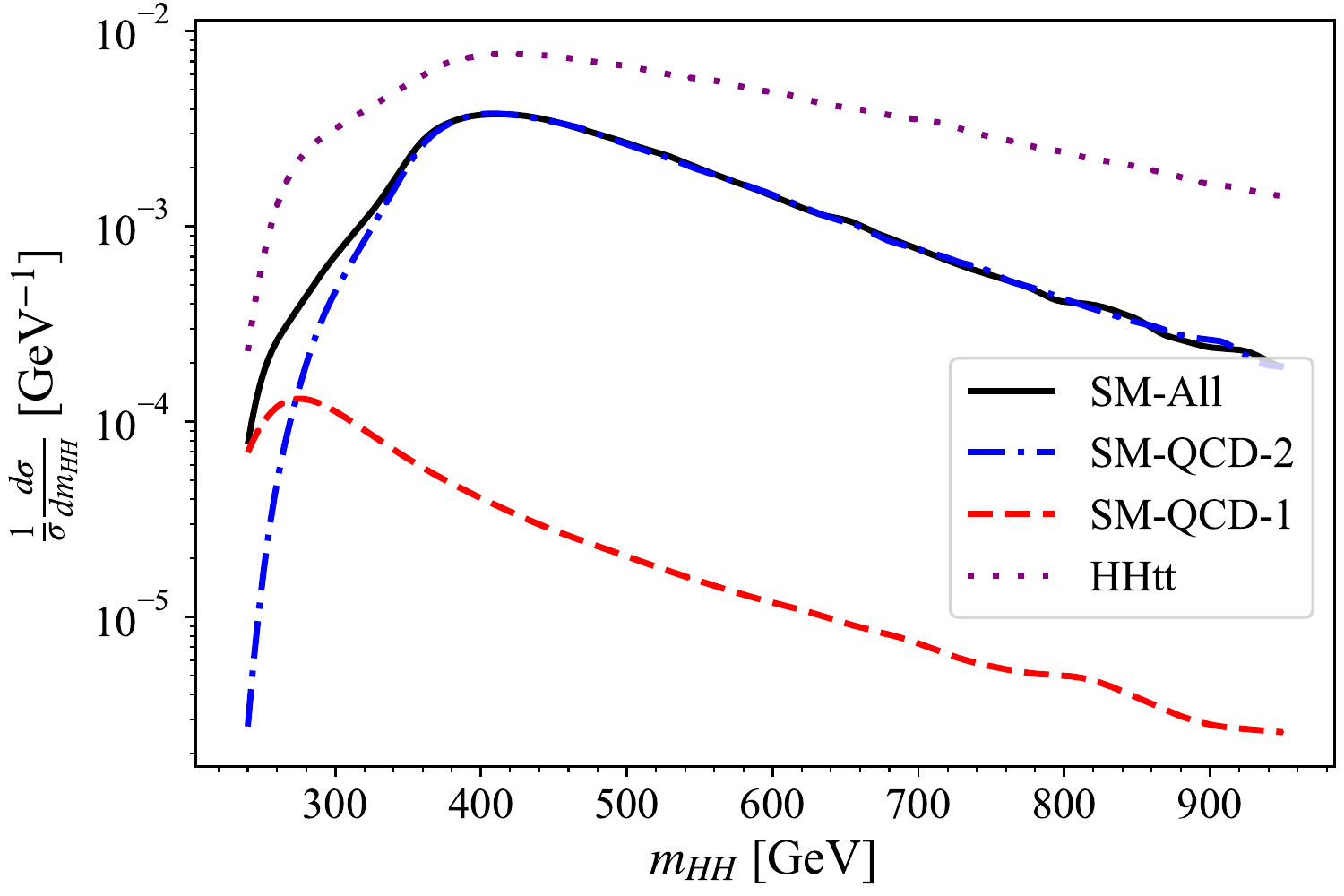}}\qquad
\subfloat[]{\includegraphics[width=.45\linewidth]{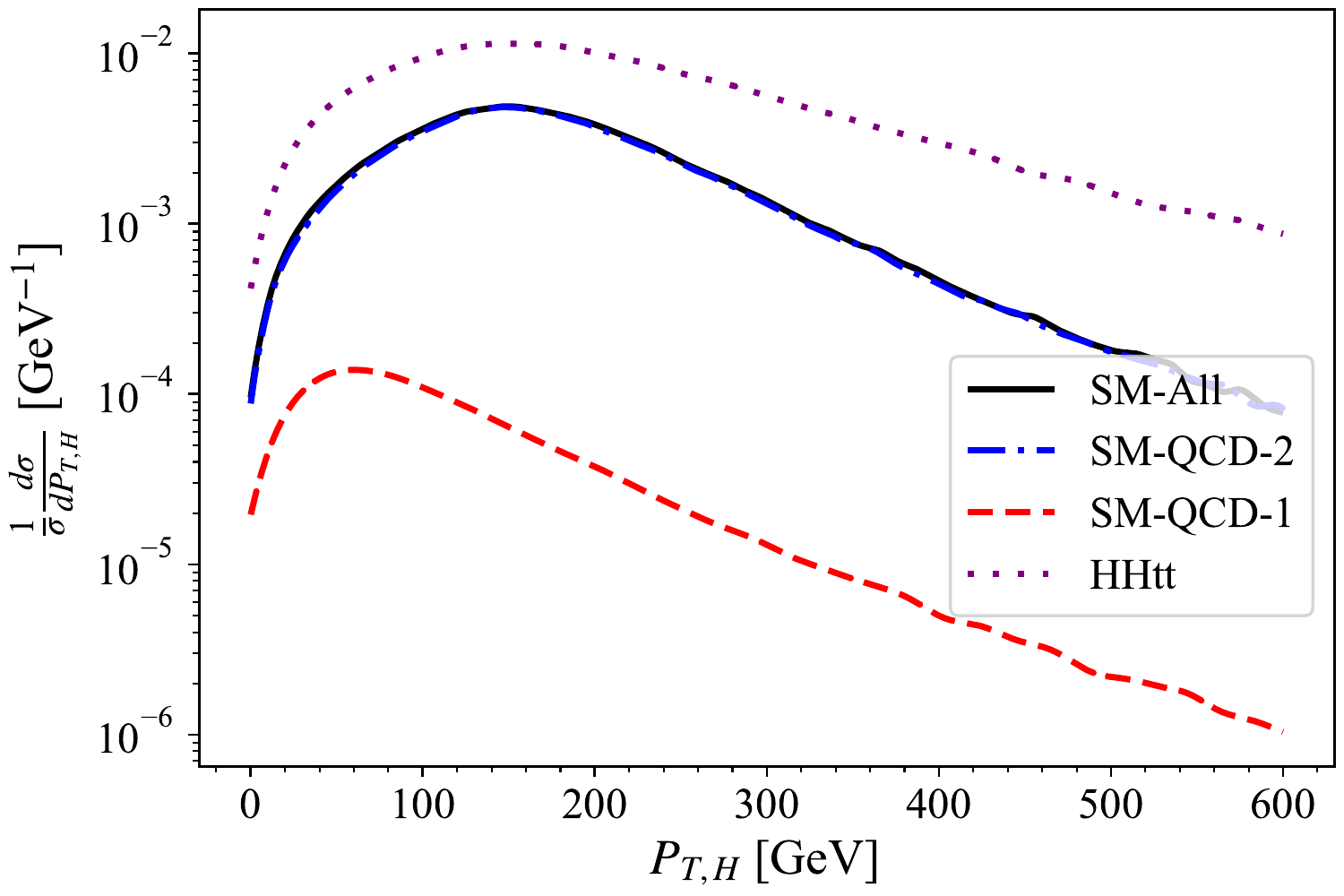}}
\caption{\label{fig:qghhq_100TeV} \emph{Individual contribution from $QCD_1$, $QCD_2$, $HHtt$ to the kinematic distributions for $qg \rightarrow HHq$ in the SM at $\sqrt{s}=100$ TeV.}}
\end{figure}


In Fig.~\ref{fig:qghhq_14_100TeV}, we show the kinematic distributions, including $m_{HH}$ and $p_T$, for SM $qg\to HHq$ in the hadron collider at CM energies $14$ and $100$ TeV. Since the dominant contributions come from one loop triangle and box diagrams shown in Fig.~\ref{fig:dia_qghhq_qcd}, which has the same loop function as $gg\to HH$, we expect Fig.~\ref{fig:qghhq_14_100TeV} to share some similarity to Fig.~\ref{fig:gghh_14_100TeV}. Comparing to $gg\to HH$ counterpart, the general shapes are very similar except they are wider. 
The maximum for invariant mass and $p_T$ distributions are still the same, $m_{HH} \sim 420$ GeV and $p_T \sim 150$ GeV respectively. For most events, the invariant mass is far above the threshold, $2m_H$ as we already discussed in Section~\ref{sec:gghh_kin}. Unlike $gg\to HH$ counterpart, we can easily see the peaks in Fig.~\ref{fig:gghh_14_100TeV} move to the right slightly and get wider as the CM energy of the hadron collider increases. This means even more events have $m_{HH} \gg 2m_H$.

Fig.~\ref{fig:qghhq_100TeV} shows the individual contribution of $qg \to HHq$ from $QCD_2$, $QCD_1$, and $HHtt$ in the $m_{HH}$ and $P_T$ distribution while comparing these contributions with the SM expectation. $QCD_2$ is the contribution from diagrams with more than one gluon coupling as shown in Fig.~\ref{fig:dia_qghhq_qcd}. $QCD_1$ contribution includes all the rest of SM diagrams for $qg \to HHq$, including tree, triangle, box, and pentagon diagrams, while $HHtt$ contribution comes from diagrams with an $HHtt$ coupling. 
As we can see, $QCD_2$ is the dominant contribution of $qg \to HHq$ since weak couplings suppress $QCD_1$ contribution. Again we can expect $gg \rightarrow HH$, $qg \rightarrow HHq$ have similar kinematic distributions and similar reactions to the parameters, $c_{3H}$, $c_{Htt}$, and $c_{HHtt}$.
We find the CM energy insensitivity of the general shapes of these distributions. Therefore, in what follows, we only present the result for $\sqrt{s}=100$ TeV. As we saw in the $gg \to HH$ channel, the contribution self-coupling to the total cross-section from diagrams that involve the trilinear Higgs is still suppressed when $c_{3H} \sim c_{HHtt}$, and $c_{HHtt}$ would have significant effects on the cross-section of the $qg \rightarrow HHq$ process.

For $QCD_1$ contribution, more events have lower invariant mass, $m_{HH}$, and transverse momentum, $p_T$. The peak of invariant mass distribution is at $m_{HH} \sim 280$ GeV, while the maximum of $p_T$ distribution is at $p_T \sim 60$ GeV.
Hence, the $QCD_1$ contribution at small invariant mass, $m_{HH} \sim 2m_H$, is more important than the $QCD_2$ contribution, which is dominant at large $m_{HH}$.

\begin{figure}[!hbt]
\centering
\subfloat[]{\includegraphics[width=.45\linewidth]{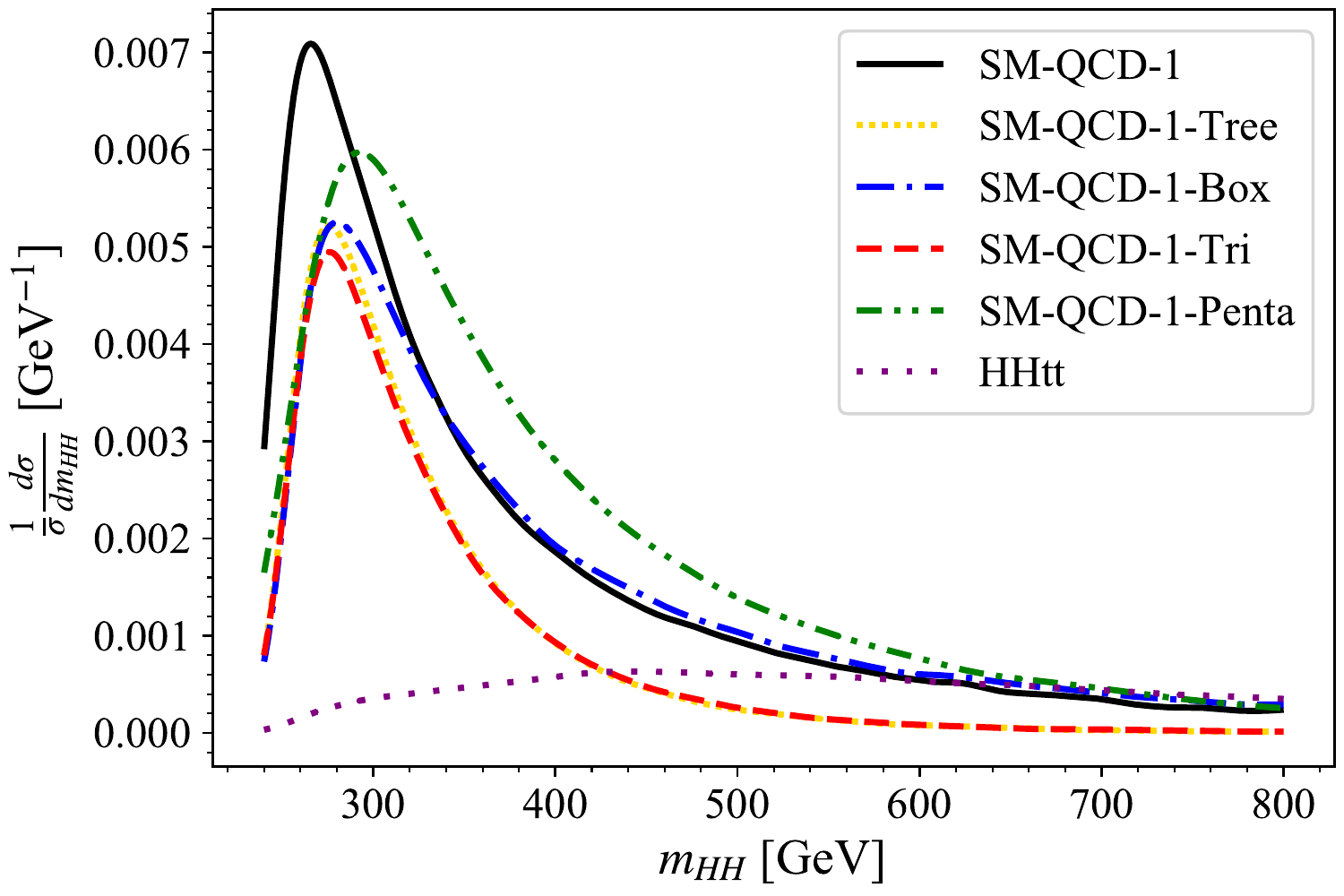}}\qquad
\subfloat[]{\includegraphics[width=.45\linewidth]{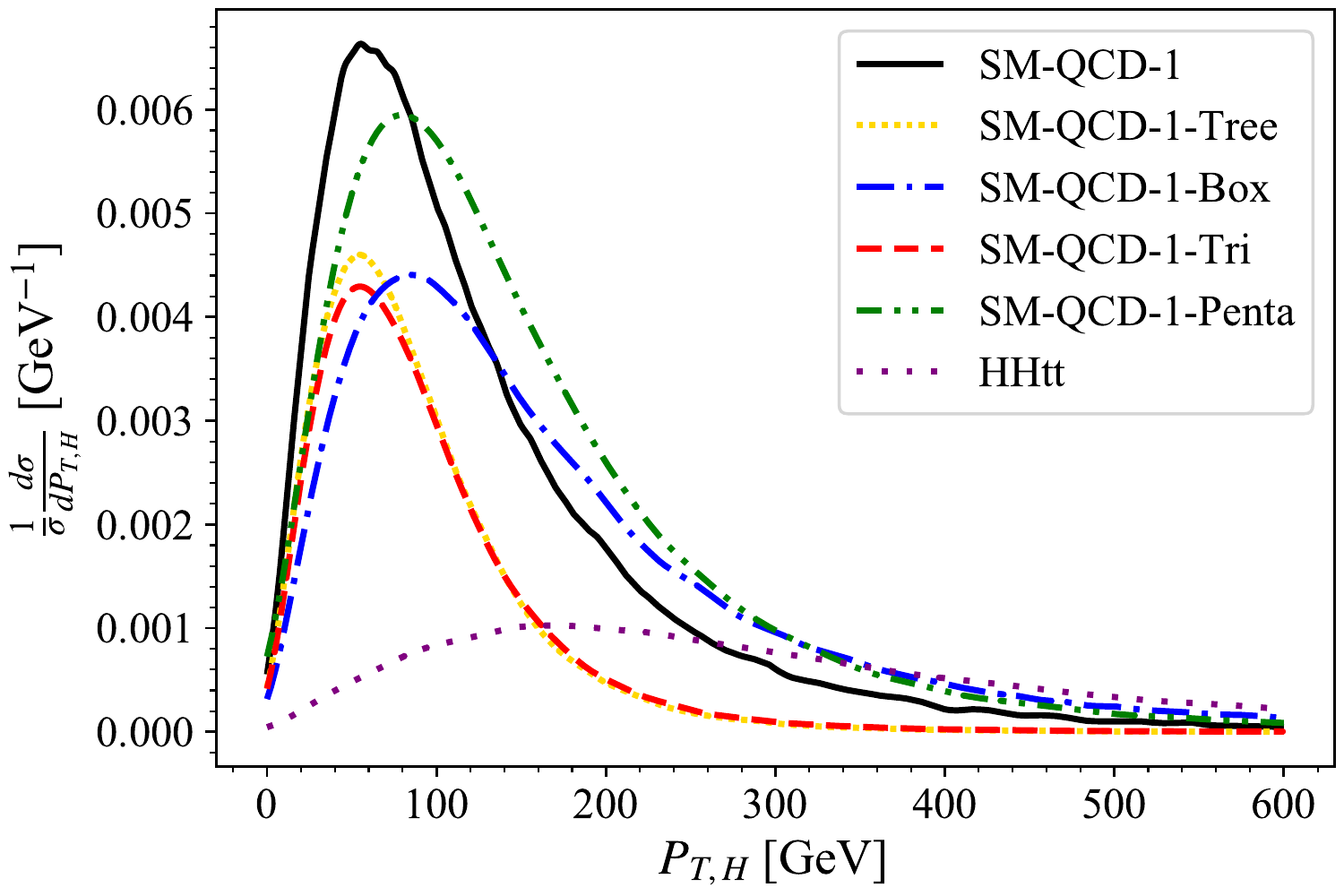}}
\caption{\label{fig:qghhq_qed_100TeV} \emph{Individual $QCD_1$ contribution from tree, triangle, box, pentagon diagrams to the kinematic distributions for $qg \rightarrow HHq$ in the SM at $\sqrt{s}=100$ TeV.}}
\end{figure}

In Fig.~\ref{fig:qghhq_qed_100TeV}, for completeness, we further divide the $QCD_1$ contribution into the individual contribution from the tree, triangle, box, pentagon, and $HHtt$ contributions from corresponding diagrams and compare them with the total $QCD_1$ contribution. We can categorize individual contributions shown in Fig.~\ref{fig:qghhq_qed_100TeV} into two groups. Group one consists of tree and box diagrams, while group two consists of the triangle and pentagon diagrams. As a result, the interference between the contribution from group one and group two is destructive, which can also be inferred from Fig.~\ref{fig:qghhq_qed_100TeV}.
The $HHtt$ contribution when $c_{HHtt}=1$ comes from diagrams with only one gluon coupling, as shown in Fig.~\ref{fig:dia_box} and Fig.~\ref{fig:dia_tri} is very small compared to the $QCD_1$ contribution, which is already small. 
However, the $HHtt$ contribution shown in Fig.~\ref{fig:qghhq_100TeV} is dominant over SM value. Therefore, most of $HHtt$ contribution to $qg \rightarrow HHq$ channel comes from diagrams shown in Fig.~\ref{fig:dia_qghhq_qcd}.

\subsection{NLO}
\begin{figure}[H]
\centering
\subfloat[]{\includegraphics[width=.45\linewidth]{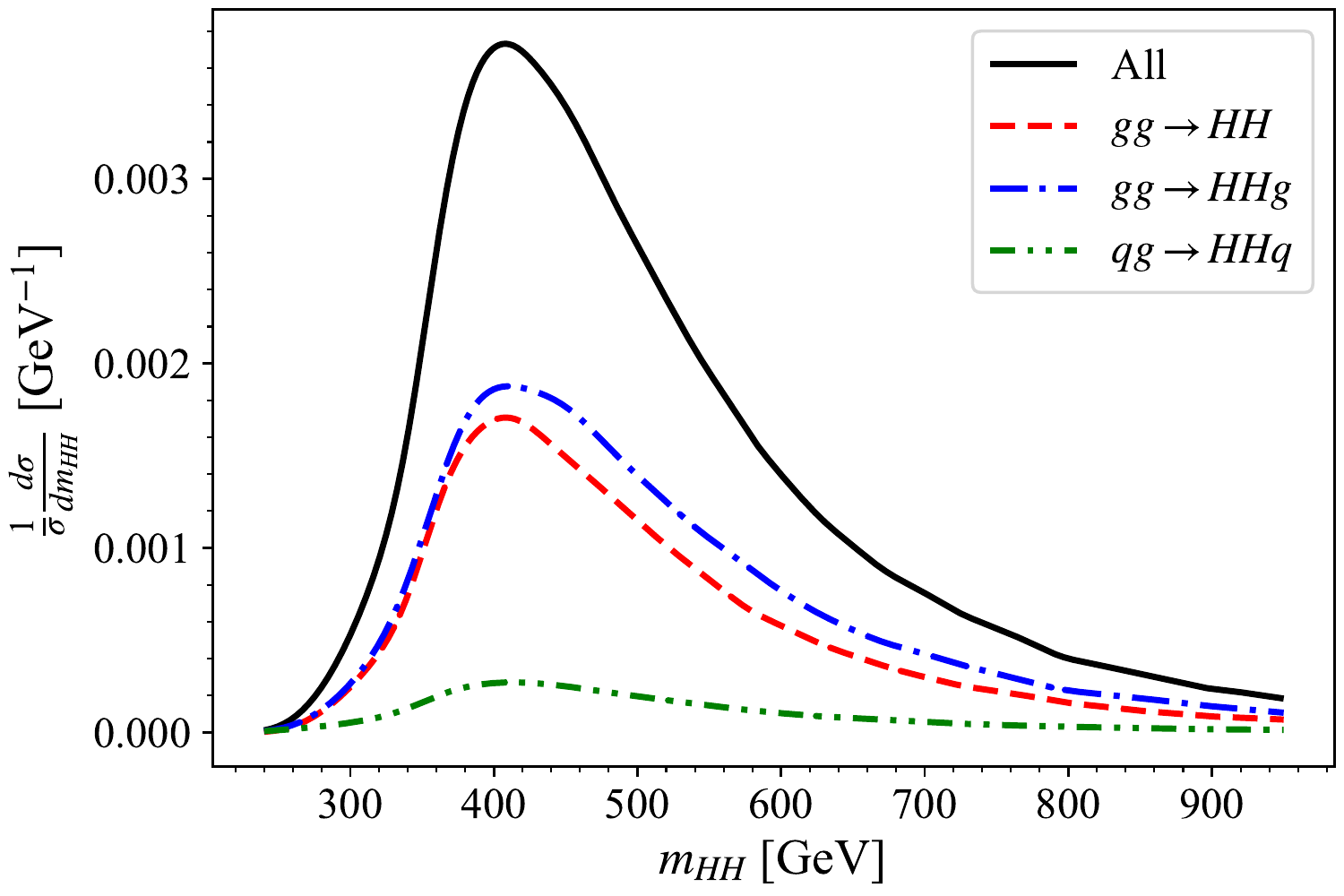}}\qquad
\subfloat[]{\includegraphics[width=.45\linewidth]{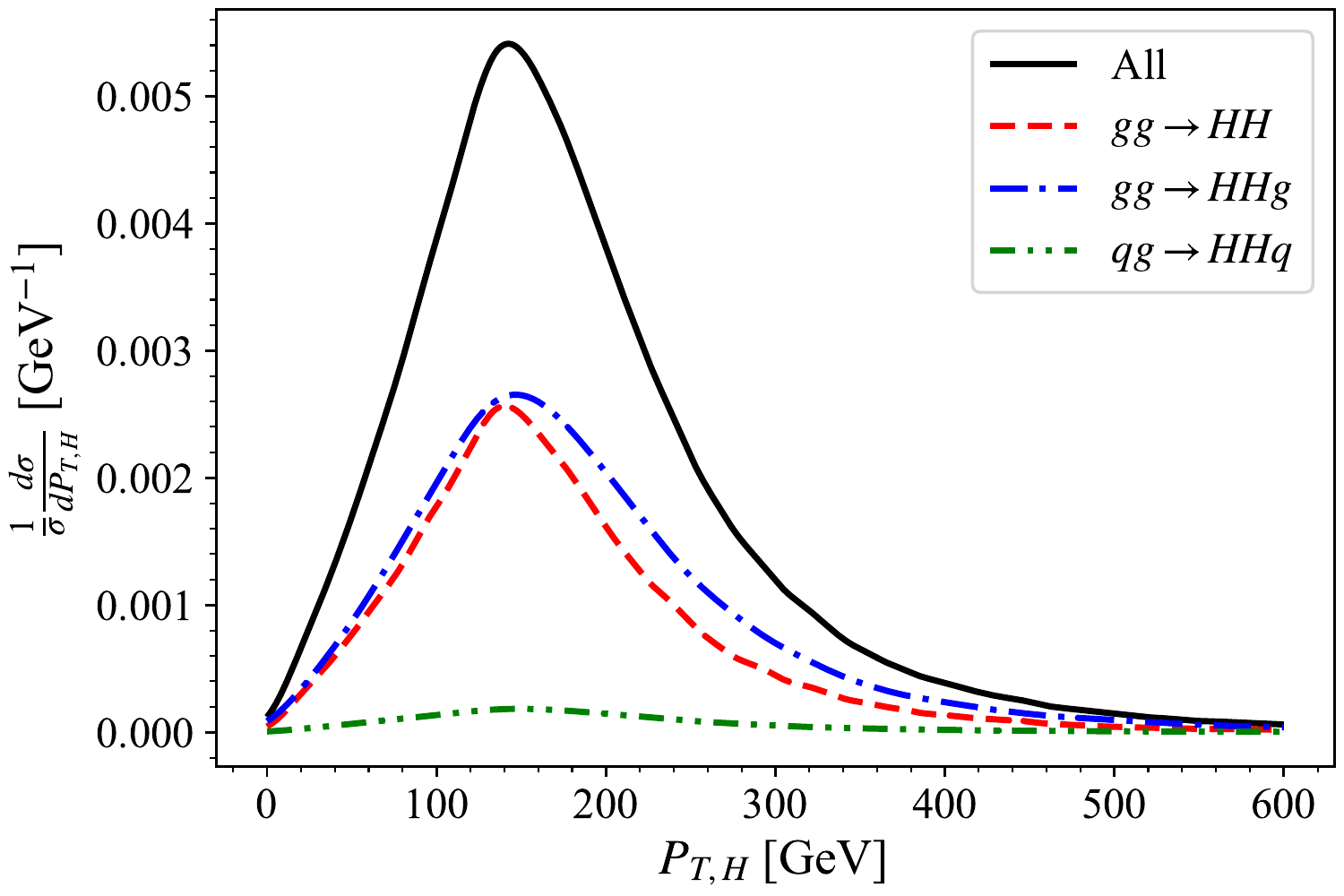}}
\caption{\label{fig:NLO_contribution_100TeV_kinematics} \emph{Individual contribution from $g g \rightarrow H H$, $g g \rightarrow H H g$, $q g \rightarrow H H q$ to the kinematic distributions for double Higgs production in the SM at $\sqrt{s}=100$ TeV.}}
\end{figure}

In Fig.~\ref{fig:NLO_contribution_100TeV_kinematics}, we show the individual SM contributions from all processes up to NLO, including $gg \rightarrow HH$, $gg \rightarrow HHg$, $qg \rightarrow HHq$ and $qq \rightarrow HHg$, while comparing them with the total NLO SM contributions in the $m_{HH}$ and $P_T$ distribution. 

We can see that $gg \rightarrow HH$ and $gg \rightarrow HHg$ contributions are equivalently dominant while $qg \to HHq$ only contributes about 10\% of the total cross-section. Similar kinematic distributions to $gg \rightarrow HH$ contribution are found in $gg \rightarrow HHg$ and $qg \rightarrow HHq$ contributions. We can therefore expect the full NLO contribution have similar kinematic distributions and similar reactions to the parameters. As expected, The peaks of invariant mass and $p_T$ are still at $m_{HH} \sim 420$ GeV and $p_T \sim 150$ GeV, respectively.

The parameters we previously used, $c_{3H}$, $c_{Htt}$ and $c_{HHtt}$, have to be generalized since the contributions from diagrams such as Fig.~\ref{fig:dia_box} do not have $HHH$, $Htt$, or $HHtt$ couplings. By following the same idea, we can create a new parameter set based on all Higgs couplings
 $c_{3H}$, $c_{Htt}$, $c_{HHtt}$, $c_{HWW}$, $c_{HZZ}$, $c_{HZZ,Htt}$, and $c_{Hbb(cc)}$ such that 
\begin{eqnarray}
\label{eq:newpar}
\mathcal{M}&=&c_{3H}\mathcal{M}_{3H} +c_{Htt}\mathcal{M}_{Htt}+c_{Htt}\mathcal{M}_{HHtt}+c_{HWW}\mathcal{M}_{HWW} \nonumber \\
&&+c_{HZZ}\mathcal{M}_{HZZ}+c_{HZZ,Htt}\mathcal{M}_{HZZ,Htt}+c_{Hbb(cc)}\mathcal{M}_{Hbb(cc)},
\end{eqnarray}
where
\begin{eqnarray}
c_{3H} &=& g_{H^3}g_{Htt}\frac{v^2}{3m_{H}^2 m_{t}}, \quad c_{HHtt} = g_{HHtt}\frac{v^2}{m_t}, \quad c_{Htt} = \left(g_{Htt}\frac{v}{m_t}\right)^2\nonumber \\ 
c_{Hbb(cc)} &=& \left(g_{Hbb(cc)}\frac{v}{m_b}\right)^2, \quad c_{HWW} = \left(g_{HWW}\frac{v}{2m_W^2}\right)^2, \quad c_{HZZ} = \left(g_{HZZ}\frac{v}{2m_Z^2}\right)^2\nonumber \\ 
c_{HWW, Htt} &=& \left(g_{HWW}g_{Htt}\frac{v^2}{2m_W^2 m_t}\right)^2. \nonumber
\end{eqnarray}
$\mathcal{M}_{x}$ is amplitude of diagrams with coupling $x$ while $\mathcal{M}_{HWW, Htt}$/$\mathcal{M}_{HZZ, Htt}$ is amplitude of diagram with $HWW/HZZ$ and $Htt$ couplings. If we adopt this new set of parameters into $gg \to HH$ channel, we can see that
\begin{equation}
c_{3H}=c_{3H}, \quad c_{Htt}=c_{Htt} ~\text{and}~ c_{HHtt}=c_{HHtt}.
\end{equation}

\begin{figure}[H]
\centering
\subfloat[]{\includegraphics[width=.45\linewidth]{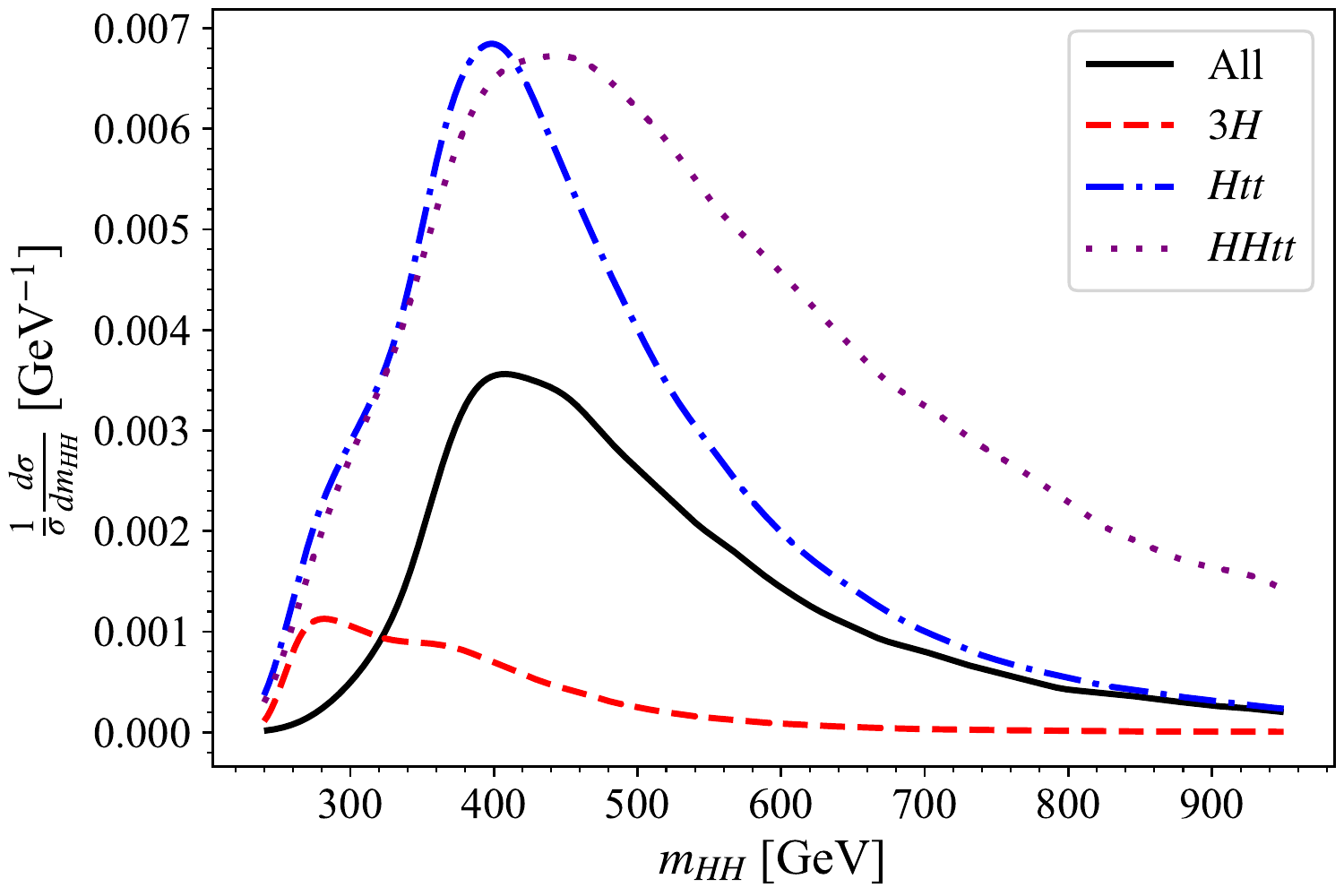}}\qquad
\subfloat[]{\includegraphics[width=.45\linewidth]{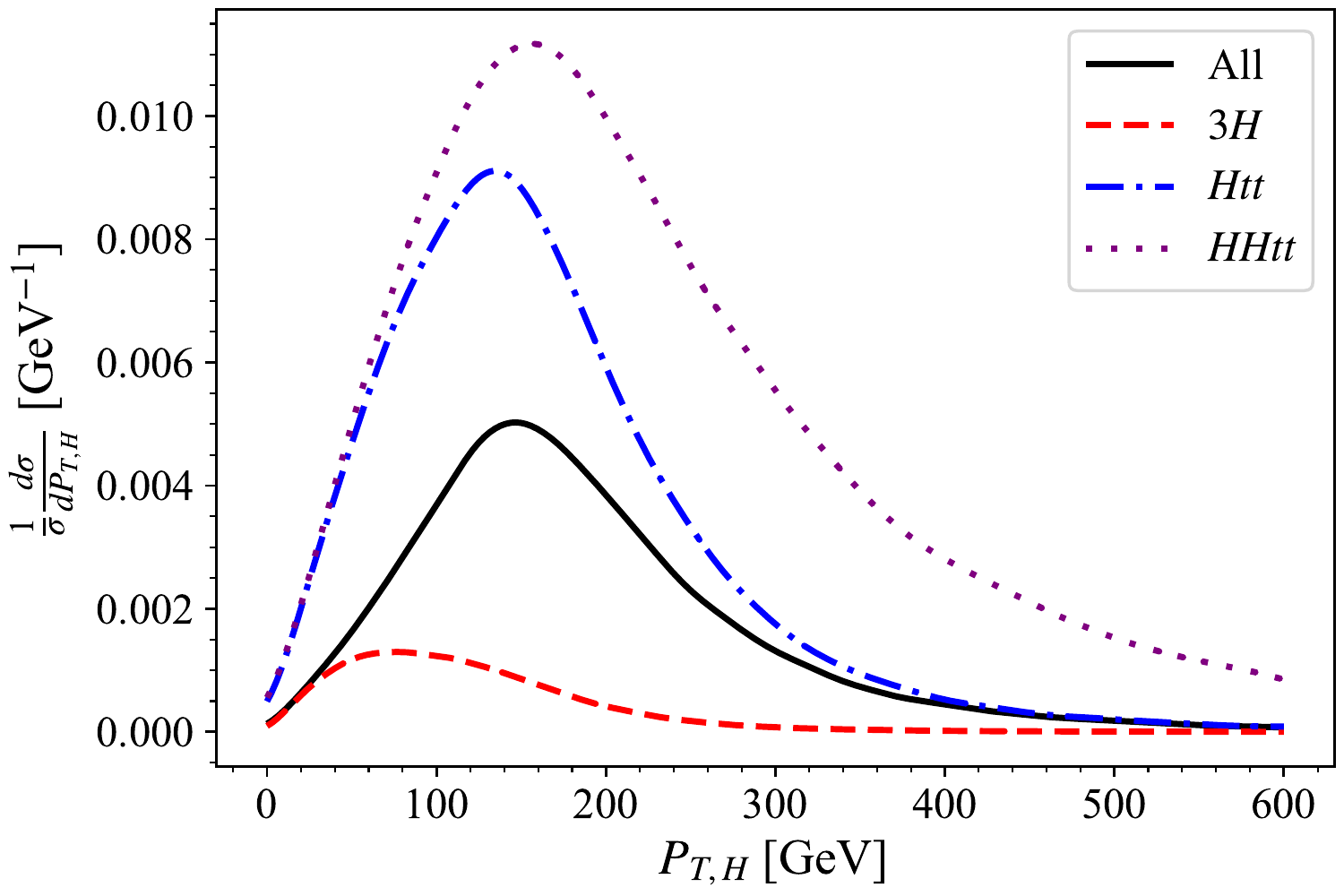}}
\caption{\label{fig:NLO_100TeV_kinematics} \emph{Individual contribution from $c_{3H}$, $c_{Htt}$ and $c_{HHtt}$ to the kinematic distributions for double Higgs production at $\sqrt{s}=100$ TeV.}}
\end{figure}

 Fig.~\ref{fig:NLO_100TeV_kinematics} shows the individual contribution from $c_{3H}$, $c_{Htt}$, $c_{HHtt}$, and the distribution of SM expectation in the $m_{HH}$ and $p_{T,H}$ distribution. The Higgs low-energy theorem breaks down since most events have an invariant mass much higher than the kinematic threshold at $2m_H$ as discussed at the end of Section~\ref{sec:loxs}.
 We note that $c_{HHtt}$ is even more dominant over $c_{3H}$ at large $m_{HH}$ in Fig.~\ref{fig:NLO_contribution_100TeV_kinematics} compare to $gg \rightarrow HH$ counterpart shown in Fig.~\ref{fig:gghh_100TeV_contributions_verify}.
Turning on a small $c_{HHtt}$ would significantly impact the measurement of $c_{3H}$ when $c_{3H}\sim c_{Htt}$ since the contribution from the Higgs trilinear coupling to the total cross-section is relatively small, and the destructive interference between the box and triangle diagrams.
These facts make a truly model-independent Higgs trilinear coupling measurement from the total rate of the Higgs pair production challenging since most of the events have $m_{HH} \gg 2m_H$, the contribution of the total cross-section from $c_{3H}$ will be suppressed.
 
\section{The $\bbgg$ Decay Channel}
\label{bbgg}
In this section, we investigate $HH \to \bbgg$ channel, which is the process that has the highest signal significance and in SM has the most sensitivity to the trilinear Higgs self-coupling as pointed out in the literature. 
Earlier studies can be found in Ref.~\onlinecite{Baur:2002rb,Baur:2002qd,Baur:2003,Baur:2004,Baglio:2012np}. Recent searches~\cite{Owen2021} for pairs of Higgs bosons in $HH \to \gamma\gamma b\bar{b}$ process have narrowed the upper bound of the Higgs pair production rate down to 4.1 times the SM expectation, created a portal of better understanding into the fundamental Higgs mechanism. We perform the partonic event generation for the signal and backgrounds by using \texttt{MadGraph5\_aMC@NLO} with the parton density functions \texttt{PDF4LHC15}~\cite{PDF4LHC}. Parameters are kept the same as what we use in Eq. \ref{eq:parameter} with $\mu_R=\mu_F=m_{HH}$. The effects of full NLO corrections for the signal, $H(b\bar{b})H(\gamma \gamma)$, are included.
We generate the background events at LO with the finite-top-mass effects and rescale them by a $K$-factor afterward (See Table~\ref{table:K_bbgg}).

We include the following major backgrounds in the analysis: the resonant processes , $t \bar{t} \gamma\gamma$ and $t \bar{t} H(\gamma\gamma)$ with $t \to b W$ subdecay , as well as $b \bar{b} H(\gamma\gamma)$ and $Z(b \bar{b} )H$, and the non-resonant processes $jj\gamma\gamma$ (with one and two fake $b$-jets), $b\bar{b}\gamma\gamma$, $b\bar{b}j\gamma$ (with one fake photon), $bj\gamma\gamma$.
The $b\bar{b}jj$ background is not included since it is negligible compared to other faked backgrounds after selection cuts. The MLM matching is applied to all background processes with at most one extra parton to avoid double-counting.
 
We generate events with exclusive cuts for signal and background processes. In what follows, the acceptance cuts are applied to each final state for each plot.
Here is the detailed event selection: We require exactly two $b$ quarks and two photons in the final state with the following cuts $p_{T,b}>30~\textrm{GeV}$, $\vert \eta_b \vert<2.5$ and $\Delta R(b,b) > 0.4$, where the distance is defined as $\Delta R=\sqrt{(\Delta\eta)^2+(\Delta \phi)^2}$. For leptons, the allowed soft transverse momentum and the allowed pseudorapidity are set to $p_{T, \ell}>20~\textrm{GeV}$ and $\vert \eta_{\ell} \vert<2.5$, respectively, to diminish the $t\bar{t} H$ background. Moreover, the select events are selected to satisfy $\vert\eta_{jet}\vert<2.5$ and $p_{T,jet}>20~\textrm{GeV}$ for QCD jets to diminish the $t\bar{t}H$ background further. The two photons has to fulfill $\Delta R(\gamma,\gamma) > 0.4$, $\vert\eta_\gamma\vert<2.5$, and$p_{T,\gamma}>30~\textrm{GeV}$. To reconstruct the Higgs bosons, the allowed invariant masses are within 25~GeV, $112.5~\textrm{GeV} \, < \,
M_{b\bar{b}} \,< \, 137.5~\textrm{GeV}$ for the $b$ quark pair, and a smaller range of 10~GeV, $120~\textrm{GeV} \, < \, M_{\gamma\gamma} \, < \, 130~\textrm{GeV}$ for the photon pair. In addition, we induce $\Delta R(\gamma,b) > 0.4$ to isolate the $b$ quarks with the photons.

\begin{figure}[t]
\centering
\includegraphics[scale=0.45, angle=0]{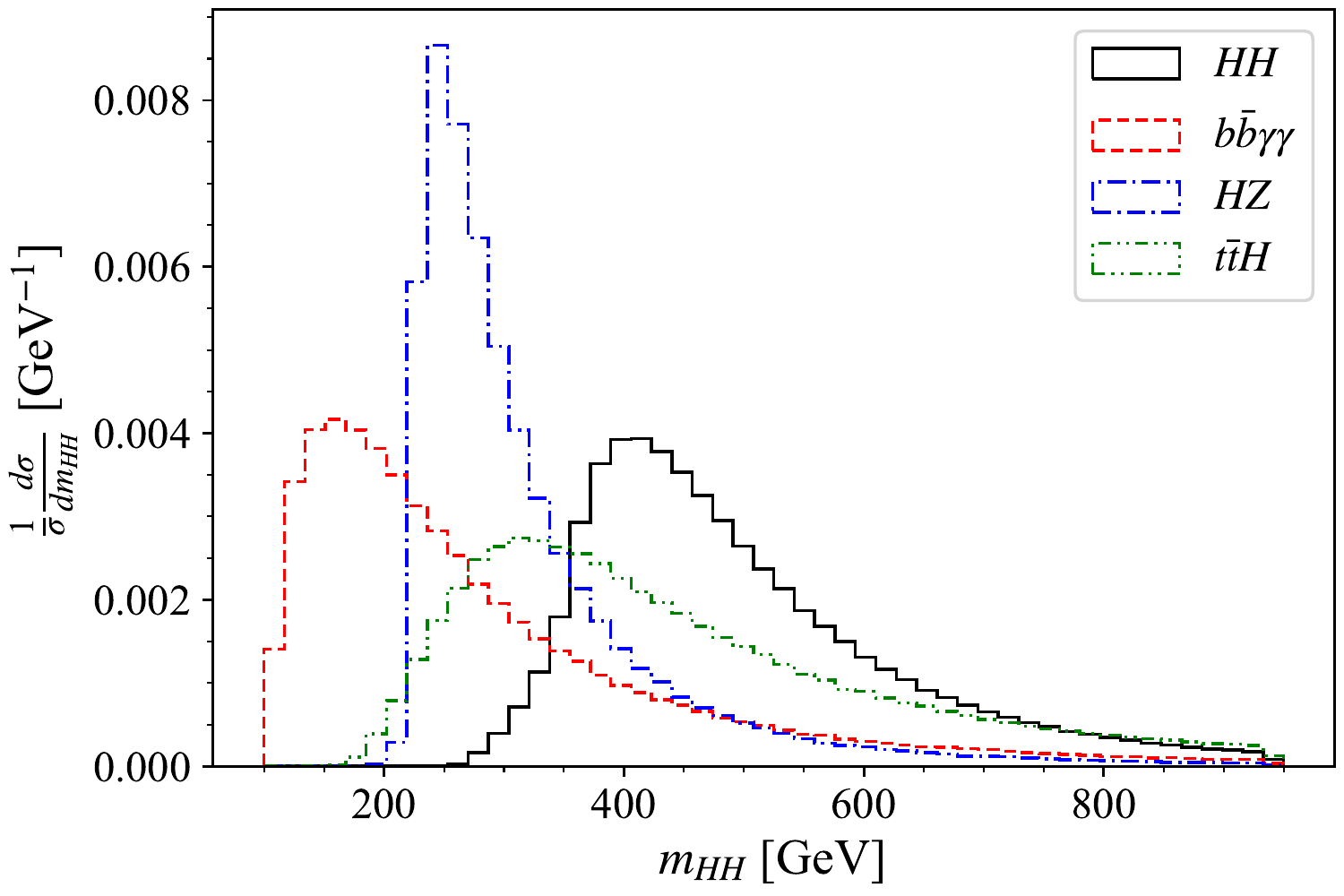}
\includegraphics[scale=0.45, angle=0]{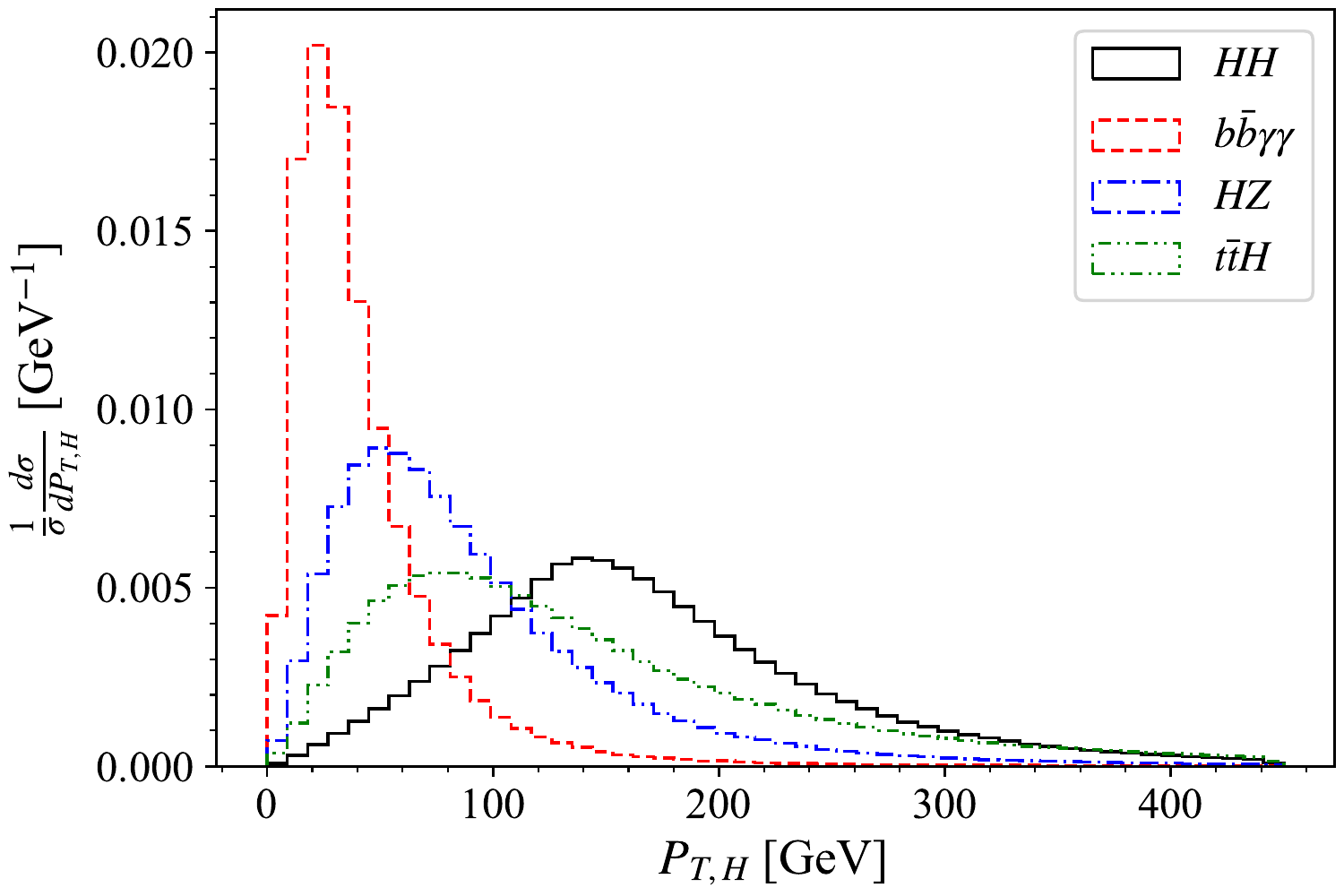}
\includegraphics[scale=0.45, angle=0]{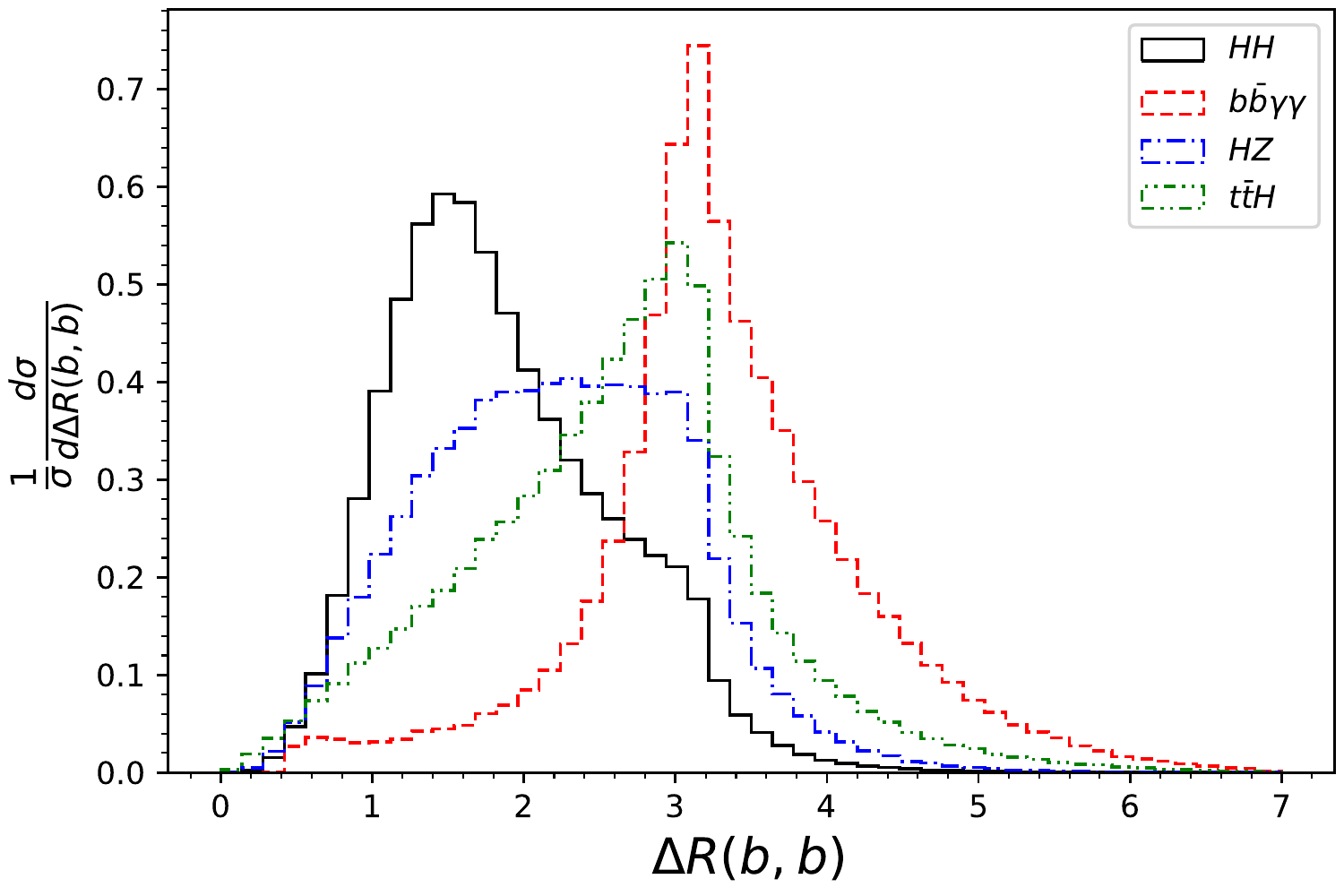}
\includegraphics[scale=0.45, angle=0]{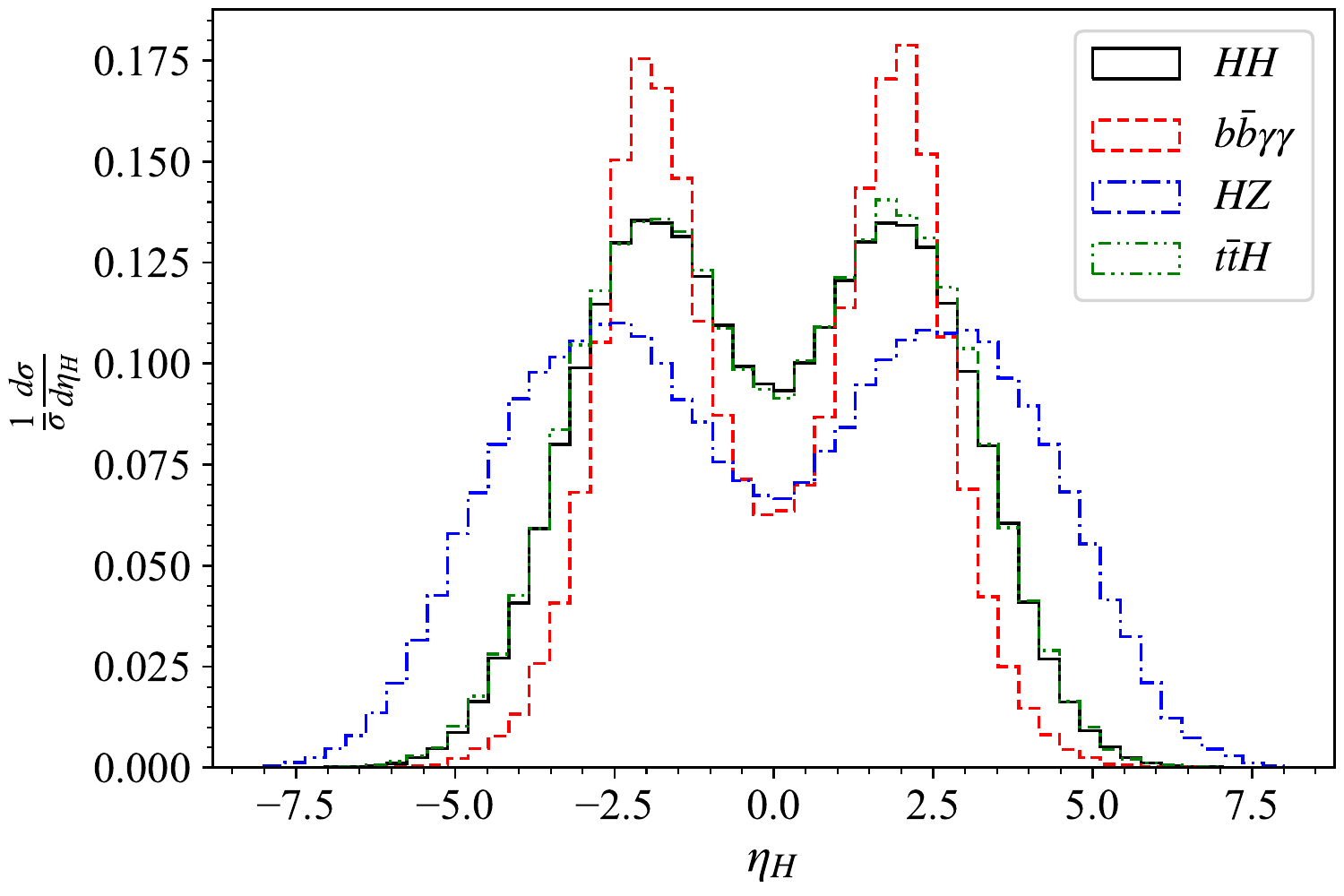}
\caption{\label{fig:distributions_bbgg} \emph{Normalized signal and backgrounds distributions of $P_{T,H}$, $m_{HH}$, $R_{bb}$ and $\eta_{H}$ in the $b\bar{b}\gamma\gamma$ channel at a $\sqrt{s}=100$ TeV pp collider.}}
\end{figure}

Besides the acceptance cuts shown above, more advanced cuts have been applied based on the distributions shown in Fig.~\ref{fig:distributions_bbgg} for the parton-level analysis. First, we select the events with a reconstructed
invariant mass of the Higgs pair that satisfy $\mhh >$ 300 GeV. Moreover, we select events that satisfy $\pth >$ 100
GeV. We also require $ \Delta R(b,b) < 2.5$ to divide the two $b$ jet and require the reconstructed Higgs boson to have the pseudorapidity $|\etah| < 2$.

For the $\bbgg$ final state, a realistic estimation of the diphoton fake rate is the most important factor of an accurate assessment for measuring the signal since the production of multijet, which is the dominant background in this case, give rise to this fake rate.

To gain more reliable results and verify if any promising feature can be found in real experiments, we include showering and hadronization effects by using the \texttt{Pythia 8}\cite{Pythia8} package~\cite{Pythia8} for the signal and background samples.
Detector simulation effects based on the current performance of ATLAS and CMS are included by using the \texttt{Delphes}~\cite{Delphes} package. We follow the parameter settings and the cut selections in Ref.~\onlinecite{1506.03302}.
We chose the operation points of $b$-tagging to have 18.8\%, 75\%, and 1\% for
charm, bottom, and light quark jets in the central region, $|\eta|<2.5$ and $P_{T,j}>50$\,GeV, respectively.
The photon identification efficiency is about 80\% for photons with $\,P_{T,\gamma}>50$\,GeV and $|\eta|<2.5$\,.\,
For the background with fake photons from misidentified jets, we assign a mistag rate of $\,f_j^{}=0.0093\exp (-P_T^{}/27)$\, as a function of $P_T^{}$ in GeV of the jet with the fake photon energy equal to the jet energy scaled by $\,0.75\pm 0.12$~\cite{TheATLAScollaboration:2013sgb}.
At $\,M_h^{}=125$\,GeV, the mass resolution is 17\,GeV for $\,h\rightarrow b\bar b$ and 2\,GeV for $\,h\rightarrow \gamma\gamma$\,. In order to be consistent with the signal, the isolated photon pair and two tagged $b$-jets in the final states are selected to satisfy $\,P_T^{}>25$\,GeV and $\,|\eta|<2.5$.

The cuts for mass-window are further applied to the invariant-masses of the two photons and two $b$-jets. For the photon pair, we impose $\,122\,\textrm{GeV}<M_{\gamma\gamma}<128\,$GeV on the invariant-mass window.
The invariant-mass window of the two $b$-jets is $\,120\,\textrm{GeV}<m_{b\bar b}<130\,$GeV.\, 

\begin{table}
 \renewcommand{\arraystretch}{1.3}
 \begin{center}
  \small
  \begin{tabular}{c|ccccccc}
   $\sqrt{s}$~[TeV] & $HH$ & $b\bar{b} \gamma\gamma$ & $t\bar{t}
   H$ & $ZH$ & $b\bar{b}H$ & $\gamma\gamma j j$ & $b\bar{b}j\gamma$ \\ \hline
   100 & 1 & 1.0 & 1.3 & 1.2 & 0.87 & 1.43 & 1.08
  \end{tabular}
  \vspace{0.2cm}
  \caption{\emph{$K$--factors for $ZH$, $b\bar{b}\gamma\gamma$ and $t\bar{t}H$
    production at $\sqrt{s}=100$~TeV~\protect\cite{Contino:2016spe}.\label{table:K_bbgg}}}
 \end{center}
\end{table}

\begin{table}
\centering
\def\arraystretch{1.5}
\begin{tabular}{c|c}
Observables  & Acceptance cuts \\ \hline
Total number $n$ of jets ($j, \gamma, l$)                           & $n<7$ in each event  \\
Pseudorapidity                        & $\eta_{b,\gamma}<2.5$ \\
Invariant mass                 & \parbox{15em}{\centering $120<m_{b \bar{b}}<130$ GeV, $122<m_{\gamma \gamma}<128$ GeV, \\ 
                                $m_{b\bar{b}\gamma\gamma}>300$ GeV} \\
Transverse momentum                 & $p_{T\gamma, b}>35$ GeV,~$p_{T\gamma\gamma, b\bar{b}}>100$ GeV   \\
$\Delta R$                               & $0.4<\Delta R(b,b)<2.0$,~ $0.4<\Delta R(\gamma,\gamma)<2.5$         
\end{tabular}
\caption{\emph{List of observables and acceptance cuts used for the analysis.}}\label{table:bbgg_cut}
\end{table}

After applying the cuts shown in Table \ref{table:bbgg_cut},the final significance we obtained is $S/\sqrt{B}=16.1$ for the integrated luminosity, $\lum=3$ ab$^{-1}$, which is close to previous studies~\cite{Baglio:2012np,1506.03302}. This strongly suggests that the $\bbgg$ channel is observable in the future upgrade of the LHC (HE-LHC) or Future-Circular-Collider (FCC).

\section{Sensitivity to effective self-couplings of Higgs bosons}
In this section, the characteristic distributions of the double Higgs production are studied for several observables with different values of effective Higgs couplings.
 
\begin{figure}[!hbt]
\centering
\includegraphics[scale=0.45, angle=0]{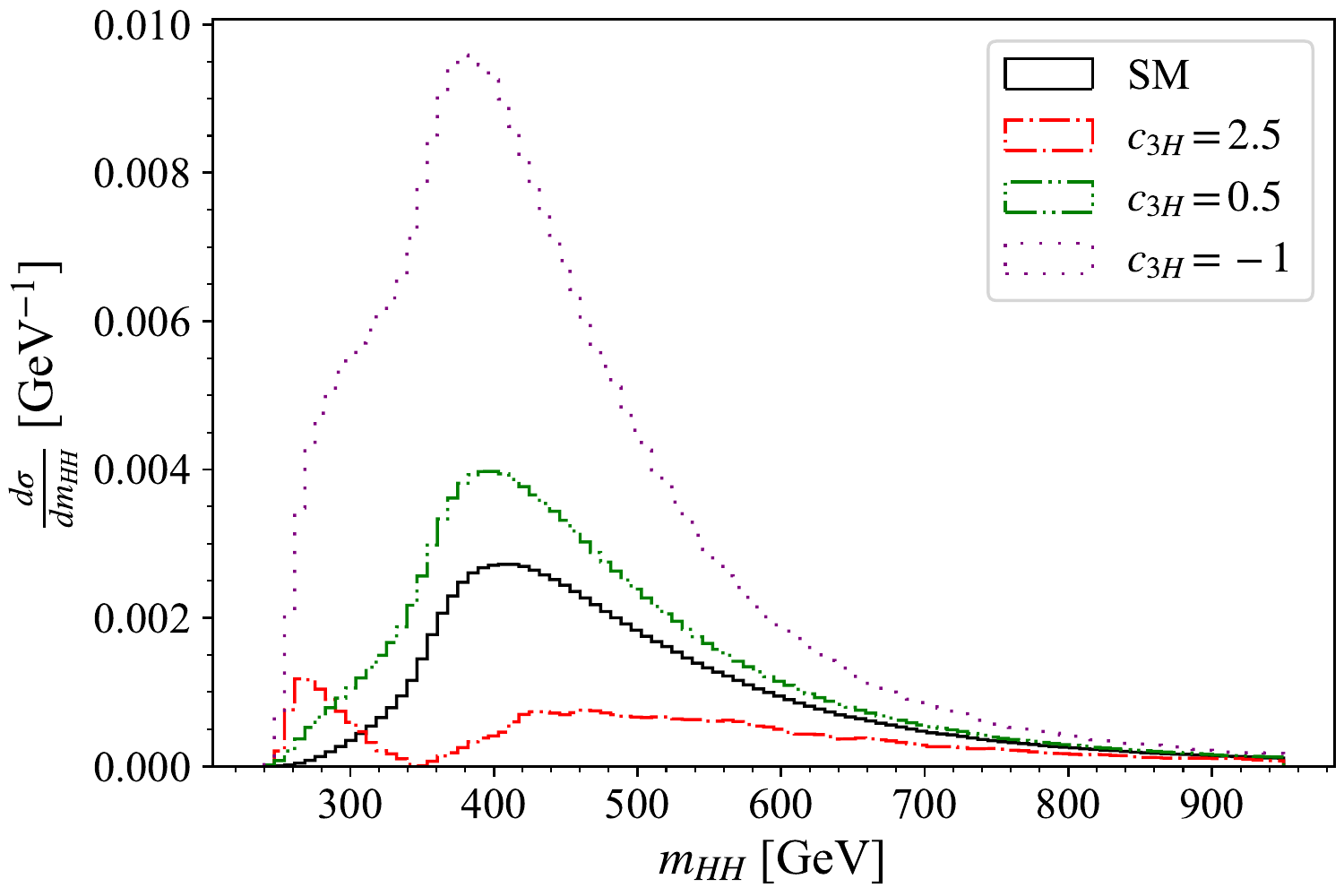}
\includegraphics[scale=0.45, angle=0]{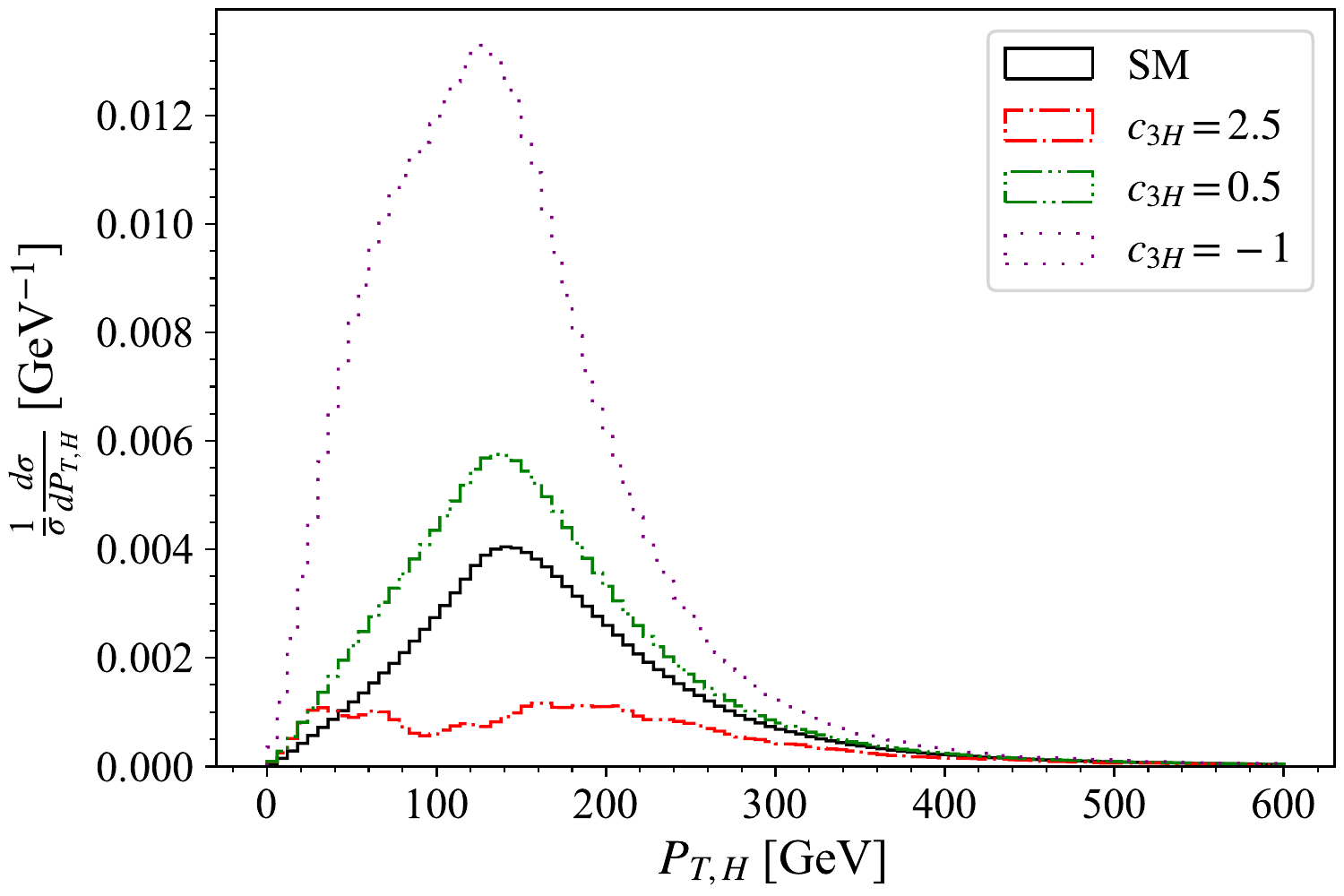}

\includegraphics[scale=0.45, angle=0]{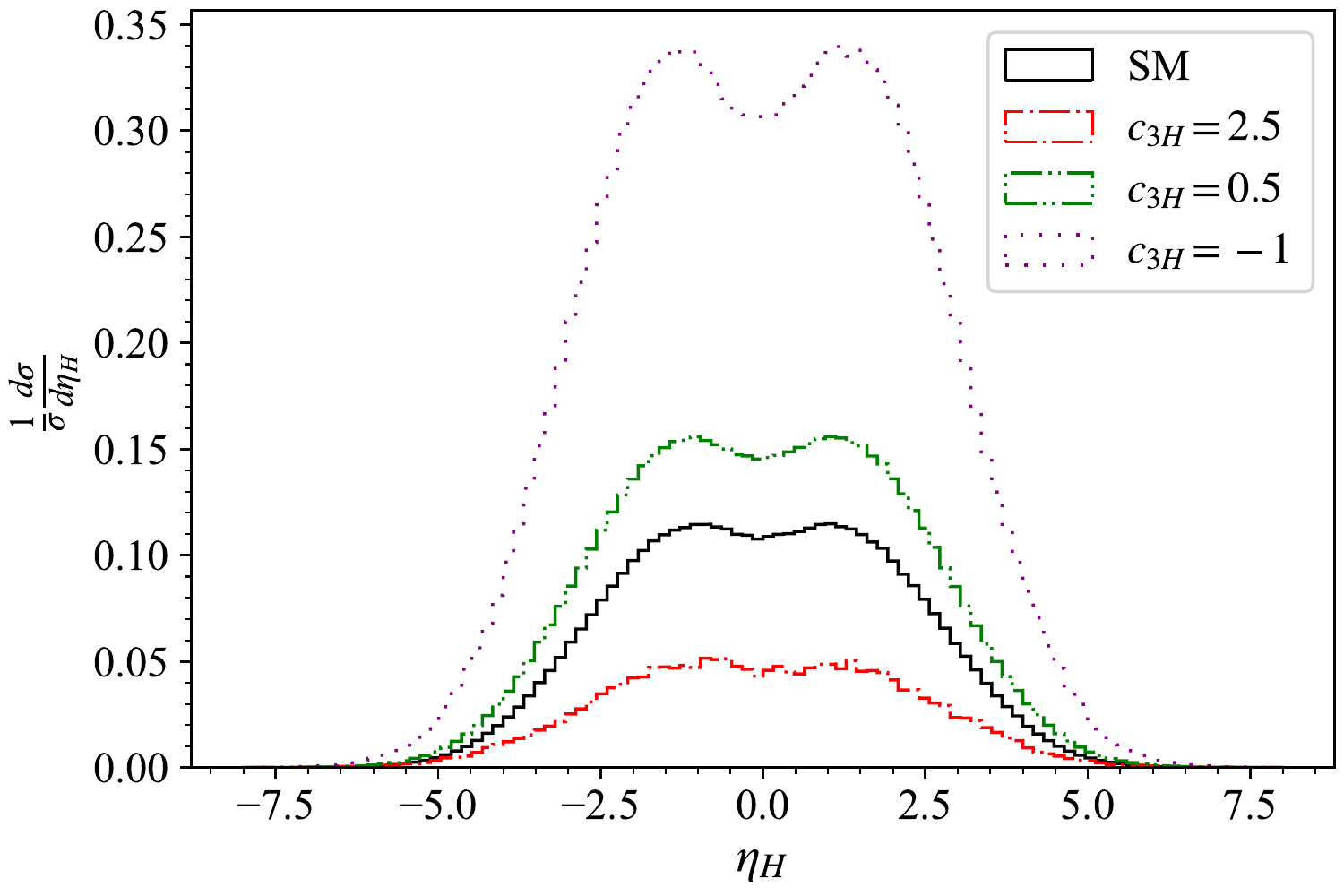}
\includegraphics[scale=0.45, angle=0]{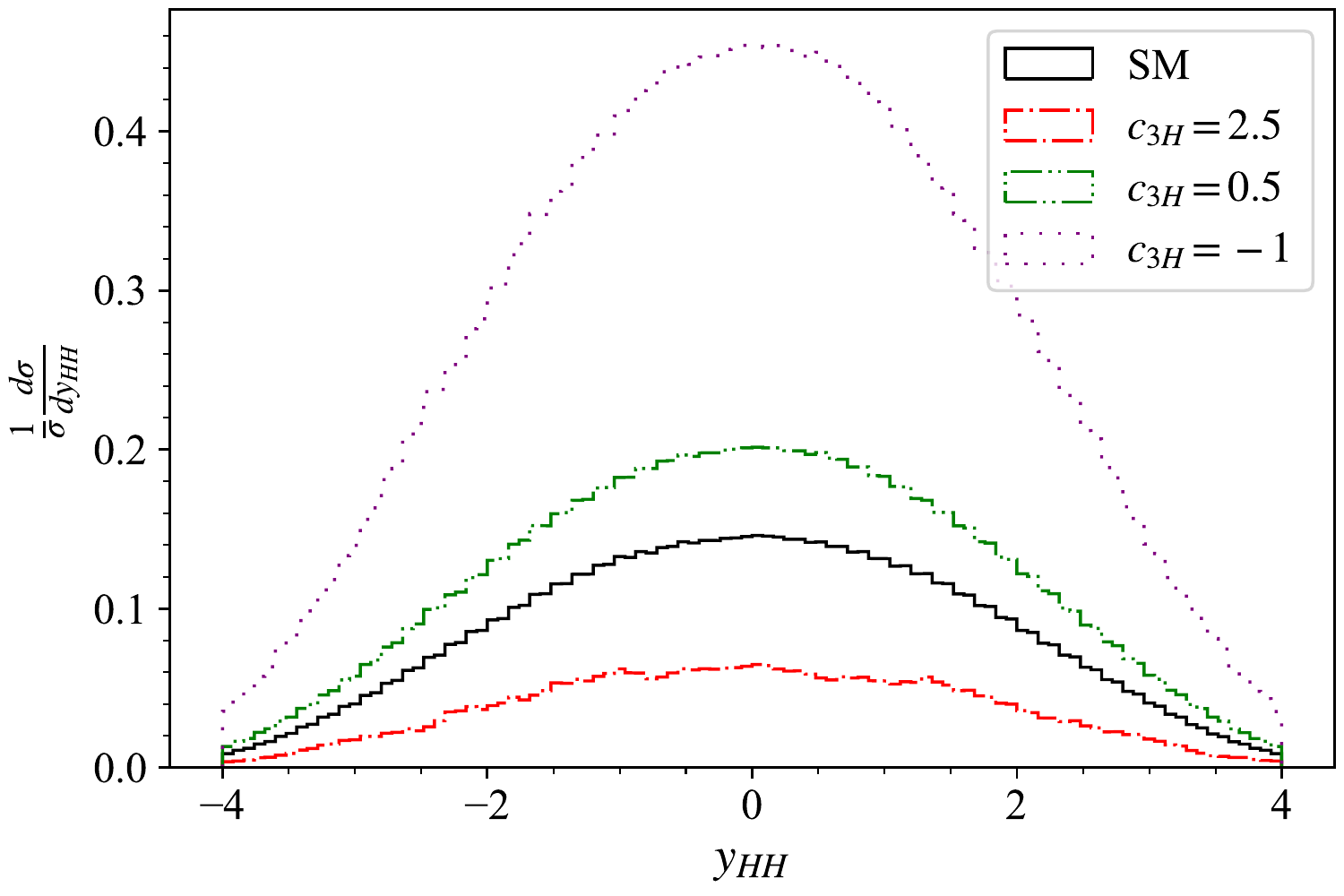}
\caption{\label{fig:NLO_couplings_tri_100TeV} \emph{Distributions of $\pth$, $\etah$, $\mhh$ and $y_{HH}$ for $c_{3H}=0.5,1,2.5,-1$.
}}
\end{figure}

 Fig.~\ref{fig:NLO_couplings_tri_100TeV}, shows the distributions of the invariant mass $\mhh$, the transverse momentum $\pth$, the pseudorapidity $\etah$, and the rapidity $y_{HH}$ of the Higgs pair with the area under the SM curve normalized to unity. Each observable distribution is shown for $c_{3H}=$ 0.5, 1, 2.5, and -1.

As in the $\pth$ distribution plot of Fig.~\ref{fig:NLO_couplings_tri_100TeV} with the distribution max at $\pth \sim 150$~GeV. The Higgs bosons from the production of inclusive Higgs pair are usually boosted. The pseudorapidity of the Higgs pair shown in the lower left of Fig.~\ref{fig:NLO_couplings_tri_100TeV} is low and has a typical symmetric distribution with the maximum close to zero due to the high transverse momentum spectrum. 
For $c_{3H}=2.5$, the interferences between the box and the triangle diagrams are destructive. This explains the dip in the $\pth$ distribution.
Comparing to a lower peak value of $M_{ZH} \gtrsim 250$~GeV for the background shown in Fig.~\ref{fig:distributions_bbgg}, the peak value is $\mhh \gtrsim 420$~GeV for the signal. Again, this destructive interference also causes a significant depletion in the signal when $c_{3H}=2.5$. For smaller $c_{3H}=0.5$, the destructive interference is diminished, and the signal is stronger than the SM expectation value for each distribution.
For $c_{3H}=-1$, the differential cross-sections for all observables are enhanced significantly since the box diagram interferes constructively with the triangle diagram when $c_{3H}$ becomes negative.
For $y_{HH}$ and $\etah$ distribution, the overall shape is the same for different
trilinear Higgs coupling values. We can infer that the
$y_{HH}$ distribution is significantly wider for the $ZH$ background than for the signal shown in Fig.~\ref{fig:distributions_bbgg}.

\begin{figure}[!hbt]
\centering
\includegraphics[scale=0.45, angle=0]{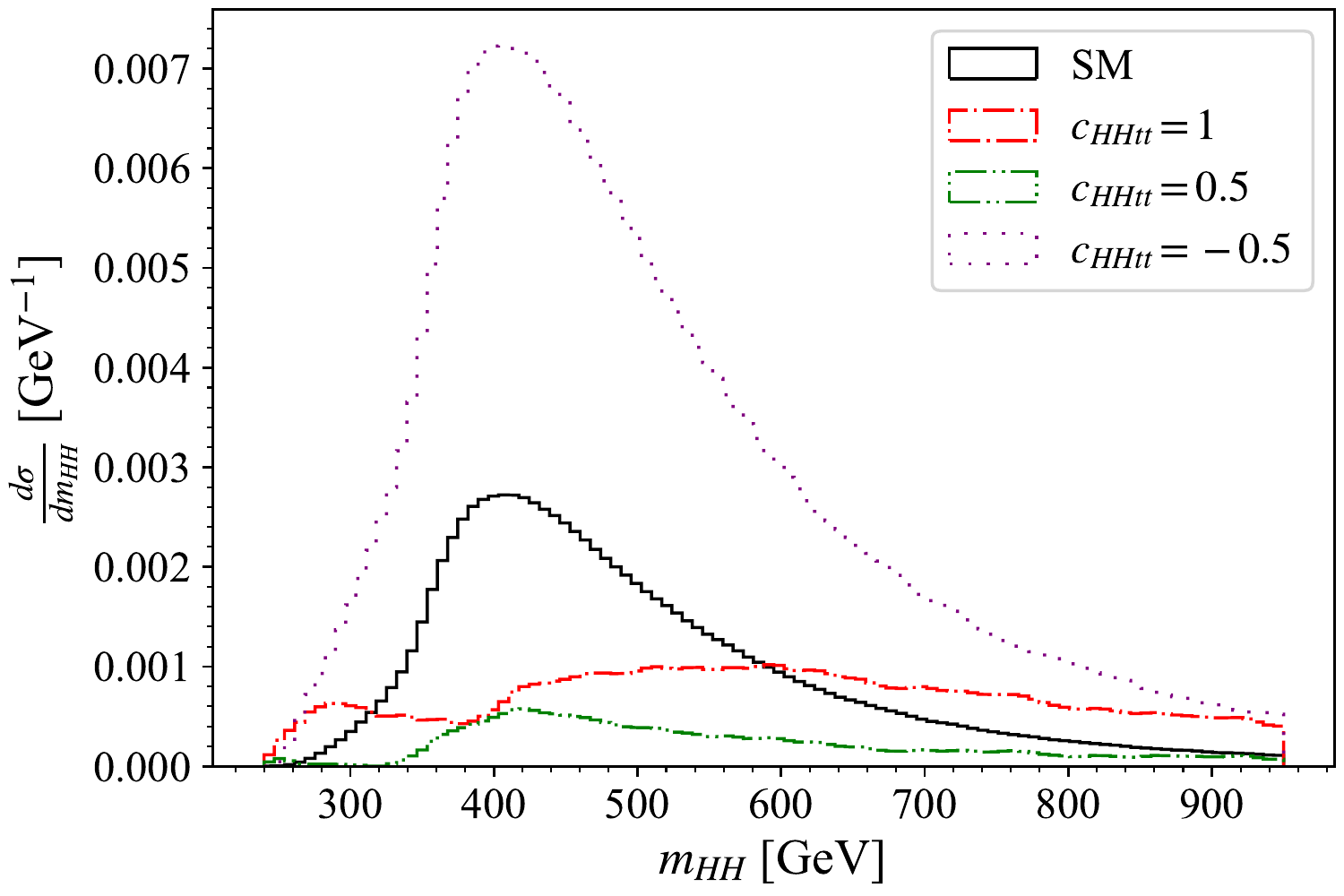}
\includegraphics[scale=0.45, angle=0]{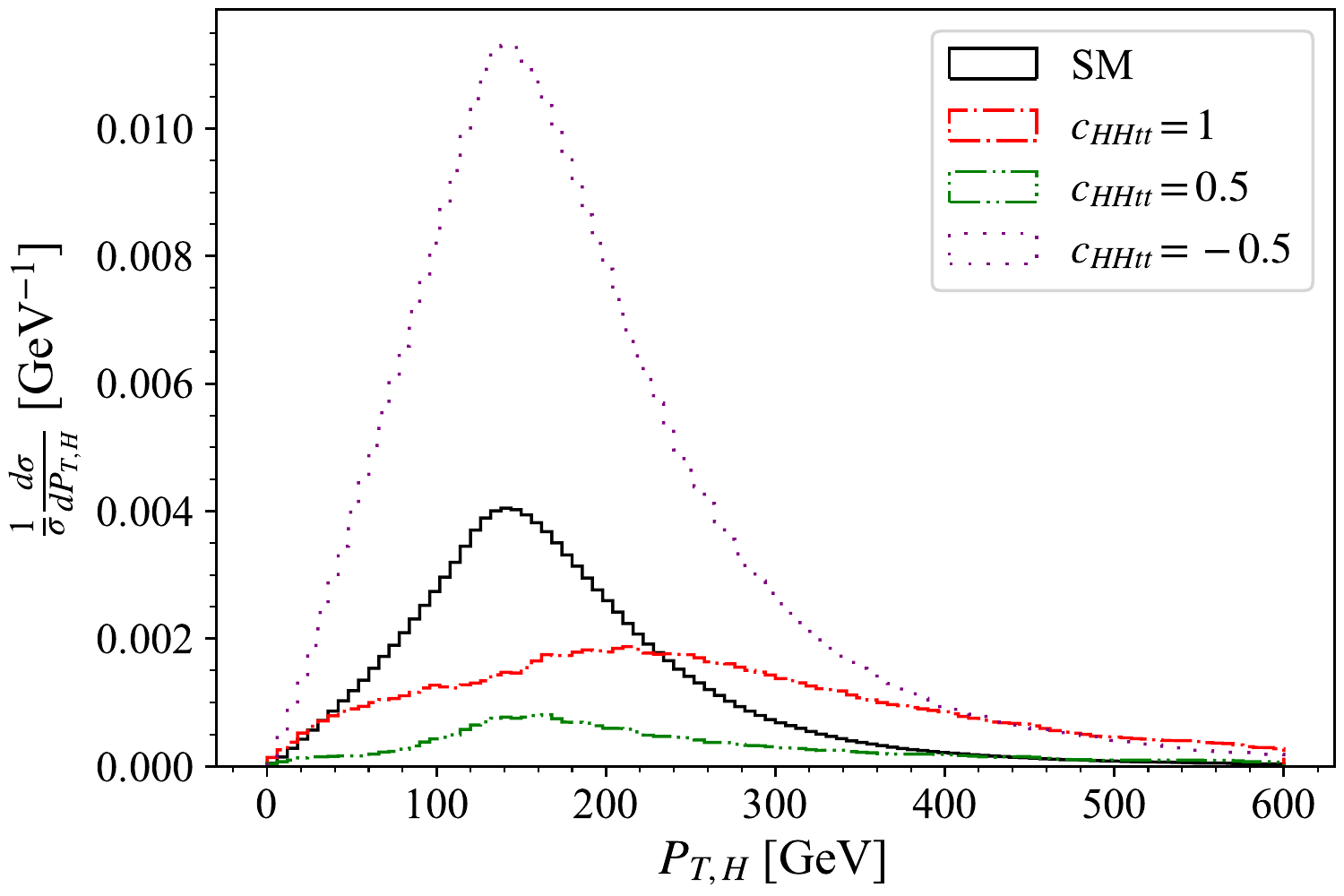}

\includegraphics[scale=0.45, angle=0]{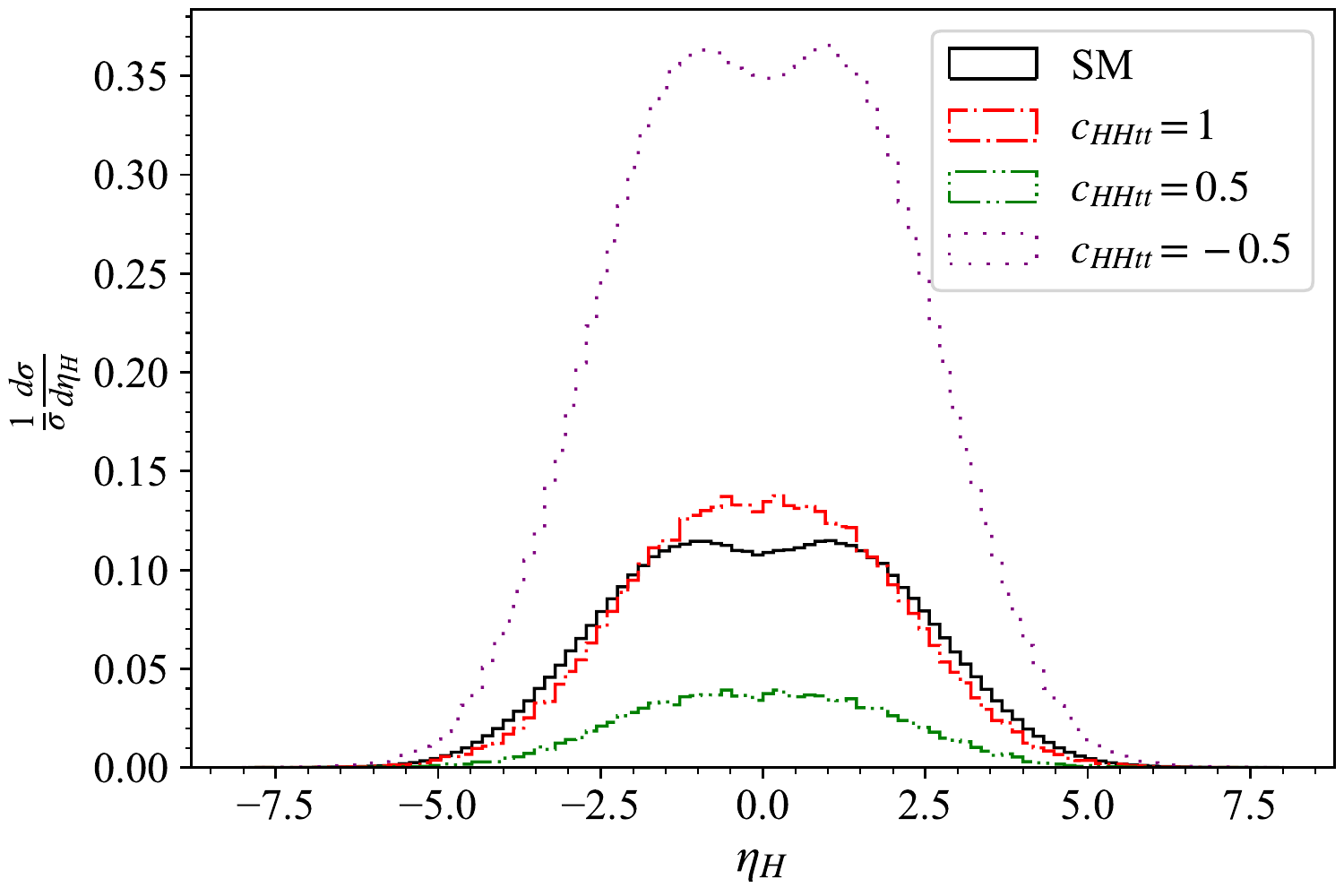}
\includegraphics[scale=0.45, angle=0]{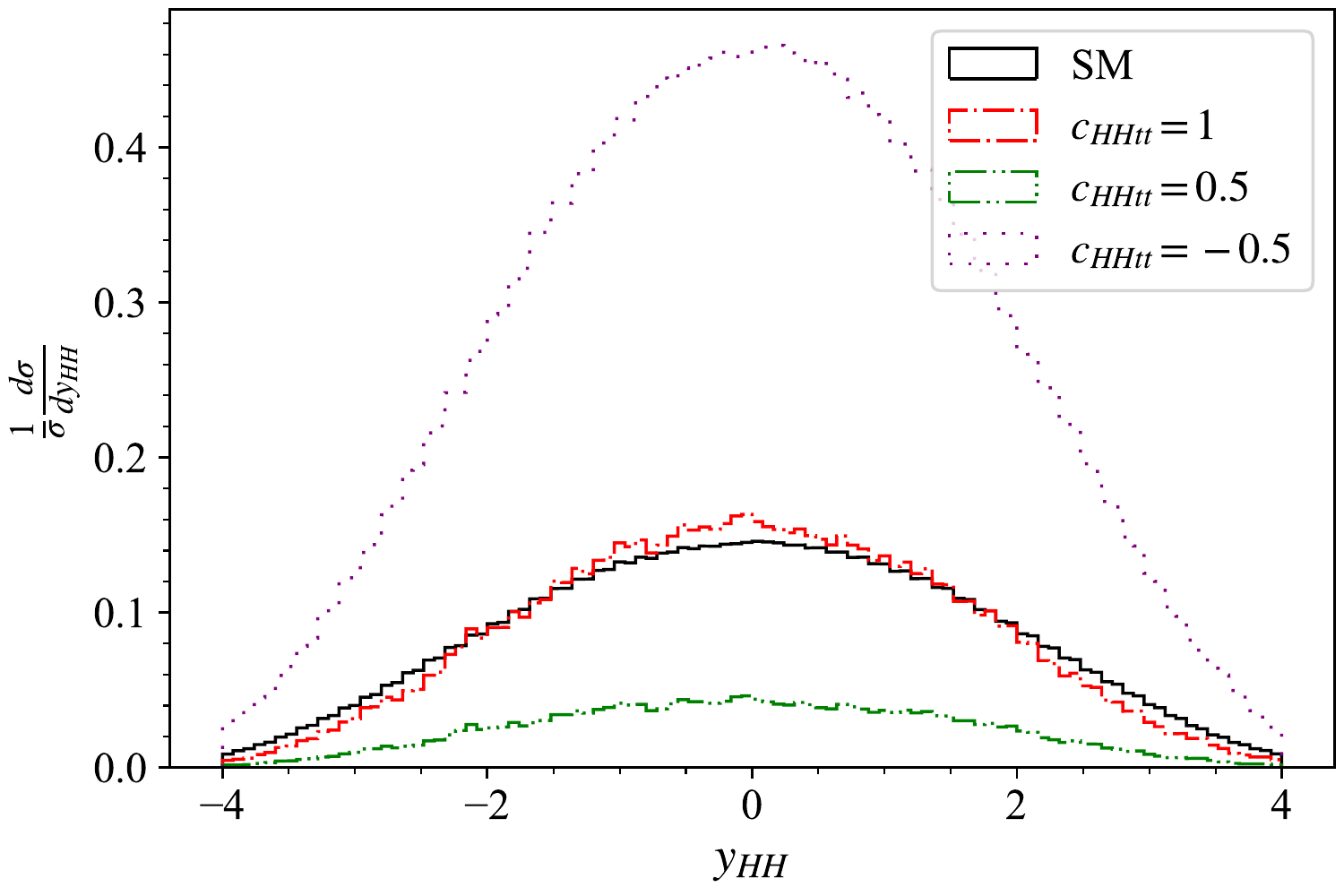}
\caption{\label{fig:NLO_couplings_nl_100TeV} \emph{Distributions of $\pth$, $\etah$,
   $\mhh$and $y_{HH}$ with $c_{HHtt}=1,0.5,0,-0.5$.}}
\end{figure}

A similar distribution analysis is shown in Fig.~\ref{fig:NLO_couplings_nl_100TeV} for $c_{HHtt}=$ -0.5, 0.5, 1 and 0, the SM value. 

For $c_{HHtt}>0$, the $HHtt$ diagrams interfere with the box diagrams destructively, which explains the dip in the distribution of $\pth$ and $\mhh$. We can see a significant depletion in the signal for $c_{HHtt}=1$, and the destructive interference further depletes the signal for $c_{HHtt}=0.5$.
Peaks for $\pth$ and $\mhh$ stay the same as SM expectations, $\pth \gtrsim 150$ and $\mhh \gtrsim 420$. For $c_{HHtt}<0$, the differential cross-sections for all observables are enhanced due to the constructive interference between the $HHtt$ diagram and the box diagram. The enhancement is large even when $c_{HHtt}$ is just -0.5.
For $y_{HH}$ and $\etah$ distribution, the overall shape is the same for different values of the $HHtt$ coupling.

%

By using the parameterization introduced in Eq.~(\ref{eq:newpar}), the total cross-section of Higgs pair production can be written as
\bea
\sigma &=& \sigma^{SM}\big[c_{3H}^2|\mathcal{M}_{3H}|^2+c_{Htt}^2|\mathcal{M}_{Htt}|^2
+c_{HHtt}^2|\mathcal{M}_{HHtt}|^2
+c_{HWW}^2|\mathcal{M}_{HWW}|^2 \nonumber \\
&& +c_{HZZ}^2|\mathcal{M}_{HZZ}|^2 +c_{HWW,Htt}^2|\mathcal{M}_{HWW,Htt}|^2+c_{Hbb}^2|\mathcal{M}_{Hbb}|^2+c_{Hcc}^2|\mathcal{M}_{Hcc}|^2 \nonumber \\
&& +2\,c_{3H}c_{Htt}|\mathcal{M}_{3H}\mathcal{M}_{Htt}|
+2\,c_{3H}c_{HHtt}|\mathcal{M}_{3H}\mathcal{M}_{HHtt}|
+2\,c_{3H}c_{HWW}|\mathcal{M}_{3H}\mathcal{M}_{HWW}|\nonumber \\ 
&&+2\,c_{3H}c_{HZZ}|\mathcal{M}_{3H}\mathcal{M}_{HZZ}|
+2\,c_{3H}c_{HWW,Htt}|\mathcal{M}_{3H}\mathcal{M}_{HWW,Htt}| \nonumber \\
&&+2\,c_{3H}c_{Hbb}|\mathcal{M}_{3H}\mathcal{M}_{Hbb}|
+2\,c_{3H}c_{Hcc}|\mathcal{M}_{3H}\mathcal{M}_{Hcc}| \nonumber \\
&&+2\,c_{Htt}c_{HHtt}|\mathcal{M}_{Htt}\mathcal{M}_{HHtt}|
+2\,c_{Htt}c_{HWW}|\mathcal{M}_{Htt}\mathcal{M}_{HWW}|+\dots
\eea
This expression is lengthy and hard to analyze the effect of changing parameters. Fortunately, $\mathcal{M}_{HWW}$, $\mathcal{M}_{HZZ}$, $\mathcal{M}_{HWW,Htt}$, $\mathcal{M}_{Hbb}$ and $\mathcal{M}_{Hcc}$ are very small compare to $\mathcal{M}_{3H}$, $\mathcal{M}_{Htt}$ and $\mathcal{M}_{HHtt}$ since they comes from $QCD_1$ contributions.
We can therefore safely drop these terms without change the overall properties, and we can write the total cross-section in the following form
\bea
\sigma &=& \sigma^{SM}\big[c_{3H}^2|\mathcal{M}_{3H}|^2+c_{Htt}^2|\mathcal{M}_{Htt}|^2
+c_{HHtt}^2|\mathcal{M}_{HHtt}|^2
+2\,c_{3H}c_{Htt}|\mathcal{M}_{3H}\mathcal{M}_{Htt}| \nonumber \\
&&
+2\,c_{3H}c_{HHtt}|\mathcal{M}_{3H}\mathcal{M}_{HHtt}|
+2\,c_{Htt}c_{HHtt}|\mathcal{M}_{Htt}\mathcal{M}_{HHtt}|]
\eea

For all contributions to the Higgs pair production including $gg \rightarrow HH$, $gg \rightarrow HHg$, $qg \to HHq$ and $q\bar{q} \rightarrow HHg$, we have
 \bea
\label{eq:hh_totalrate}
\sigma = \sigma^{SM} [1.8590~ c_{Htt}^2 + 0.21485~ c_{3H}^2 + 2.9524~ c_{HHtt}^2-1.0739~ c_{Htt} c_{3H} \\
- 4.1431~ c_{Htt} c_{HHtt} +1.2271~ c_{3H} c_{HHtt} ].\nonumber 
\eea 
\begin{figure}
\centering
\subfloat[]{\includegraphics[width=.45\linewidth]{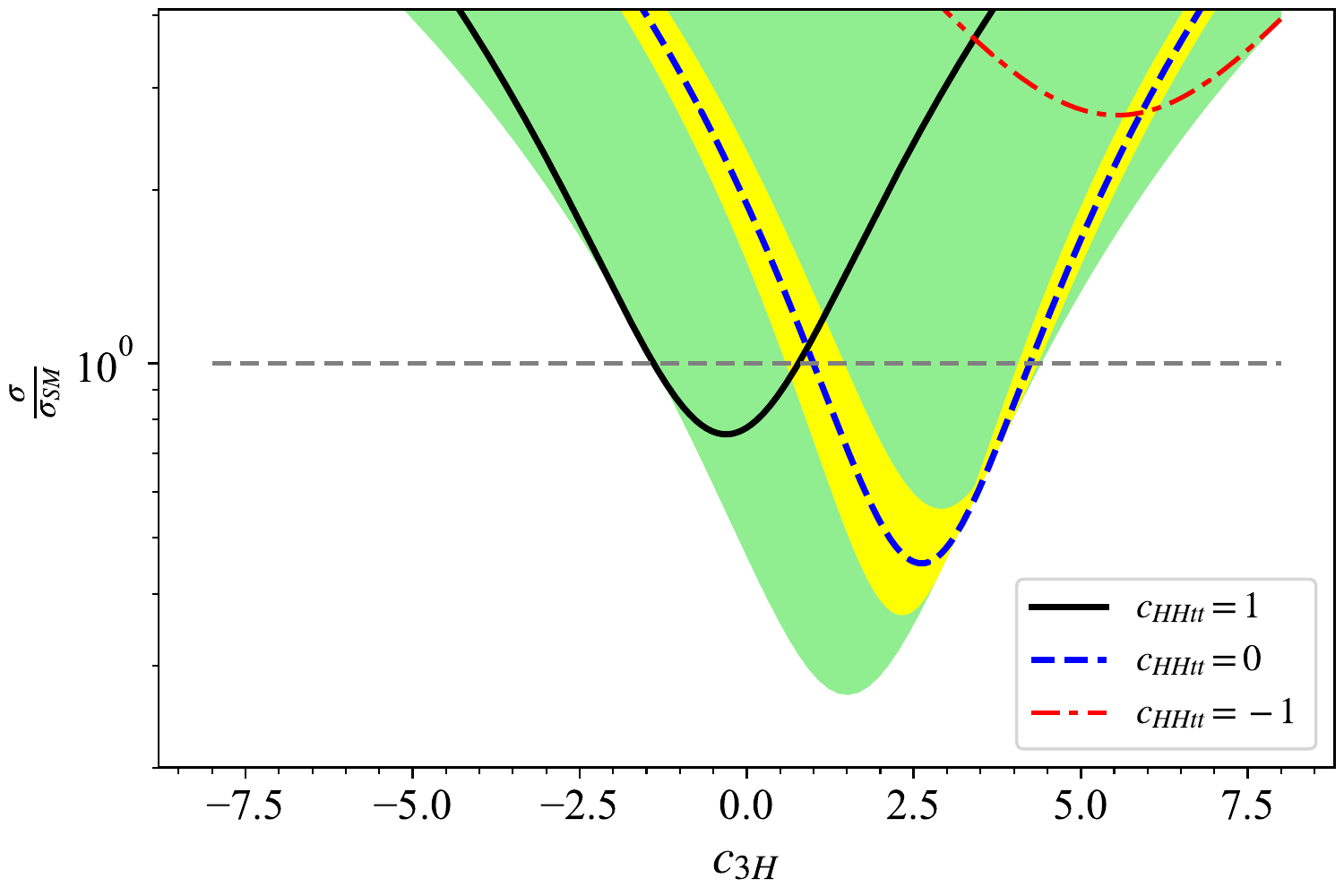}}\qquad
\subfloat[]{\includegraphics[width=.45\linewidth]{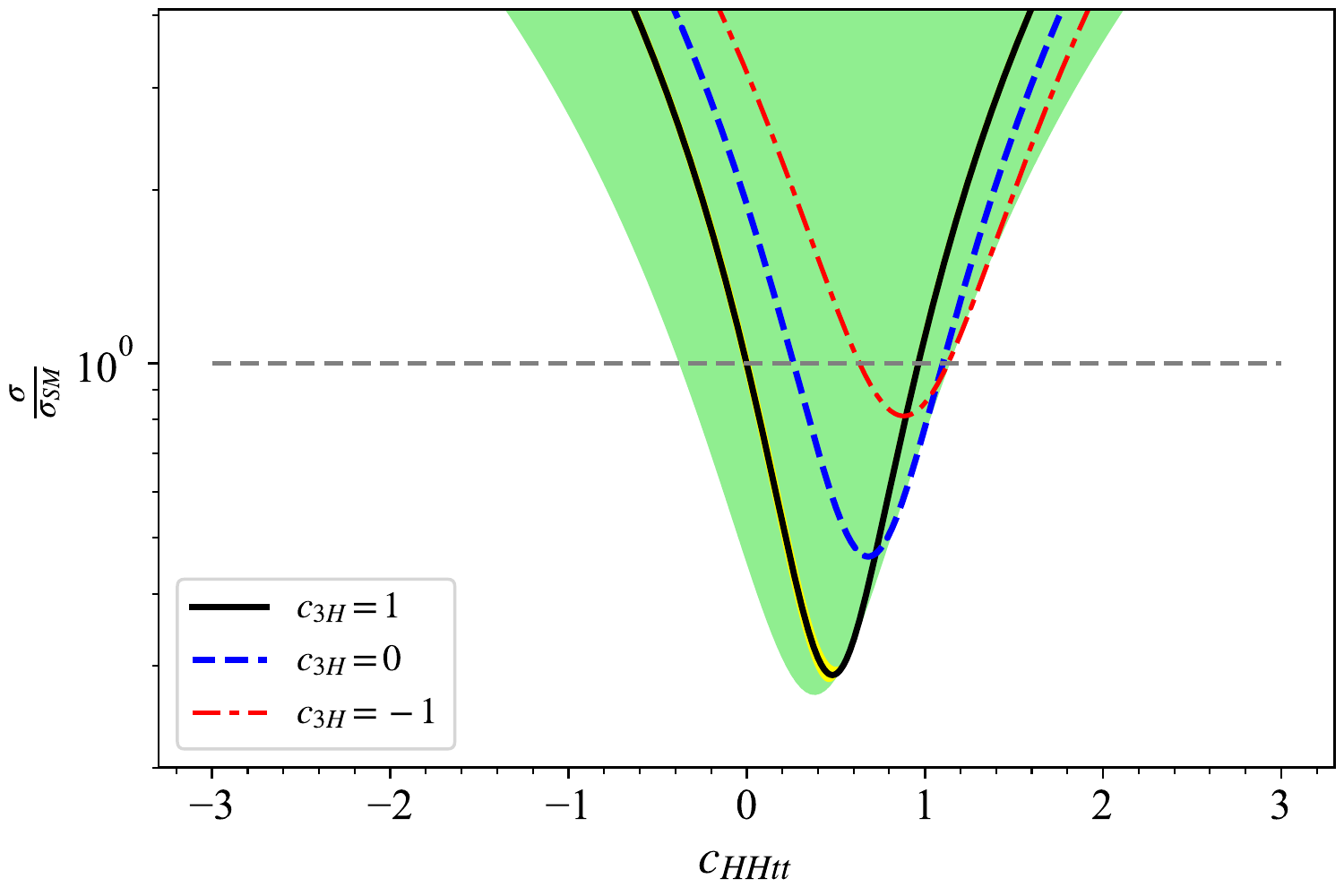}}
\vspace*{0.5cm}
\subfloat[]{\includegraphics[width=.45\linewidth]{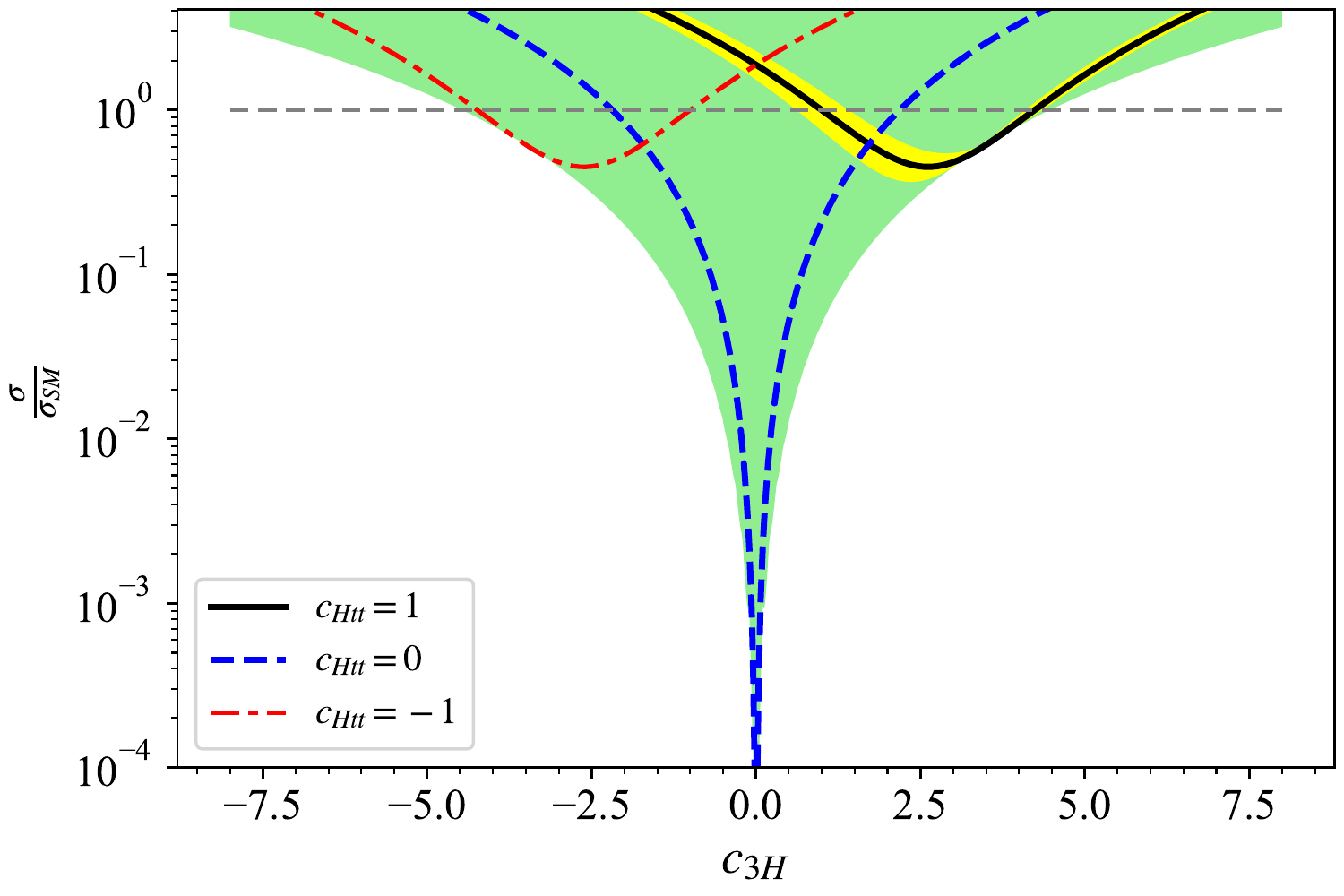}}
\caption{\label{fig:hh_xs_1} \emph{\textup{(a)} The ratio of $\sigma/\sigma^{SM}$, with varying $c_{HHtt}$ and $c_{3H}$ while fixing $c_{Htt}$ at unity, are shown as the green region. The yellow band denotes the region where $c_{HHtt}$ is within $\pm 0.1$ of its expected SM value. $c_{HHtt}$ is allowed to vary from -3 to 3 and $c_{3H}$ is allowed to vary from -8 to 8. The SM rate is the dashed horizontal line.
\textup{(b)} Same as \textup{(a)}, but with $c_{HHtt}$ along the horizontal axis. \textup{(c)} $c_{HHtt}=0$ with $c_{Htt}$ and $c_{3H}$ varying from $-3$ to $3$ and $-8$ to $8$ respectively. The yellow band denotes the region where $c_{Htt}$ is within $\pm 0.1$ of its expected SM value.}}
\end{figure}
A similar result for the LO $gg \rightarrow HH$ at $\sqrt{s}=100$ is calculated in Ref.~\onlinecite{Chen:2014xra}.
For the LHC with $\sqrt{s}=14$, similar numerical coefficients are found. We stress that the coefficient of $c_{3H}^2$ is around one order of magnitude lower than those of $c_{Htt}^2$ and $c_{HHtt}^2$, which agrees with the observation made in Ref.~\onlinecite{Contino:2012xk}.
 Fig.~\ref{fig:hh_xs_1} visualize Eq.~(\ref{eq:hh_totalrate}) by showing new physics effects in the total rate of Higgs pair production to the SM expectation ratio. Recently, the upper bound of the Higgs pair production rate was set to $4.1$ times the SM value~\cite{Owen2021}. In the following discussion, the parameters are allowed to vary between $-3$ and $3$ for $c_{Htt}$ and $c_{HHtt}$, while $c_{3H}$ is allowed to vary between $-8$ and $8$.
In Fig.~\ref{fig:hh_xs_1}(a), $c_{Htt}$ is fixed to unity, its SM value, while $c_{HHtt}$ is allowed to vary. The green region shows the resulting total rate variation, and a strong enhancement can be found on all allowed regions. When $c_{3H}\lesssim -1.5$ or $c_{3H}\gtrsim 4.1$, the production rate is always enhanced. The red dash-dotted, blue-dashed, and black-solid curves represent three reference cases in the plot for $c_{HHtt}=-1$, $c_{HHtt}= 0$, and $c_{HHtt} = 1$, respectively. The yellow band shows where $c_{HHtt}$ is within $\pm 0.1$ of its expected SM value. 
Even with vanishing or negative, opposite sign to the SM expectation, trilinear Higgs boson coupling, we can see a large area of the parameter space in $c_{Htt}$ and $c_{3H}$, which reproduce the same cross-section of the Higgs pair production as in the SM. For the case that $c_{HHtt}$ and $c_{Htt}$ are both close to their SM value, the small area around two intersections of the yellow band and the gray dashed line indicates two possible regions of parameter space that allows $c_{3H}$ to produce the expected SM cross-section value. We can set the allowed limit for SM parameter $c_{3H}$ by finding where the SM curve, blue-dashed curve, reaches the observed limit of production cross-section, $4.1$ times the Standard Model prediction. Our result agrees with the finding of~\cite{Owen2021}, $-1.5<c_{3H}^{SM}<6.7$, and we can easily see that $c_{HHtt}$ parameter greatly expand the allowed limit of $c_{3H}$ to $-5<c_{3H}<8$
In Fig.~\ref{fig:hh_xs_1}(b), $c_{Htt}$ is still fixed to unity, but with $c_{HHtt}$ along the horizontal axis. The production cross-section is always enhanced when $c_{HHtt}\lesssim -0.4$ or $c_{HHtt}\gtrsim 1.1$. The yellow band for $c_{3H}$ within $\pm 0.1$ of its expected SM value is very narrow due to the fact that $c_{3H}$ contribution is very small compare to $c_{Htt}$ and $c_{HHtt}$ contributions when $c_{3H}=1$ (See Fig.~\ref{fig:NLO_100TeV_kinematics}). Again, two possible regions of parameter space allow $c_{HHtt}$ to produce the expected SM cross-section value when $c_{3H}$ and $c_{Htt}$ are both close to their SM value.

In Fig.~\ref{fig:hh_xs_1}(c), we fix $c_{HHtt}$ to be its SM value, zero, and $c_{Htt}$ and $c_{3H}$ are both allowed to vary. The black-solid curve in Fig.~\ref{fig:hh_xs_1}(c) is for the SM $c_{Htt}$ that corresponds to the blue-dashed curve in Fig.~\ref{fig:hh_xs_1}(a). The minimum $\sigma / \sigma^{SM}$ ratio occurs at $c_{3H}\lesssim 2.5$ and has a value around $0.45$, which agrees with the finding of~\cite{Baglio:2020ini}. Notice that zero production cross-section can only occur trivially when three parameters are all zero, and it becomes a minimum point when we vary $c_{HHtt}$ since each contribution can not cancel each other at every phase space point, as we can see in Fig.~\ref{fig:NLO_couplings_tri_100TeV}.
The SM total rate for the Higgs pair production could be again reproduced by a large area of the parameter space in $c_{3H}$ and $c_{Htt}$.

\begin{figure}
\centering
\subfloat[]{\includegraphics[width=.45\linewidth]{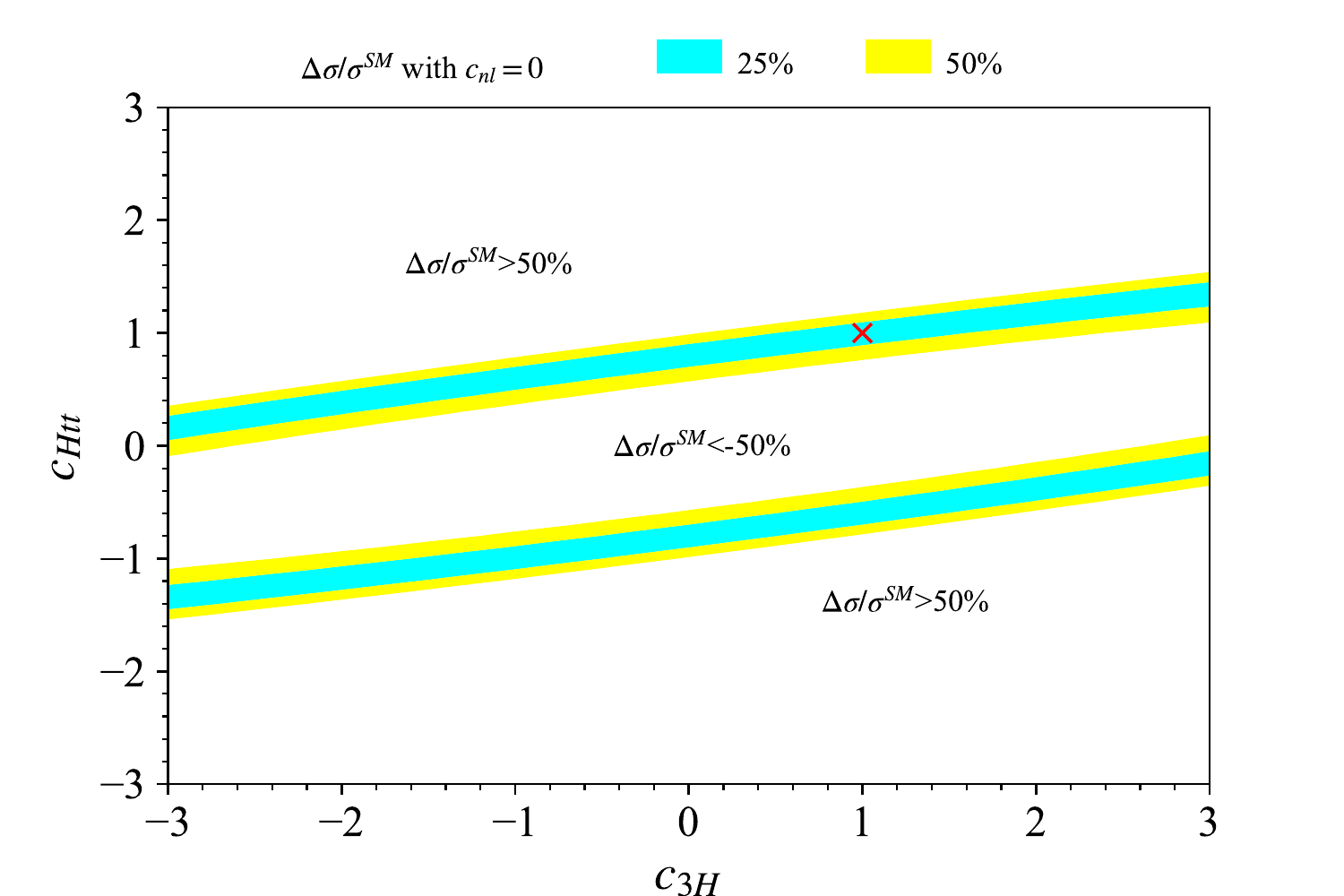}}\qquad
\subfloat[]{\includegraphics[width=.45\linewidth]{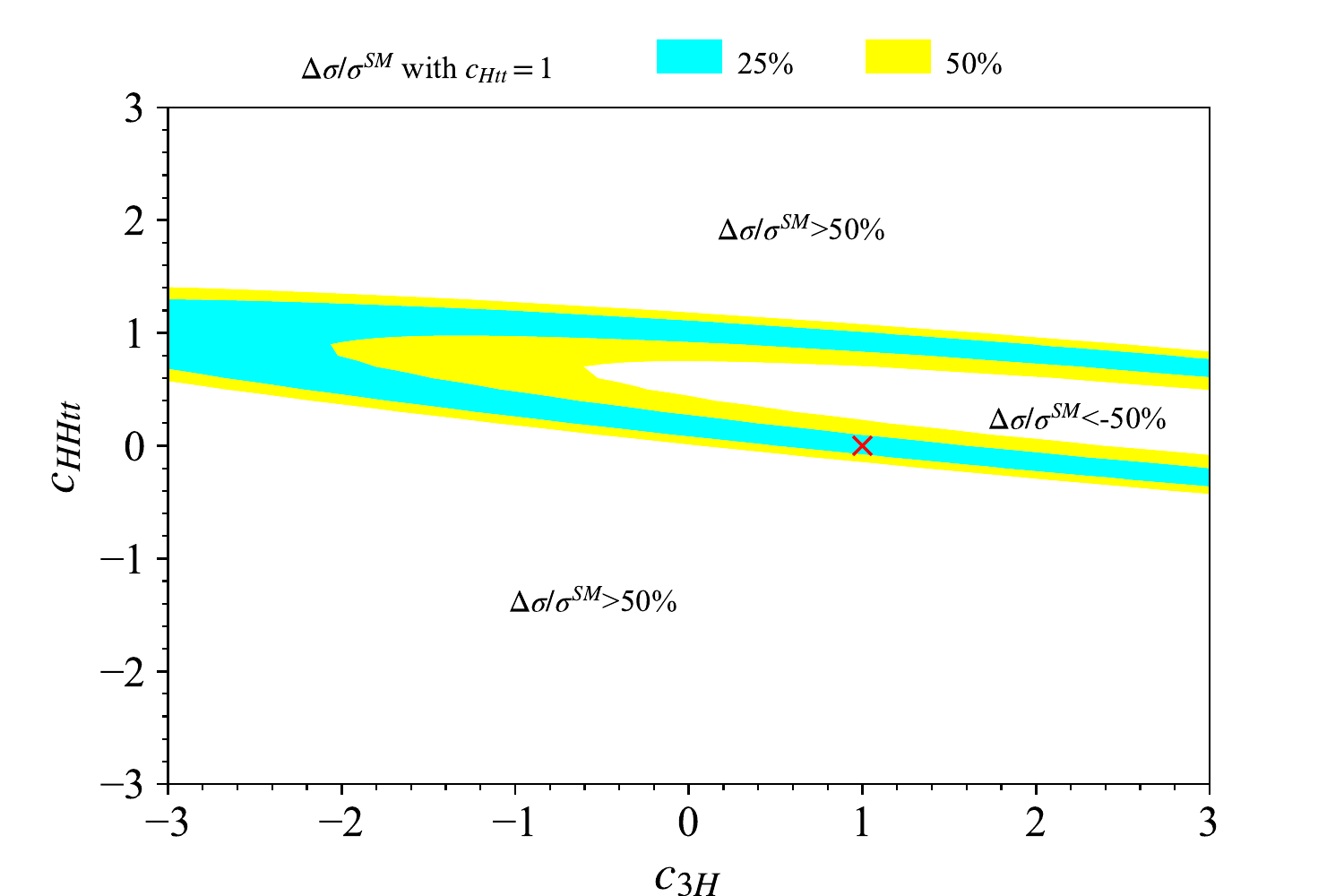}}
\vspace*{0.5cm}
\subfloat[]{\includegraphics[width=.45\linewidth]{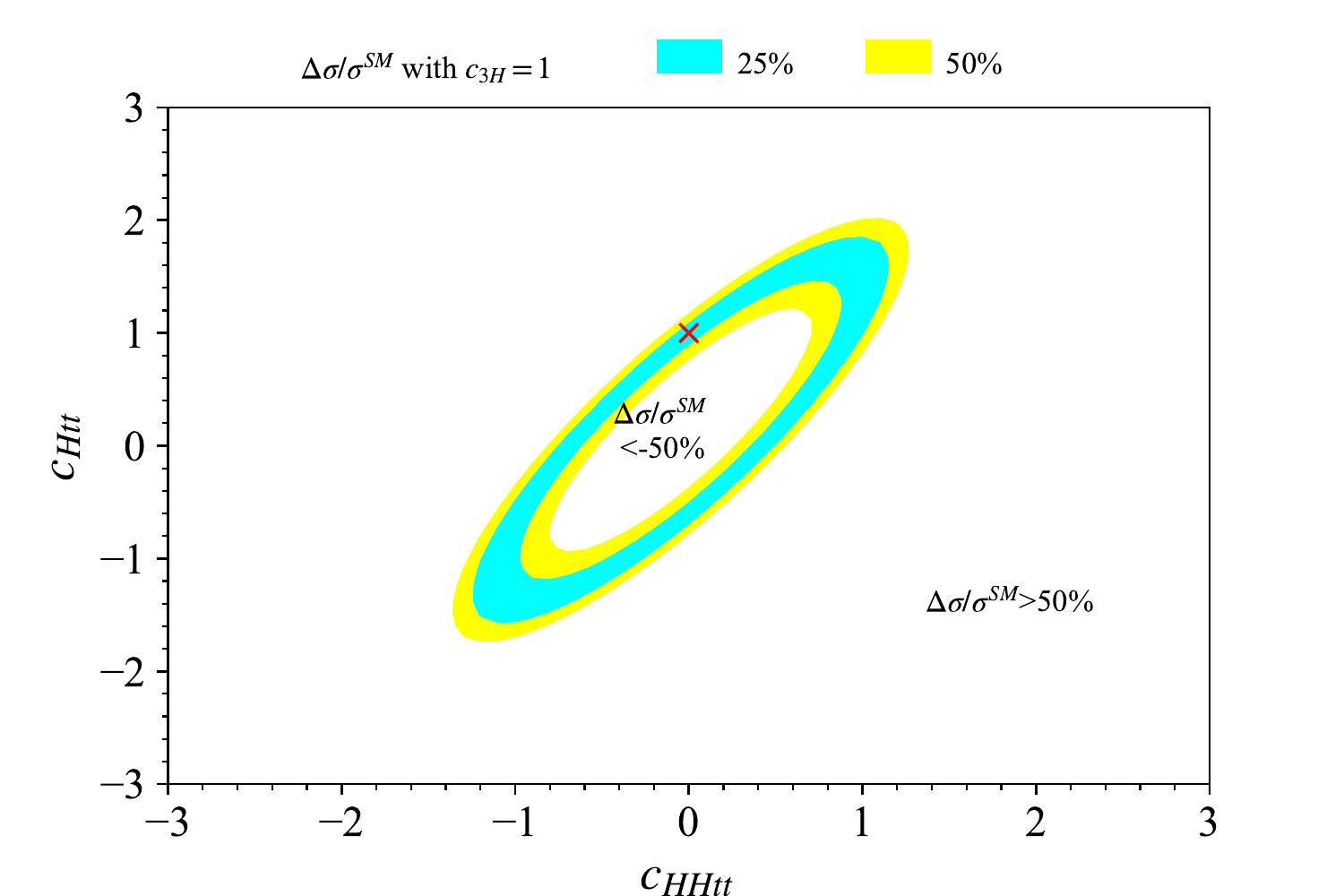}}
\caption{\label{fig:xs_cut}{\em Cross-section contour plot for $gg\to hh\to \gamma\gamma b\bar{b}$ channel after including the veto cuts in Table \ref{table:bbgg_cut}. The parameter space that match the expected SM values within $25\%$ and $50\%$ are indicated by cyan and yellow areas, respectively. The red cross marks the SM value.
}}
\end{figure}

We have discussed that it is possible to discover the Higgs pair production in a $100$ TeV proton-proton collider which was already shown in Ref.~\onlinecite{Baglio:2012np,Yao:2013ika,1506.03302}. Then, we study how the event selections affect the extraction of new physics effects in the Higgs pair production. In what follows, the event selections listed in Table \ref{table:bbgg_cut} was imposed again.
%
For the signal analysis, full simulations are performed for parameters within the range $-3< c_{3H, Htt, HHtt}<3$. Then, as the partonic case, we can fit the number of selected signal events by a similar function shown in Eq.~(\ref{eq:hh_totalrate}).
The contributions of different diagrams would cause different selection efficiency due to the fact that the kinematic distributions are different for each diagram. However, we can still factor out the parameters $c_{Htt}$, $c_{3H}$ and $c_{HHtt}$ during the calculations, and this will again give a simple parameterization:

\bea
\label{eq:hh_totalrate_cut}
\sigma = \sigma^{SM} [3.1265~ c_{HHtt}^2 + 1.5332~ c_{Htt}^2 +0.072904~ c_{3H}^2-3.7322~ c_{Htt} c_{HHtt} \\
- 0.60614~ c_{3H} c_{HHtt} +0.81739~ c_{3H} c_{Htt} ].\nonumber 
\eea 


In Fig.~\ref{fig:xs_cut}, we consider constraints on $c_{Htt}$, $c_{3H}$ and $c_{HHtt}$ from measurements of the total cross-section at CM energy of $100$ TeV with contour lines go along 25\% and 50\% deviations from the SM value. For each plot in Fig.~\ref{fig:xs_cut}, we vary two of $c_{3H},~c_{Htt}$ and $c_{HHtt}$ and fix the rest to the SM value. For example, in Fig.~\ref{fig:xs_cut}(a), $c_{Htt}$ and $c_{3H}$ are allowed to vary within the interval $(-3,3)$, while $c_{HHtt}=0$ as in the SM.

The cyan and yellow bands represent the parameter spaces that match the result of SM within $25\%$ and $50\%$, respectively. 
We see that the sensitivity of cross-section to $c_{3H}$ is low. The same insensitivity remains in Fig.~\ref{fig:xs_cut}(b), where we set $c_{Htt}$ to unity, its SM value.

In Fig.~\ref{fig:xs_cut}(c), where $c_{3H}=1$ takes the SM value and $c_{Htt}$, $c_{HHtt}$ are varying within the interval $(-3,3)$. Moreover, due to the fact that the triangle diagrams interfere with box diagrams destructively, increasing $c_{HHtt}$ can offset any effect of increasing $c_{Htt}$.
Therefore, to explain the various contributions of new physics in double Higgs production, total cross-section alone is not sufficient, and further studies for additional kinematic information are needed.

\begin{figure}[H]
\centering
\subfloat[]{\includegraphics[width=.45\linewidth]{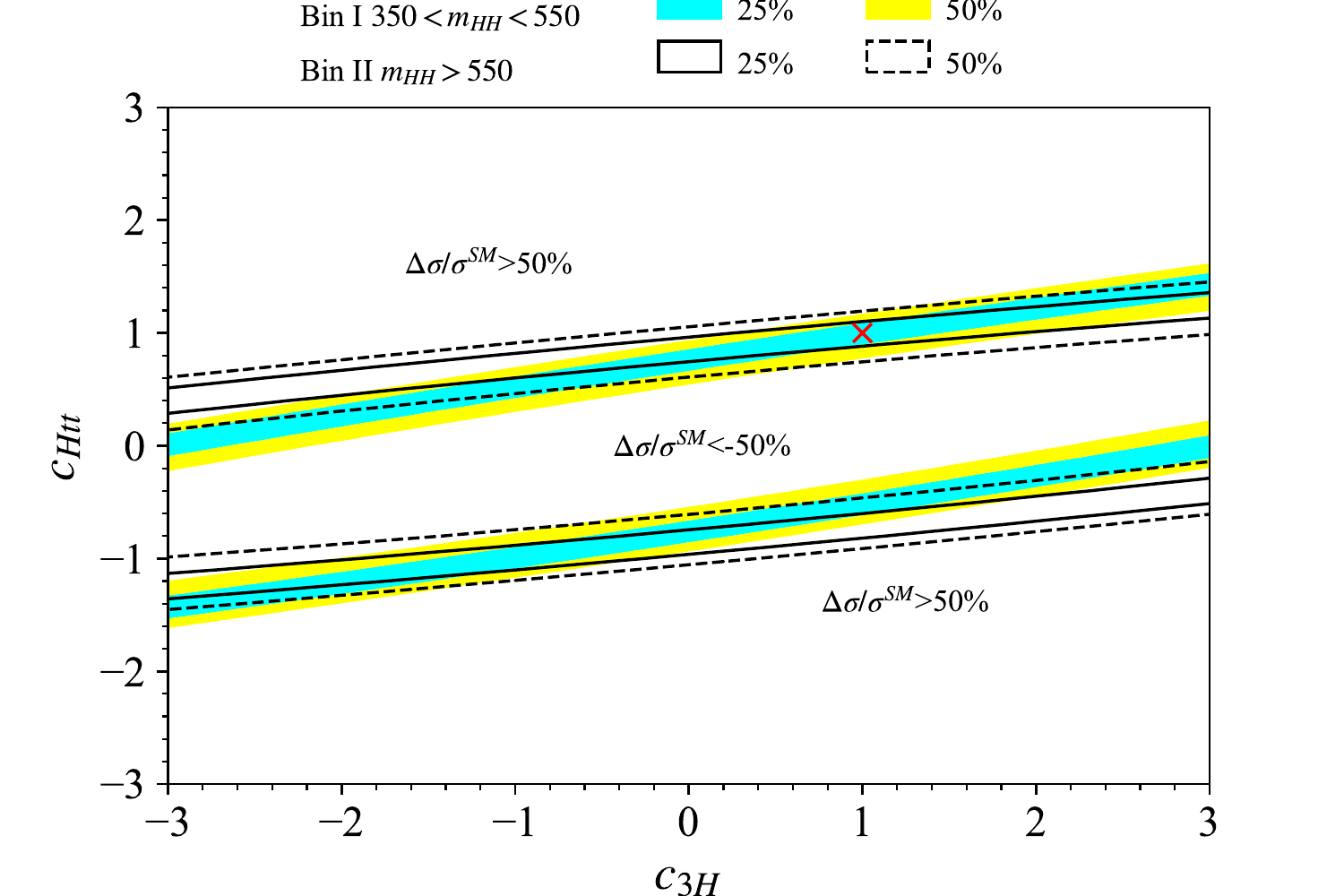}}\qquad
\subfloat[]{\includegraphics[width=.45\linewidth]{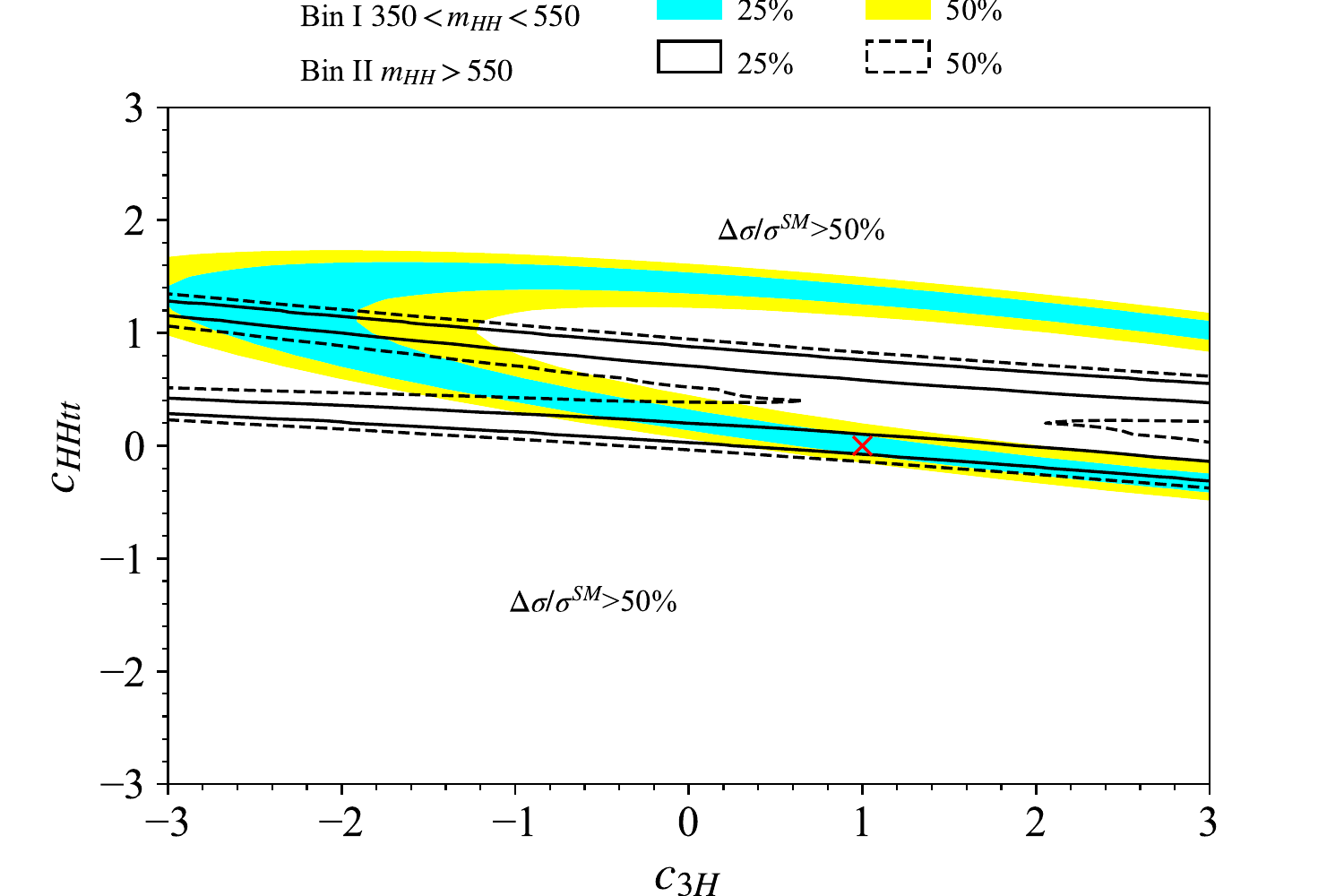}}
\subfloat[]{\includegraphics[width=.45\linewidth]{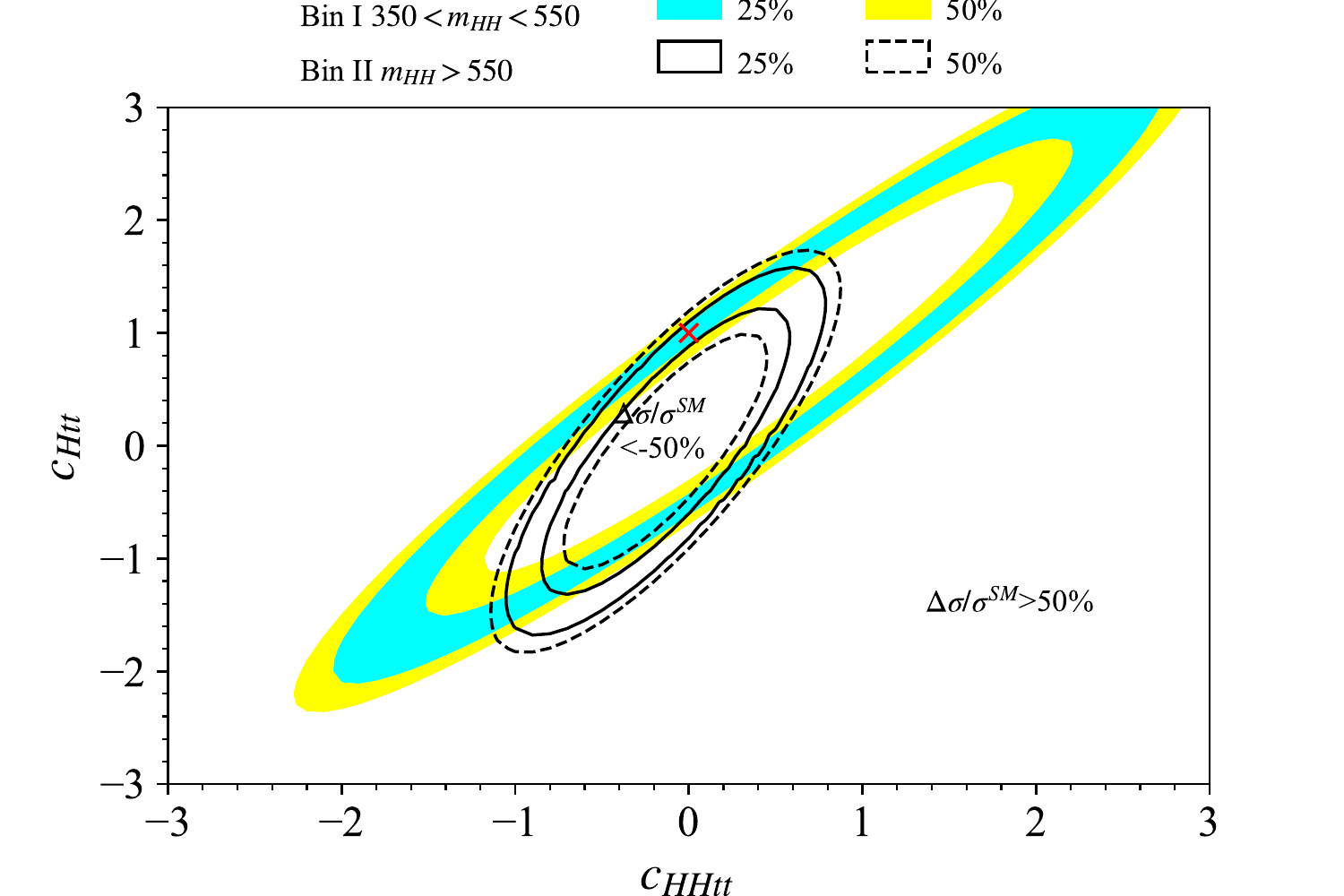}}
\vspace*{0.5cm}

\caption{\label{fig:mhhcut}{\em Contour plots for the cross-sections of a high energy and a low energy bin. Bin I: $350~\textrm{GeV} < m_{HH} < 550~\textrm{GeV}$ and Bin II: $m_{HH}>550~\textrm{GeV}$. The cross-section matching the value of SM within $25\%$ is shown as the cyan region for Bin I and the region between two solid-black curves for Bin II. The cross-section that is within $25-50\%$ of SM expectation is shown as the yellow region for Bin I and the region between solid and dashed curves for Bin II. The red cross mark the SM value.
}}
\end{figure}

As we have already seen, the contributions of $c_{Htt}$, $c_{3H}$ and $c_{HHtt}$ have very different distributions of $p_T^h$ and $m_{HH}$. The $c_{3H}$ component peaks at low $m_{HH}$, the $c_{Htt}$ peaks at a higher $m_{HH}$, and the $c_{HHtt}$ shifts the $m_{HH}$ distribution to even larger values. (See Fig.~\ref{fig:NLO_100TeV_kinematics}). Following the analysis in Ref.~\onlinecite{Chen:2014xra}, we divide the $m_{HH}$ and $p_T$ distributions into a low bin and a high bin, and the differential cross-section in each bin is used to constrain $c_{Htt}$, $c_{3H}$, and $c_{HHtt}$. We note that fitting the two $p_T$ bins and the two $m_{HH}$ bins give quite similar constraints, which are consistent with the results of~\cite{Chen:2014xra}. Therefore, in the following, we only show the contour plots of the constraints from fitting the two $m_{HH}$ bins. 
From Fig.~\ref{fig:NLO_contribution_100TeV_kinematics}, the following two $m_{HH}$ bins are chosen in our analysis.
\bea
\textrm{Bin\ \ I}&:& 350~\textrm{GeV} \le m_{HH} \le 550~\textrm{GeV} \nonumber \\
\textrm{Bin\ II}&:& 550 ~\textrm{GeV} \le m_{HH} \nonumber
\eea
For Bin I and Bin II, the parameterizations of the cross-sections with respect to $c_{Htt}$, $c_{3H}$ and $c_{HHtt}$ are given in Eq.~(\ref{eq:hh_totalrate_cut}).
\bea
\label{eq:hh_totalrate_cut_2}
\sigma_{I} &=& \sigma_{I}^{SM} [2.1837~ c_{HHtt}^2 + 1.6984~ c_{Htt}^2 +0.10647~ c_{3H}^2-3.6533~ c_{Htt} c_{HHtt} \nonumber \\
&&- 0.80491~ c_{3H} c_{HHtt} +0.71334~ c_{3H} c_{Htt} ].\nonumber \\
\sigma_{II} &=& \sigma_{II}^{SM} [4.2030~ c_{HHtt}^2 + 1.3446~ c_{Htt}^2 +0.03458~ c_{3H}^2-3.8224~ c_{Htt} c_{HHtt} \nonumber \\
&&- 0.37915~ c_{3H} c_{HHtt} +0.93621~ c_{3H} c_{Htt} ].
\eea
Fig.~\ref{fig:mhhcut} shows the constraints from the differential cross-section, which lie within 25\% and 50\% of SM expectations in each bin. Again, two of $c_{Htt}$, $c_{3H}$, and $c_{HHtt}$ are allowed to vary, while the other is fixed at the SM value. 
In Fig.~\ref{fig:mhhcut}(a), where $c_{HHtt}$ is fixed while $c_{3H}$ and $c_{Htt}$ are allowed to vary, this set of contours has the largest overlap among all three sets of contours. In Fig.~\ref{fig:mhhcut}(b), where $c_{3H}=1$, we see only a small overlap between the contour from Bin I and Bin II, and the degeneracies in $c_{3H}$ and $c_{HHtt}$ are broken effectively by the measurements in the two bins.
In Fig.~\ref{fig:mhhcut}(c), where $c_{Htt}$ is allowed to vary, along with $c_{HHtt}$, we see the non-overlapping region becomes larger than in Fig.~\ref{fig:mhhcut}(b). However, the change of $c_{Htt}$ from its SM value is expected to be small due to the precise weak interaction measurements already done. Therefore, the relation present in Fig.~\ref{fig:mhhcut}(c) may not be as useful as in Fig.~\ref{fig:mhhcut}(b). 
Our results shown in Fig.~\ref{fig:mhhcut}(a) and Fig.~\ref{fig:mhhcut}(c) are similar to the finding of~\cite{Chen:2014xra} by using LO with Higgs effective theory corrections to calculate the cross-section. For the contour shown in Fig.~\ref{fig:mhhcut}(b), on the other hand, we have a much larger non-overlapping region near SM expectation compare to the finding of~\cite{Chen:2014xra}. Therefore full NLO calculations are required due to the effects of event selections on $gg\to HHg$ and $qg \to HHq$ channels.
Nonetheless, we note that some degeneracy remains when the differential cross-sections in the low and high bins meet the expected SM values. When it comes to constraining $c_{3H}$, the situation worsens. However, a significant improvement in constraining $c_{3H}$ from using the measurement of total cross-section alone can still be achieved by including the kinematic information from both low and high $m_{HH}$ bins.

\section{Conclusion}
\label{chap:conclusion}
In this work, we investigated the use of the kinematic distribution to reveal the new physics effects in the Higgs pair production, including LO channel, $gg\rightarrow HH$, and all NLO channels, $g g \rightarrow HHg$, $q g \rightarrow HHq$, and $q q \rightarrow HHg$. We showed that three dimensionless coefficients, $c_{Htt}$, $c_{3H}$, and $c_{HHtt}$, can be used to parameterize the differential cross-section with various new physics effects. We investigated the interactions of different contributions in the $p_T$ spectra and the invariant mass spectra of the Higgs pair. We then numerically study the constraints of these parameters in a 100 TeV proton-proton collider under planning by finding the best fit for the differential rates in a low and a high $m_{HH}$ as well as $p_T$ bins. The constraints from low and high bins ended up being very similar to those from the two $p_T$ bins. Finally, it was found that we can constrain $c_{Htt}$ and $c_{HHtt}$ effectively, despite some degeneracy persists. Moreover, the coefficient $c_{3H}$, which directly reveals the effect of trilinear Higgs self-coupling, is less constrained. This is roughly consistent with the earlier result in the LO $gg\rightarrow HH$ channel~\cite{Chen:2014xra}, full NLO calculations that more effectively break the degeneracy in $c_{3H}$ and $c_{HHtt}$ are therefore required.

Nonetheless, the extra kinematic information from the two invariant mass bins still gives much better results than the total cross-section alone.

Measurements of the trilinear Higgs coupling should be a top priority in upcoming research programs on the Higgs boson, since only the properties of Higgs self-interaction of the 125 GeV Higgs boson have not been thoroughly tested experimentally. Recent searches~\cite{Owen2021} for pairs of Higgs bosons in $HH \to \gamma\gamma b\bar{b}$ process set a limit on the double Higgs production rate of 4.1 times the SM value and created a portal to better understanding the fundamental Higgs mechanism. 

The work is far from done. Much more work needs to be done in the phenomenology of double Higgs production. The Matrix Element Method based multivariate analysis~\cite{Kondo:1988yd}, which has been applied to the Higgs discovery in the $4\ell$ channel \cite{Gao:2010qx,Gao:2010qx_2,Gao:2010qx_3,Gao:2010qx_4,Gao:2010qx_5,Gao:2010qx_6} and the top quark analyses~\cite{Dalitz:1991wa,Dalitz:1991wa_2,Dalitz:1991wa_3,Dalitz:1991wa_4,Dalitz:1991wa_5} can be performed to exploit the full kinematic information in the future . Also, the recent search for the Higgs pair performed by the ATLAS collaboration, which applied multivariate analysis based on Boosted Decision Trees (BDT) to event selection in the SM process, obtained the best limit for the double Higgs boson production currently. The same technique can be applied to searches for new physics in the future.


\bibliography{citations1}

\end{document}